\documentstyle[pra,aps,amssymb,amsmath,graphicx]{revtex}

\def\c{\cdot}
\def\cn{\;}
\def\vvec{\underline}

\def\Gr{{\mathcal G}}
\def\eps{\varepsilon}
\def\adj{^\dagger}
\def\ktau{(\vec{k},\tau)}
\def\komega{(\vec{k},i\omega{_n})}
\def\kaomega{(\vec{k},\omega)}

\renewcommand\vec[1]{{\mathbf #1}}
\def\mat[#1][#2]{{\vspace{2ex}^{#1}\vspace{-2ex}\underline{\underline{#2}}}}
\def\Gr{{\mathcal G}}
\def\gmat[#1][#2]{\hspace{.3ex}{\frac{}{}}^{#1} #2}
\def\vv[#1][#2][#3]{{\hspace{.1ex}_{#1}^{#2}%
    \underline{#3}}}
\def\kink{e_{\vec{k}}}

\def\Im{ {\mathfrak Im} }
\def\arctan{ {\mathrm arctan} }

\begin{document}
\draft \title{The structure of the perturbation series of the spin-1
  Bose gas at low temperatures}
\author{P\'eter Sz\'epfalusy$^{(1,2)}$ and Gergely Szirmai$^{(1)}$}
\address{$^{(1)}$Department of Physics of Complex Systems, Roland E\"otv\"os
University, P\'azm\'any P\'eter s\'et\'any 1/A, Budapest, H-1117,
\\ $^{(2)}$Research Institute for Solid State Physics and Optics of the
Hungarian Academy of Sciences, Budapest, P.O.Box 49, H-1525}
\date\today

\maketitle

\begin{abstract}
  The properties of Green's functions and various correlation
  functions of density and spin operators are considered in a
  homogeneous spin-1 Bose gas in different phases. The dielectric
  formalism is worked out and the partial coincidence of the
  one-particle and collective spectra is pointed out below the
  temperature of Bose-Einstein condensation. As an application the
  formalism is used to give two approximations for the propagators and
  the correlation functions and the spectra of excitations including
  shifts and widths due to the thermal cloud.
\end{abstract}

\pacs{PACS numbers: 03.75.Fi, 67.40.Db, 05.30.Jp}

\section{Introduction}
\label{sec:intro}

The recent realization of BEC in an optical trap has opened a new area
of research \cite{Stamper-Kurn2001a,SKea,Sea1,Sea2,Miesea}. The
theoretical investigation of spinor systems is growing rapidly these
days concerning the ground state structure and symmetry braking
\cite{HG1,LPB,HYip,CH}, the various phases in traps \cite{IMO1,HG2}, the
different vortex states \cite{Yip1,SVK}, the exploration of collective
excitations \cite{Ho2,OM,MS1,Zhou2} and a number of other problems
\cite{KU,GM,Puea1,Puea2,CYH,HYin,Ueda,Zhou1,RZOK,ABK}. Compared to the BEC
realized in magnetic traps \cite{Aea2,Dea}, where the spin degree of
freedom of the particle field is frozen, such systems, since the
spinor nature of the particles is preserved, have a wider variety of
excitations including spin density waves, transverse spin waves or
quadrupolar spin waves.

It is a well established property of scalar Bose-Einstein condensed
systems that the one-particle and the density correlation function
spectra coincide. To treat this problem consistently the dielectric
formalism has proved to be particularly useful, which has been worked
out first for homogeneous systems \cite{Griffin,SzK,MW,KSz,PG,WG} and
recently generalized and applied to trap systems
\cite{BSz,Rea1,Rea2,FRSzG}. In this paper this formalism is extended
to gases with a spinor Bose-Einstein condensate.  To make the
presentation more transparent only homogeneous systems will be
considered. The new feature is the appearance of correlation functions
including spin fluctuations besides the density autocorrelation
function and a variety of new one-particle Green's functions. Their
perturbation series are analysed simultaneously.  After suitable
rearrangement of the expansions it has been achieved that certain
one-particle Green's functions and correlation functions have common
denominators leading to the coincidence of their spectra, though with
different spectral weights. Examples are the Green's functions
corresponding to a spin transfer zero and the density correlation
function, furthermore those Green's functions and correlation
functions of the spin operators, which can be characterized by spin
transfers $+1$ or $-1$. Exceptions are found in the polar phase
(occurring when the interaction in the spin channel is repulsive),
namely those correlation functions of the spin operators which create
zero or $\pm2$ spin transfers do not possess condensate induced
intermixing with the one-particle Green's functions.The general theory
is illustrated in the Bogoliubov theory (valid at very low
temperatures) and in the random phase approximation (RPA). Within the
RPA we do not include exchange processes, that already in the case of
the scalar Bose-Einstein condensed systems has led to involved
calculations \cite{FRSzG}. Their extension to spinor Bose-Einstein gas
will be presented in a separate paper. As a matter of fact the
relative importance of the exchange processes is less in the present
case since the direct terms are enlarged by summations over spin
variables.

Though the RPA (as a mean field theory) looses its validity near the
phase transition it is enlightening to investigate it in the
transition region. It turns out that, while the transition to the
polar phase is continuous when decreasing the temperature, it becomes
weakly first order for ferromagnetic ordering due to the small (for
this coupling attractive) spin dependent part of the interaction. It
is worth recalling that in case of the scalar Bose-Einstein
condensation the transition is second order in this RPA-Hartree model,
but including exchange contributions makes it to a first order one and
the effect is then not small, see Ref. \cite{FRSzG} and references
therein.

The paper is organized as follows. In the second section, after giving
the specifications of the Hamiltonian and introducing the canonical
transformation to induce the Bose-Einstein condensation, the normal
and anomalous one-particle Green's functions (matrices in the spin
variables) and the different correlation functions of the particle
number density and spin density operators are defined. Section
\ref{sec:formalism} is the backbone of the paper. It starts with a
summary of symmetry properties of the functions introduced in Section
\ref{sec:based}. Then their general structure is analyzed within the
framework of perturbation expansion. We proceed by classifying the
processes according to the spin transfers involved. Then in the spirit
of the dielectric formalism the proper irreducible graphs are
separated. The main results are the description of condensate induced
intermixing of one-particle and collective modes and the specification
of the conditions when it occurs. Sections \ref{sec:bogoapp} and
\ref{sec:RPA} are devoted to approximate calculations. In Section
\ref{sec:RPA} corrections consisting of damping terms and frequency
shifts as well are given in RPA to the Bogoliubov approximation
discussed in Section \ref{sec:bogoapp}. We present in Section
\ref{sec:RPA} also results for static properties in RPA. Section
\ref{sec:conc} contains the summary and further discussion.

\section{Basic equations and definitions}
\label{sec:based}

\subsection{The effective Hamiltonian of the spin-1 Bose gas}
\label{sec:effHam}

In a gas containing spin-1 particles the wavefunction is a
three-component spinor. To be more specific we use the basis set of
$\widehat{F}_z$ eigenvectors in the spin space, which leads to the
wavefunction
\begin{equation}
  \label{eq:wavefunction}
  \psi(\vec{r})=\left(
    \begin{array}{c}
      \psi_+(\vec{r})\\
      \psi_0(\vec{r})\\
      \psi_-(\vec{r})
    \end{array}\right).
\end{equation}
In this representation the spin operators are
\begin{equation}
  \label{eq:spinops}
  F_x=\frac{1}{\sqrt{2}}\left [ \begin{array}{c c c}
      0&1&0\\
      1&0&1\\
      0&1&0
    \end{array}\right ], \qquad
  F_y=\frac{1}{\sqrt{2}}\left [ \begin{array}{c c c}
      0&-i&0\\
      i&0&-i\\
      0&i&0
    \end{array}\right ], \qquad
  F_z=\left [ \begin{array}{c c c}
      1&0&0\\
      0&0&0\\
      0&0&-1
    \end{array}\right ],
\end{equation}
and the corresponding raising and lowering operators:
\begin{equation}
  \label{eq:raisop}
    F_+=\sqrt{2}\left [ \begin{array}{c c c}
      0&1&0\\
      0&0&1\\
      0&0&0
    \end{array}\right ],\quad
  F_-=\sqrt{2}\left [ \begin{array}{c c c}
      0&0&0\\
      1&0&0\\
      0&1&0
    \end{array}\right ].
\end{equation}

We consider a system of spin-1 particles in a box with periodic
boundary conditions and without an external potential. The interaction
between the particles is an s-wave scattering. Such a rotationally invariant
scattering interaction can be described by a potential
\begin{equation}
  \label{eq:pot2}
  V(\vec{r}_1-\vec{r}_2)=\delta(\vec{r}_1-\vec{r}_2)
  [c_n+c_s\vec{F}_1\c\vec{F}_2],
\end{equation}
where the constants $c_n$ and $c_s$ is connected to the s-wave
scattering lengths ($g_0$ and $g_2$) in the total hyperfine spin
channel zero and two with $c_n=(g_0+2g_2)/3$ and $c_s=(g_2-g_0)/3$
\cite{Ho2}.

In the second quantized formalism we introduce annihilation and
creation operators $a_r(\vec{k})$ and $a_r^\dagger(\vec{k})$ which
destroy and create one particle states of plane waves with momentum
$\vec{k}$ and spin projection $r$. These operators are bosonic in our case, 
so they satisfy the commutation relations:
\begin{subequations}
  \label{eqs:commrel}
  \begin{align}
    [a_r(\vec{k}), a_s^\dagger(\vec{k'})]=\delta_{r,s}\delta
    (\vec{k}-\vec{k'}),\label{eq:commrel1}\\
    [a_r(\vec{k}), a_s(\vec{k'})]=[a_r^\dagger(\vec{k}), a_s^\dagger
    (\vec{k'})]=0.\label{eq:commrel2}
  \end{align}
\end{subequations}
In this formalism the grand canonical Hamiltonian of the system reads
as:
\begin{equation}\label{eq:ham1}
  {\mathcal H}=\sum_{\vec{k}}(e_{\vec{k}}-\mu)a_r^\dagger(\vec{k})
  a_r(\vec{k})+\frac{1}{2}\mathop{\sum_{\vec{k}_1+\vec{k}_2=}}_{=\vec{k}_3
    +\vec{k}_4}a^\dagger_{r'}(\vec{k}_1)a^\dagger_r(\vec{k}_2)V^{r's'}_{rs}
  a_s(\vec{k}_3)a_{s'}(\vec{k}_4),
\end{equation}
where $e_{\vec{k}}=\hslash^2k^2/2M$ stands for the kinetic energy of a particle
and $\mu$ is the chemical potential of the system and
\begin{equation}\label{eq:pot3}
  V^{r's'}_{rs}=c_n\delta_{rs}\delta_{r's'}+c_s(\vec{F})_{rs}(\vec{F})_{r's'}
\end{equation}
according to \eqref{eq:pot2}. Here and in the following the convention
of summing over repeated indices is applied except when stated
otherwise.

There are two types of systems depending on the sign of $c_s$.
If it is less than zero then the spins prefer parallel alignment which
leads to a macroscopic magnetization in the presence of a
Bose-Einstein condensate. If $c_s$ is greater than zero then the
energetically favorable state is when $\left<\vec{F}\right>=0$. The
former case is called {\it ferromagnetic case} and later is called
{\it polar case} \cite{Ho2}.

\subsection{Description of the symmetry breaking}
\label{sec:symbrk}

In the Bose-Einstein condensed phases some field operators have
anomalous averages reflecting a broken gauge symmetry: $\left< a_r (0)
\right> = \sqrt{N_0} \zeta_r$ and $\left< a_r \adj(0) \right> =
\sqrt{N_0} \zeta_r\adj$ where $N_0$ is the number of particles in the
condensate and $\zeta_r$ is the normalized spinor of the condensate
\cite{Ho2}.  For the \textit{polar case} one can take $\zeta_r$ as
$(0,1,0)^T$ and for the \textit{ferromagnetic case} $\zeta_r$ can be
taken as $(1,0,0)^T$ where the superscript $T$ denotes the operation
of transposition.  The averaging is made over a symmetry breaking
ensemble.  To consider this symmetry breaking, one can introduce a new
set of bosonic annihilation and creation operators with a canonical
transformation:
\begin{subequations}\label{eqs:cantr}
  \begin{align}
    b_r(\vec{k})=a_r(\vec{k})-\delta_{\vec{k},0}
    \sqrt{N_0}\zeta_r,\label{eq:cantr1}\\
    b_r^\dagger(\vec{k})=a_r^\dagger(\vec{k})-\delta_{\vec{k},0}
    \sqrt{N_0}\zeta_r\adj\label{eq:cantr2}.
  \end{align}
\end{subequations}
The relation between the chemical potential and condensate density can
be derived from the requirement that $\left< b_r (\vec{k}) \right> =
\left< b_r\adj(\vec{k})\right>=0$ Here and from now on the averages
will be made over the grand canonical ensemble with a density matrix
$\rho={\mathcal Z}^{-1}e^{\beta {\mathcal H}}$ where ${\mathcal
  Z}=\mathrm{Tr}\ e^{\beta {\mathcal H}}$ is the grand canonical
partition function. Substituting the canonical transformation into
\eqref{eq:ham1} one obtains the Hamiltonian in terms of the new
operators:
\begin{equation}\label{eq:ham2}
  \begin{split}
    {\mathcal H}=\sum_{\vec{k}}(e_{\vec{k}}-\mu_0)b^\dagger_r(\vec{k})
    b_r(\vec{k})-\mu\sqrt{N_0}\big[\zeta_r\adj b_r(0)+b_r\adj(0)\zeta_r\big]
    -\mu N_0
    +\frac{1}{2}\sum b^\dagger_{r'}(\vec{k}_1)b^\dagger_r(\vec{k}_2)
    V^{r's'}_{rs}b_s(\vec{k}_3)b_{s'}(\vec{k}_4)\\
    +\frac{\sqrt{N_0}}{2}\sum b^\dagger_{r'}(\vec{k}_1)b^\dagger_r(\vec{k}_2)
    V^{r's'}_{rs} b_s(\vec{k}_3)\zeta_{s'}\delta_{\vec{k}_4,0}+
    \frac{\sqrt{N_0}}{2}\sum b^\dagger_{r'}(\vec{k}_1)b^\dagger_r(\vec{k}_2)
    V^{r's'}_{rs}\zeta_{s}\delta_{\vec{k}_3,0}b_{s'}(\vec{k}_4)\\
    +\frac{\sqrt{N_0}}{2}\sum b^\dagger_{r'}(\vec{k}_1)\zeta_r\adj
    \delta_{\vec{k}_2,0}V^{r's'}_{rs}b_s(\vec{k}_3)b_{s'}(\vec{k}_4)
    +\frac{\sqrt{N_0}}{2}\sum\zeta_{r'}\adj\delta_{\vec{k}_1,0}b^\dagger_r
    (\vec{k}_2)V^{r's'}_{rs} b_s(\vec{k}_3)b_{s'}(\vec{k}_4)\\
    +\frac{N_0}{2}\sum b^\dagger_{r'}(\vec{k}_1)b^\dagger_r(\vec{k}_2)
    V^{r's'}_{rs}\zeta_s\delta_{\vec{k}_3,0}\zeta_{s'}\delta_{\vec{k}_4,0}
    +\frac{N_0}{2}\sum b^\dagger_{r'}(\vec{k}_1)\zeta_r\adj\delta_{\vec{k}_2,0}
    V^{r's'}_{rs}b_s(\vec{k}_3)\zeta_{s'}\delta_{\vec{k}_4,0}\\
    +\frac{N_0}{2}\sum\zeta_{r'}\adj\delta_{\vec{k}_1,0}b^\dagger_s(\vec{k}_2)
    V^{r's'}_{rs}b_s(\vec{k}_3)\zeta_{s'}\delta_{\vec{k}_4,0}
    +\frac{N_0}{2}\sum b^\dagger_{r'}(\vec{k}_1)\zeta_r\adj\delta_{\vec{k}_2,0}
    V^{r's'}_{rs}\zeta_{s}\delta_{\vec{k}_3,0}b_{s'}(\vec{k}_4)\\
    +\frac{N_0}{2}\sum\zeta_{r'}\adj\delta_{\vec{k}_1,0}b^\dagger_{r}
    (\vec{k}_2)V^{r's'}_{rs}\zeta_{s}\delta_{\vec{k}_3,0}b_{s'}(\vec{k}_4)
    +\frac{N_0}{2}\sum\zeta_{r'}\adj\delta_{\vec{k}_1,0}\zeta_r\adj
    \delta_{\vec{k}_2,0}V^{r's'}_{rs}b_s(\vec{k}_3)b_{s'}(\vec{k}_4)\\
    +\frac{N_0^{3/2}}{2}\zeta_{r'}\adj\zeta_r\adj V^{r's'}_{rs}\zeta_s
    b_{s'}(0)+\frac{N_0^{3/2}}{2}\zeta_{r'}\adj\zeta_r\adj V^{r's'}_{rs}
    b_{s}(0)\zeta_{s'}+\frac{N_0^{3/2}}{2}\zeta_{r'}\adj b^\dagger_r(0)
    V^{r's'}_{rs}\zeta_s\zeta_{s'}\\
    +\frac{N_0^{3/2}}{2}b^\dagger_{r'}(0)\zeta_{r}\adj V^{r's'}_{rs}\zeta_s
    \zeta_{s'}+\frac{N_0^2}{2}\zeta_{r'}\adj\zeta_{r}\adj V^{r's'}_{rs}
    \zeta_{s}\zeta_{s'}+\sum_{\vec{k}}(\mu_0-\mu)b^\dagger_r(\vec{k})
    b_r(\vec{k}).
  \end{split}
\end{equation}
Here we added and subtracted the term $\sum_{\vec{k}} \mu_0
b^\dagger_r(\vec{k}) b_r(\vec{k})\quad (\mu_0\leq0)$ to avoid the
difficulty of being $\mu$ positive in the condensed phases, which
would lead to a singularity in the unperturbed propagator.

It should be stressed that with the canonical transformation
\eqref{eqs:cantr} one defines bare quasiparticle states $\prod_i
b_{r_i}\adj (\vec{k}_i) \left|0\right>$. The time evolution of these
quasiparticle states are determined by the Hamiltonian \eqref{eq:ham2}
which contains terms with different number of creation and destruction
operators leading to the nonconservation of the total number of
quasiparticles.

\subsection{Green's functions and correlation functions}
\label{sec:definitions}

First we define the Green's functions as
\begin{equation}
  \label{eq:grdef}
      \Gr^{rs}_{\gamma\delta}\ktau=-\Big<T_\tau\big[b^\gamma_r\ktau
    b^{\delta\adj}_s(\vec{k},0)\big]\Big>,
\end{equation}
with
\begin{equation}
  \label{eq:grop}
      b^\gamma_r(\vec{k})=\left\{
      \begin{array}{l l}
        b_r(\vec{k}) & ,\gamma=1\\
        b_r\adj(-\vec{k}) & ,\gamma=-1.
      \end{array}\right.
\end{equation}
The Greek indices introduced here are for distinguishing between the
normal and anomalous Green's functions. Automatic summation over
repeated Greek indices is understood as for the Roman ones. Here and
from now on $\tau$ is the imaginary time and $T_\tau$ is the $\tau$
ordering operator \cite{FW}. In this symmetry breaking system because
of the fact that the Hamiltonian Eq. \eqref{eq:ham2} contains terms
with two creation or two annihilation operators anomalous Green's
functions arise (with $\gamma\delta=-1$). The expression
\eqref{eq:grdef} is the generalization of the well known normal and
anomalous Green's functions in case of a complex scalar field
\cite{FW,Griffin}. The propagators \eqref{eq:grdef} are periodic in
$\tau$ with period $\beta\hslash$ so their Fourier series can be
defined as
\begin{equation}
  \label{eq:grfour}
      \Gr^{rs}_{\gamma\delta}\komega=\frac{1}{\beta\hslash}\int_0^{\beta
        \hslash}d\tau e^{i\omega_n\tau}\Gr^{rs}_{\gamma\delta}\ktau,
\end{equation}
where $\omega_n=2n\pi/\beta\hslash$ is the Bose discrete Matsubara
frequency. The spectrum of the one-particle elementary excitations can
be determined as poles of the analytic continuations of the Green's
functions.

The collective excitations in the system can be described with the
correlation functions of the following operators:
\begin{subequations}
  \label{eqs:densops}
  \begin{align}
    n(\vec{k})&=\sum_{\vec{q}}a^\dagger_r(\vec{k}+\vec{q})a_r(\vec{q}),\\ 
   {\mathcal{F}}_z(\vec{k})&=\sum_{\vec{q}}a^\dagger_r(\vec{k}+\vec{q})
   (F_z)_{rs}a_s(\vec{q}),\\
   {\mathcal{F}}_\pm(\vec{k})&=\sum_{\vec{q}}a^\dagger_r(
   \vec{k}+\vec{q})(F_\pm)_{rs}a_s(\vec{q}),\\
   {\mathcal{F}}^Q_\pm(\vec{k})&=\sum_{\vec{q}}a^\dagger_r(\vec{k}+
   \vec{q})(F_\pm^2)_{rs}a_s(\vec{q}),\\
   \sigma_{rs}(\vec{k})&=\sum_{\vec{q}}a\adj_r(\vec{k}+\vec{q})a_s(\vec{q}).
  \end{align}
\end{subequations}
Here $n(\vec{k})$ is the particle density operator ${\mathcal
  F}_z(\vec{k})$ is the spin z component density operator, ${\mathcal
  F}_\pm(\vec{k})$ is the density operator of the $\pm 1$ spin raising
or lowering operator and ${\mathcal F}^\pm_Q(\vec{k})$ is the density
of the $\pm 2$ spin raising or lowering operators and at last
$\sigma_{rs}(\vec{k})$ is the general density operator from which the
other ones can be easily calculated as e.g.
$n(\vec{k})=\sigma_{++}(\vec{k})+\sigma_{00}(\vec{k})+\sigma_{--}(\vec{k})$
or ${\mathcal
  F}_z(\vec{k})=\sigma_{++}(\vec{k})-\sigma_{--}(\vec{k})$. The
different collective excitations can be found as poles of the
analytical continuations of the corresponding correlation functions
(defined for $\vec{k}\neq0$), given by
\begin{subequations}
  \label{eqs:dc}
  \begin{align}
    &D_{nn}\ktau=-\Big< T_\tau\big[ n\ktau n\adj(\vec{k}, 0)
    \big]\Big>,\label{eq:d1}\\
    &D_{zz}\ktau=-\Big< T_\tau\big[ {\mathcal{F}}_z\ktau
    {\mathcal{F}}_z\adj(\vec{k}, 0)\big]\Big>,\label{eq:d2}\\
    &D_{nz}\ktau=-\Big< T_\tau\big[ n\ktau{\mathcal
      {F}}_z\adj(\vec{k}, 0)\big]\Big>,\label{eq:d3}\\
    &D_{\pm\pm}\ktau=-\Big< T_\tau\big[ {\mathcal{F}}_\pm
    \ktau{\mathcal{F}}_\pm\adj(\vec{k},0)\big]\Big>,\label{eq:d4}\\
    &D^Q_{\pm\pm}\ktau=-\Big< T_\tau\big[ {\mathcal{F}}^Q_\pm
    \ktau{\mathcal{F}}^{Q\dagger}_\pm(\vec{k},0)\big]\Big>.\label{eq:d5}
  \end{align}
  Note that $n\adj(\vec{k})=n(-\vec{k})$, ${\mathcal{F} }_z\adj
  (\vec{k}) = {\mathcal{F}}_z(-\vec{k})$, ${\mathcal{F} }_\pm\adj
  (\vec{k}) = {\mathcal{F}}_\mp(-\vec{k})$ and ${\mathcal{F} }^{Q
    \dagger}_\pm (\vec{k}) = {\mathcal{F}}^Q_\mp(-\vec{k})$. A general
  correlation function can also be defined as
  \begin{equation}\label{eq:d6}
    D^{sr}_{r's'}\ktau=-\Big<T_\tau\big[\sigma_{rs}\ktau
    \sigma_{s'r'}(-\vec{k},0)\big]\Big>,
  \end{equation}
\end{subequations}
from what the above ones can be calculated as:
\begin{subequations}
  \label{eq:propkif}
  \begin{align}
    &D_{nn}\ktau=\sum_{r,s}D^{rr}_{ss}\ktau,\label{propkifnn}\\
    &D_{zz}\ktau=\sum_{r,s}rsD^{rr}_{ss}\ktau,\label{propkifzz}\\
    &D_{nz}\ktau=\sum_{r,s}sD^{rr}_{ss}\ktau,\label{propkifnz}\\
    &D_{++}\ktau=2\left[D^{0+}_{+0}\ktau+D^{0+}_{0-}\ktau+D^{-0}_{+0}
      \ktau+D^{-0}_{0-}\ktau\right],\label{propkif++}\\
    &D_{--}\ktau=2\left[D^{0-}_{-0}\ktau+D^{0-}_{0+}\ktau+D^{+0}_{-0}
      \ktau+D^{+0}_{0+}\ktau\right],\label{propkif--}\\
    &D^Q_{++}\ktau=4D^{-+}_{+-}\ktau,\label{propkif2++}\\
    &D^Q_{--}\ktau=4D^{+-}_{-+}\ktau.\label{propkif2--}
  \end{align}
\end{subequations}

All of these correlation functions are also periodic in $\tau$ with
period $\beta\hslash$, so their Fourier transforms can be defined
analogously to Eq.  \eqref{eq:grfour} and the appearing Matsubara
frequencies are also the same as for the Green's functions.

We will also apply anomalous correlation functions
\begin{subequations}
  \label{eqs:anom}
  \begin{align}
    &A^{sr}_{a\alpha}\ktau=-\Big<T_\tau\big[\sigma_{rs}\ktau
    b^{\alpha\adj}_a(\vec{k},0)\big]\Big>,\label{eq:anom1}\\
    &A^{a\alpha}_{r's'}\ktau=-\Big<T_\tau\big[b^{\alpha}_a\ktau
    \sigma_{s'r'}(-\vec{k},0)\big]\Big>,\label{eq:anom2}
  \end{align}  
\end{subequations}
which have zero value in the symmetric phase (i.e. for a noncondensed
system).

Though the one-particle Green's functions are the autocorrelation
functions of the order parameter field operator, we preserve in the
following the term correlation function for the other correlation
functions introduced above.

\section{General formalism}
\label{sec:formalism}

In this section the general formalism is worked out. First symmetry
properties of the functions defined in the previous section will be
discussed. Next perturbation theory will be presented and then we will
turn our attention to relations valid in all orders of perturbation
theory.

\subsection{Symmetry properties}
\label{sec:symprop}

Due to the rotational symmetry in the coordinate space the Green's
functions \eqref{eq:grdef} and all the correlation functions
\eqref{eqs:dc} and \eqref{eqs:anom} depend only on the modulus of the
momentum.

For the Green's functions the following symmetry properties can be
derived. Since (in absence of a magnetic filed) the Green's functions
\eqref{eq:grdef} are real and the cyclic property of trace
\begin{equation}\label{eq:sym1}
  \begin{split}
    \Gr^{rs}_{\gamma\delta}\ktau=-\Big<T_\tau\big[b^\gamma_r\ktau
    b^{\delta
      \adj}_s(\vec{k},0)\big]\Big>=-\Big<T_\tau\big[b^\gamma_r\ktau
    b^{\delta
      \adj}_s(\vec{k},0)\big]\adj\Big>\\
    =-\Big<T_\tau\big[b^\delta_s\ktau
    b^{\gamma\adj}_r(\vec{k},0)\big]\Big>=
    \Gr^{sr}_{\delta\gamma}\ktau
  \end{split}
\end{equation}
holds. Furthermore the time displacement symmetry of Hamiltonian
systems, the $b^\gamma_r (\vec{k}) = {b^{-\gamma}_r}\adj (-\vec{k})$
relation and that after $T_\tau$ the order of the bosonic operators
are irrelevant lead to
\begin{equation}\label{eq:sym2}
  \begin{split}
    \Gr^{rs}_{\gamma\delta}\ktau=-\Big<T_\tau\big[b^\gamma_r\ktau
    b^{\delta
      \adj}_s(\vec{k},0)\big]\Big>=-\Big<T_\tau\big[b^{\delta\adj}_s(
    \vec{k},0)b^\gamma_r\ktau
    \big]\Big>\\
    =-\Big<T_\tau\big[b^{\delta\adj}_s(\vec{k},-\tau
    ) b^\gamma_r(\vec{k},0) \big]\Big>=
    -\Big<T_\tau\big[b^{-\delta}_s(-\vec{k},-\tau)
    b^{-\gamma\adj}_r(-\vec{k},0)
    \big]\Big>\\
    =\Gr^{sr}_{-\delta,-\gamma}(-\vec{k}, -\tau).
  \end{split}
\end{equation}
On the basis of the above equalities one finds
\begin{equation}\label{eq:sym3}
  \Gr^{rs}_{\gamma\delta}\ktau=\Gr^{rs}_{\gamma\delta}(k,\tau)=
  \Gr^{sr}_{\delta\gamma}(k,\tau)=\Gr^{sr}_{-\delta,-\gamma}(k,-\tau)
  =\Gr^{rs}_{-\gamma,-\delta}(k,-\tau).
\end{equation}

For the generalized density correlation functions of Eq.
\eqref{eq:d6} the following symmetry relations can be derived:
\begin{equation}
  \label{eq:ds1}
  \begin{split}
    -D^{sr}_{r's'}\ktau=\Big<T_\tau\big[\sigma_{rs}\ktau\sigma_{s'r'}
    (\vec{-k},0)\big]\adj\Big>=\Big<T_\tau\big[\sigma\adj_{s'r'}(-\vec{k},0)
    \sigma\adj_{rs}(\vec{k},-\tau)\big]\Big>\\
    =\Big<T_\tau\big[\sigma_{r's'}\ktau\sigma_{sr}(\vec{-k},0)\big]\Big>=
    -D^{s'r'}_{rs}\ktau
  \end{split}
\end{equation}
and
\begin{equation}\label{eq:ds2}
  \begin{split}
    -D^{sr}_{r's'}\ktau=\Big<T_\tau\big[\sigma_{rs}\ktau\sigma_{s'r'}
    (\vec{-k},0)\big]\Big>=\Big<T_\tau\big[\sigma_{s'r'}(\vec{-k},0)
    \sigma_{rs}\ktau\big]\Big>\\
    =\Big<T_\tau\big[\sigma_{s'r'}(-\vec{k},-\tau)\sigma_{rs}
    (\vec{k},0)\big]\Big>=-D^{r's'}_{sr}(-\vec{k},-\tau),
  \end{split}
\end{equation}
where we further used that $\sigma_{rs}\adj (\vec{k}) = \sigma_{sr}
(-\vec{k})$.  Combining Eqs. \eqref{eq:ds1} and \eqref{eq:ds2}
together one arrives at
\begin{equation}
  \label{eq:ds3}
  D^{sr}_{r's'}\ktau=D^{sr}_{r's'}(k,\tau)=D^{s'r'}_{rs}(k,\tau)=
  D^{r's'}_{sr}(k,-\tau)=D^{rs}_{s'r'}(k,-\tau).
\end{equation}

For the anomalous correlation functions the two symmetry properties
can be derived in the same way. The first one cames form the fact that
the expectation values here are also real
\begin{equation}
  \label{eq:ansym1}
    A^{sr}_{a\alpha}\ktau=-\Big<T_\tau\big[\sigma_{rs}\ktau
    b^{\alpha\adj}_a(\vec{k},0)\big]\adj\Big>=-\Big<T_\tau\big[
    b^{\alpha}_a\ktau\sigma_{sr}(-\vec{k},0)\big]\Big>=
    A^{a\alpha}_{rs}\ktau.
\end{equation}
The second one can derived using the invariance of the trace under
cyclic permutations and that the order is irrelevant behind an
ordering operator:
\begin{equation}
  \label{eq:ansym2}
  A^{sr}_{a\alpha}(-\vec{k},-\tau)=-\Big<T_\tau\big[\sigma_{rs}
  (-\vec{k},-\tau)b^{\alpha\adj}_a(-\vec{k},0)\big]\Big>=-\Big<T_\tau\big[
  b^{\alpha\adj}_a(-\vec{k},\tau)\sigma_{rs}(-\vec{k},0)\big]\Big>=
  A^{a,-\alpha}_{sr}\ktau.
\end{equation}
Combining Eq. \eqref{eq:ansym1} and Eq. \eqref{eq:ansym2} together
results in (also using that the momentum dependence comes from
$k=|\vec{k}|$)
\begin{equation}
  \label{eq:ansym3}
  A^{sr}_{a\alpha}\ktau=A^{sr}_{a\alpha}(k,\tau)
  =A_{rs}^{a\alpha}(k,\tau)=
  A_{sr}^{a,-\alpha}(k,-\tau)=A^{rs}_{a,-\alpha}
  (k,-\tau).
\end{equation}

\subsection{Perturbation theory}
\label{sec:perturb}

To calculate the averages of the grand canonical ensemble with the
Hamiltonian \eqref{eq:ham2} one can use the methods of the
finite temperature many body physics \cite{FW}.

\begin{figure}[ht!]
  \begin{center}
    \includegraphics*[30mm,260mm][190mm,275mm]{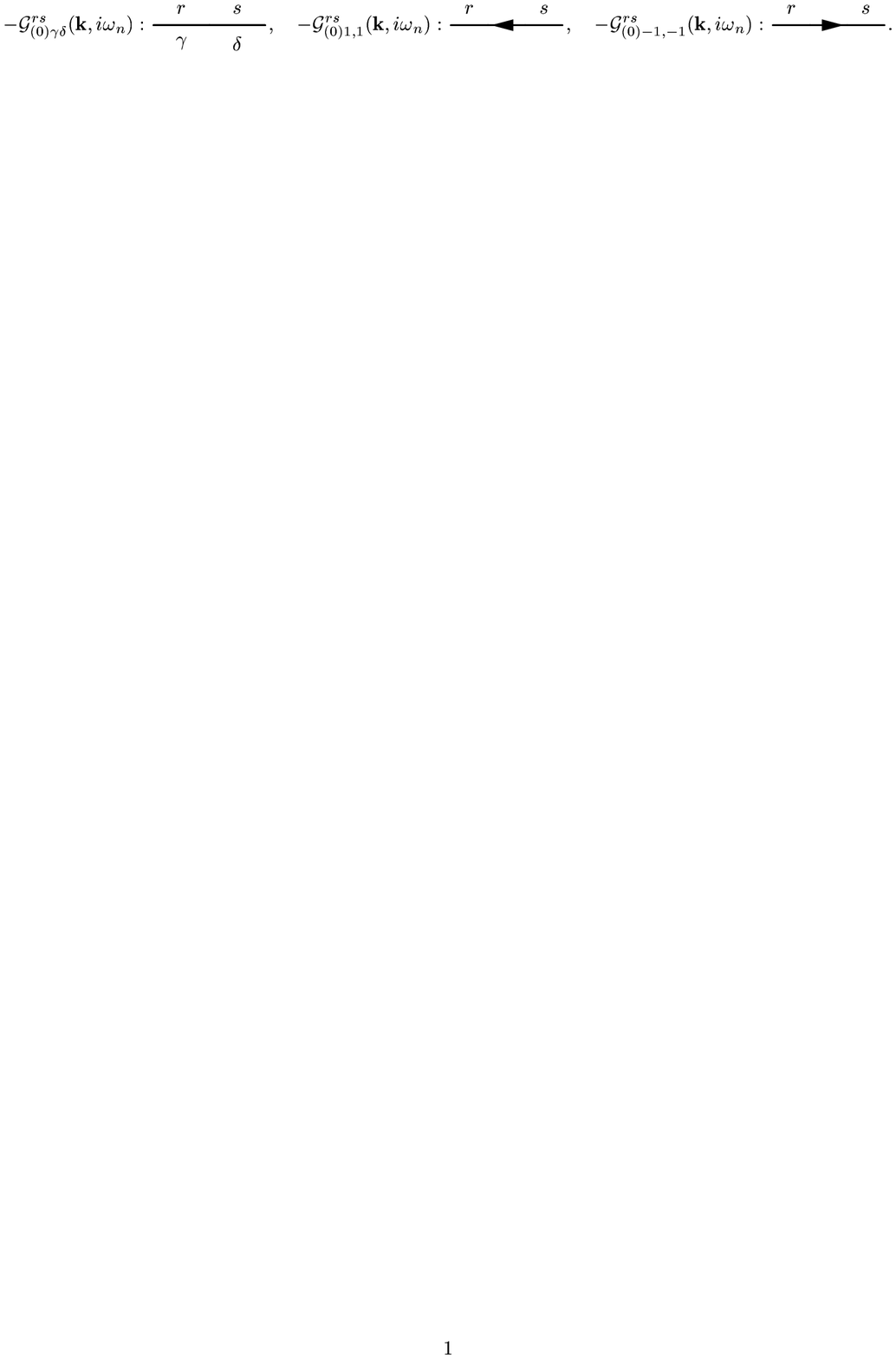}
    \caption{The graphical representation of the free propagators}
    \label{fig:freeprop}
  \end{center}
\end{figure}

\begin{figure}[ht!]
  \begin{center}
    \includegraphics*[30mm,250mm][190mm,275mm]{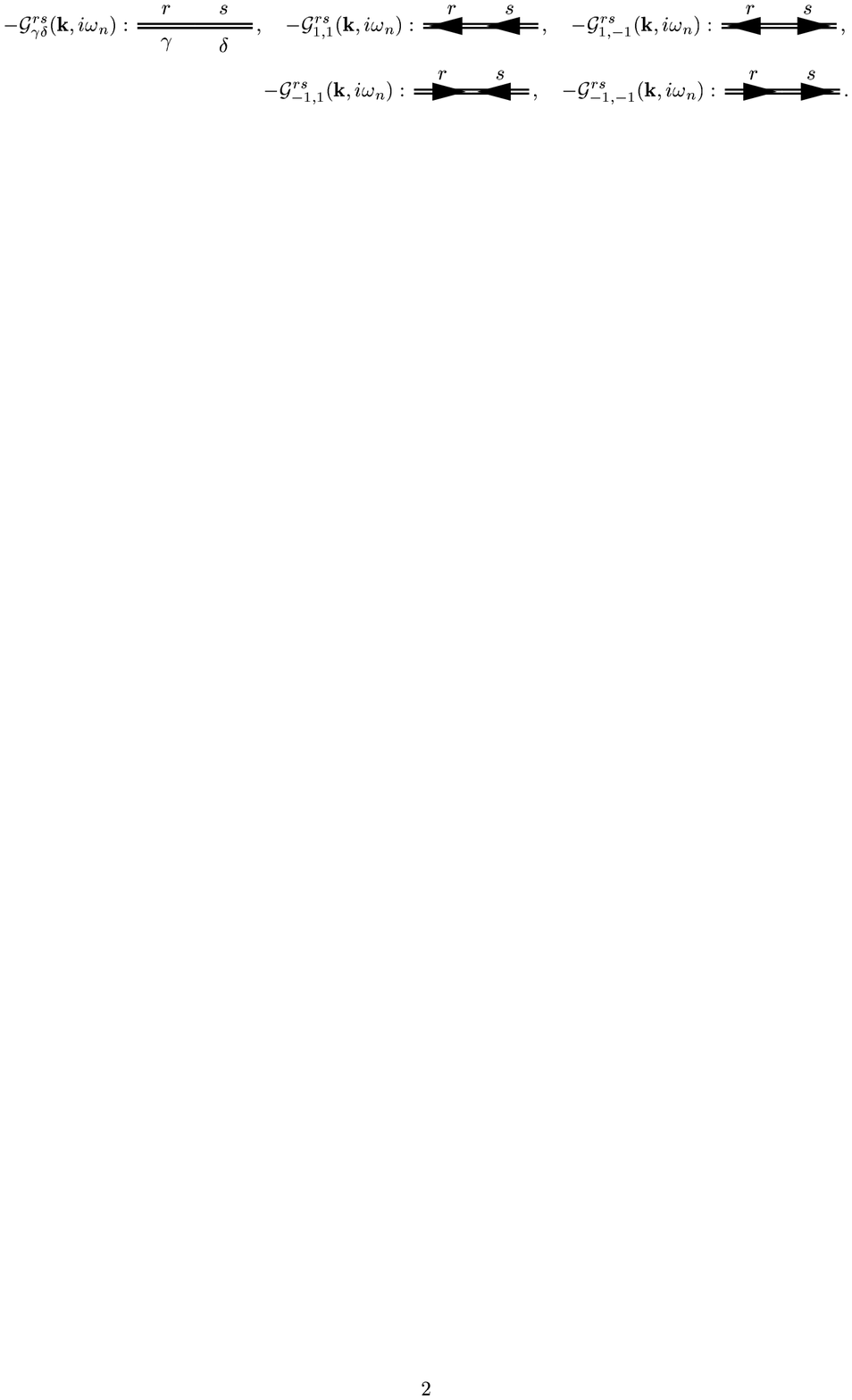}
    \caption{The graphical representation of the full propagators}
    \label{fig:exprop}
  \end{center}
\end{figure}

\begin{figure}[ht]
  \begin{center}
    \includegraphics*[30mm,250mm][190mm,275mm]{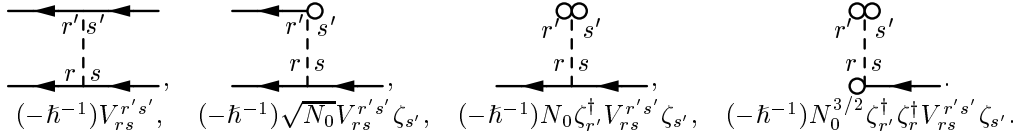}
  \end{center}
  \caption{The Feynman graphs of a few interaction processes involving zero,
    one, two and three condensate atoms}
  \label{fig:inter}
\end{figure}

\begin{figure}[ht!]
  \begin{center}
    \includegraphics*[30mm,250mm][190mm,275mm]{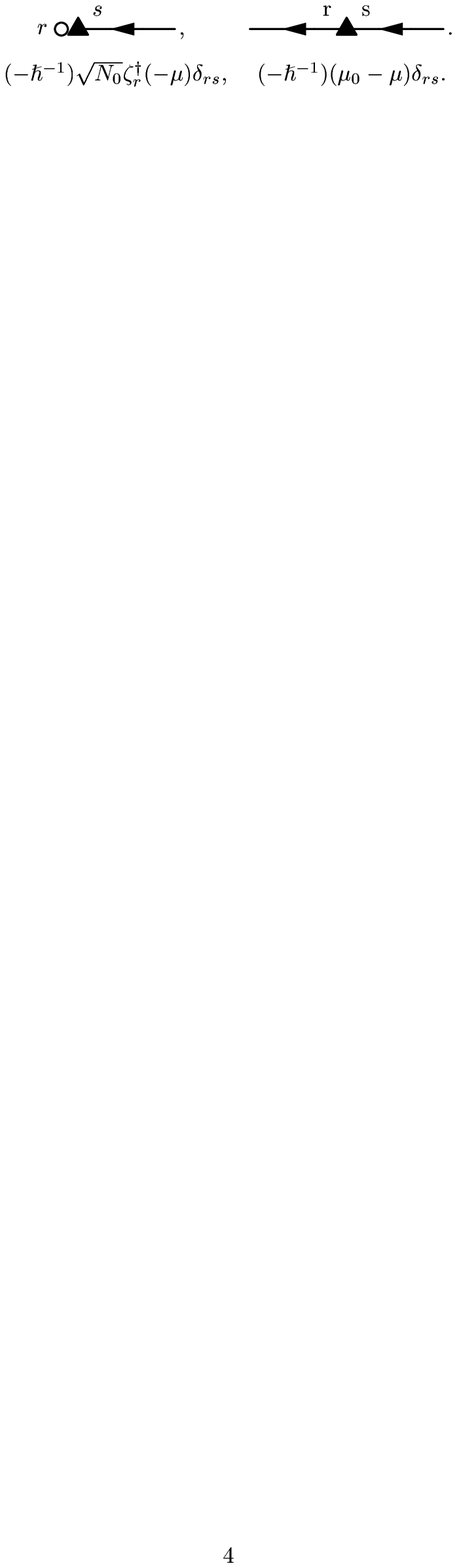}
  \end{center}
  \caption{The Feynamn graphs of the non interaction processes}
  \label{fig:vert}
\end{figure}

The Hamiltonian of the non interacting system is the first term in Eq.
\eqref{eq:ham2}:
\begin{equation}
  \label{eq:hamnon}
  {\mathcal H}_0=\sum_{\vec{k}}(e_{\vec{k}}-\mu_0)b^\dagger_r(\vec{k})
  b_r(\vec{k}),
\end{equation}
which defines free Green's functions:
\begin{equation}
  \label{eq:freeprop}
  \Gr^{rs}_{(0)\gamma\delta}\komega=\frac{\delta_{rs}\delta_{\gamma\delta}}
  {\gamma i\omega_n-\hslash^{-1}(e_{\vec{k}}-\mu_0)},
\end{equation}
This will be symbolized as a line or a line with an arrow if the Greek
indices are specified as seen in Fig. \ref{fig:freeprop}. The full
Green's functions will be symbolized with double lines as shown in
Fig.  \ref{fig:exprop}. The terms in the Hamiltonian Eq.
\eqref{eq:ham2} without any filed operator can be disregarded for our
purposes. Among the interaction terms there is one containing four
field operators and corresponding to a scattering of two
quasi-particles which are noncondensed before and after the collision
as well.  There are four interaction terms containing three field
operators and six terms containing two field operators, corresponding
to scattering processes involving both condensate and noncondensate
atoms. There are interaction terms containing one field operator
describing scattering processes involving three condensate and one
noncondensate atoms.  These interaction terms will be graphically
represented as shown in Fig. \ref{fig:inter}.  There is one remaining
term with one field operator and the last term with two operators
corresponding to vertices with one and two legs as seen in Fig.
\ref{fig:vert}.

\begin{figure}[ht!]
  \begin{center}
    \includegraphics*[30mm,215mm][190mm,275mm]{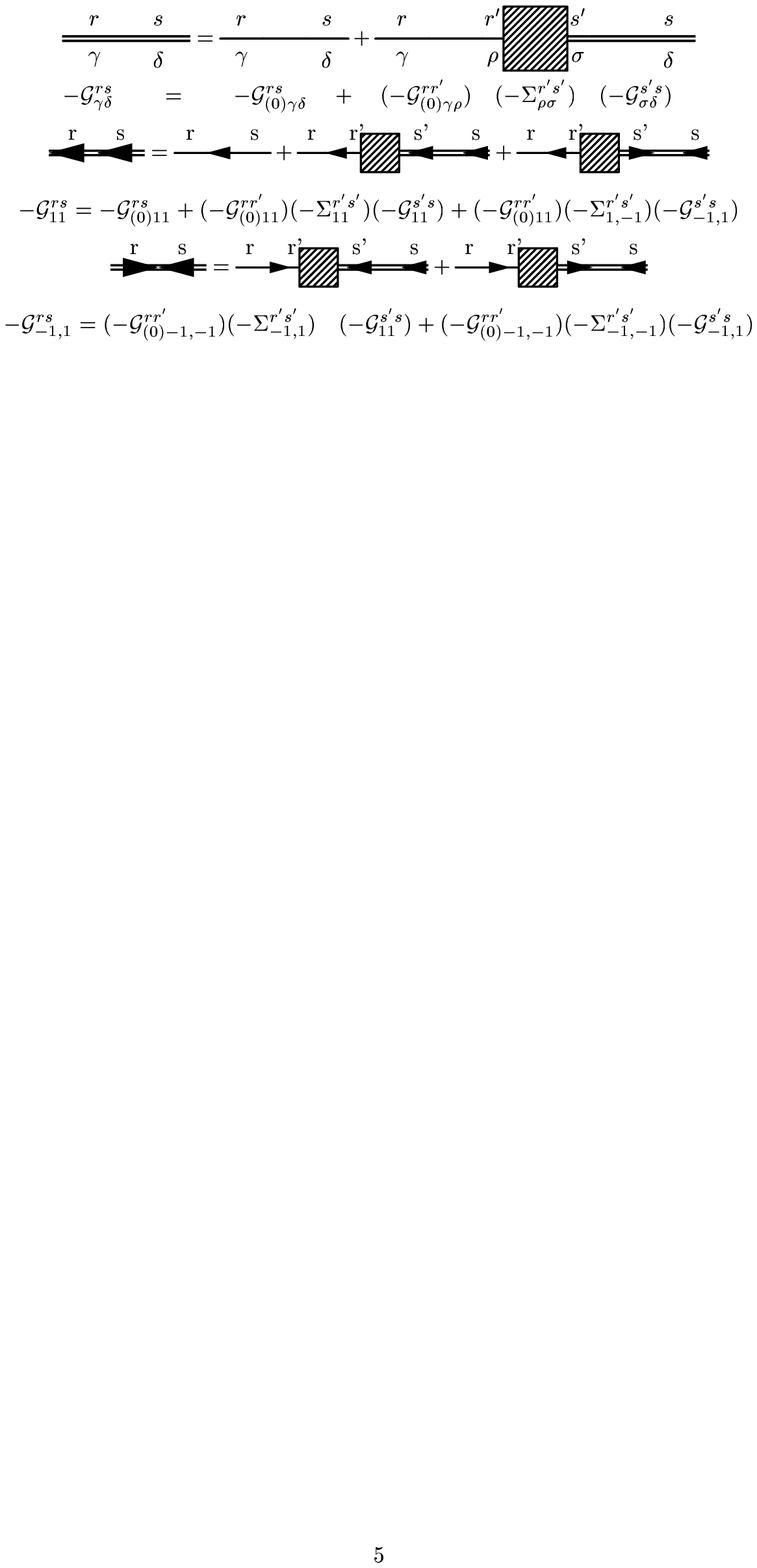}
    \caption{The graphical symbolization of the Dyson-Beliaev equations,
      the hatched squares represent the one-particle irreducible graphs}
    \label{fig:DyBel}
  \end{center}
\end{figure}

The Green's functions can be expanded to perturbation series in the
usual way. Rearranging these series one arrives at the generalized
Dyson-Beliaev equations \cite{FW,Griffin,SzSz1}:
\begin{equation}
  \label{eq:dyson}
  \Gr^{rs}_{\gamma\delta}\komega=\Gr^{rs}_{(0)\gamma\delta}\komega+\Gr^{rr'}
  _{(0)\gamma\rho}\komega\Sigma^{r's'}_{\rho\sigma}\komega\Gr^{s's}_{\sigma
    \delta}\komega,
\end{equation}
where $\Sigma^{rs}_{\gamma\delta}$ is the self-energy, the
contribution of those graphs which are one-particle irreducible
(cannot be split to two by cutting a single one-particle line) and
connect to two external lines with indices $(r,\gamma)$ and
$(s,\delta)$. This equation is graphically represented in Fig.
\ref{fig:DyBel}. The symmetry properties of the Green's functions of
Eq. \eqref{eq:sym3} stand for the self-energies as well.

\begin{figure}[ht]
  \begin{center}
    \includegraphics*[30mm,235mm][190mm,275mm]{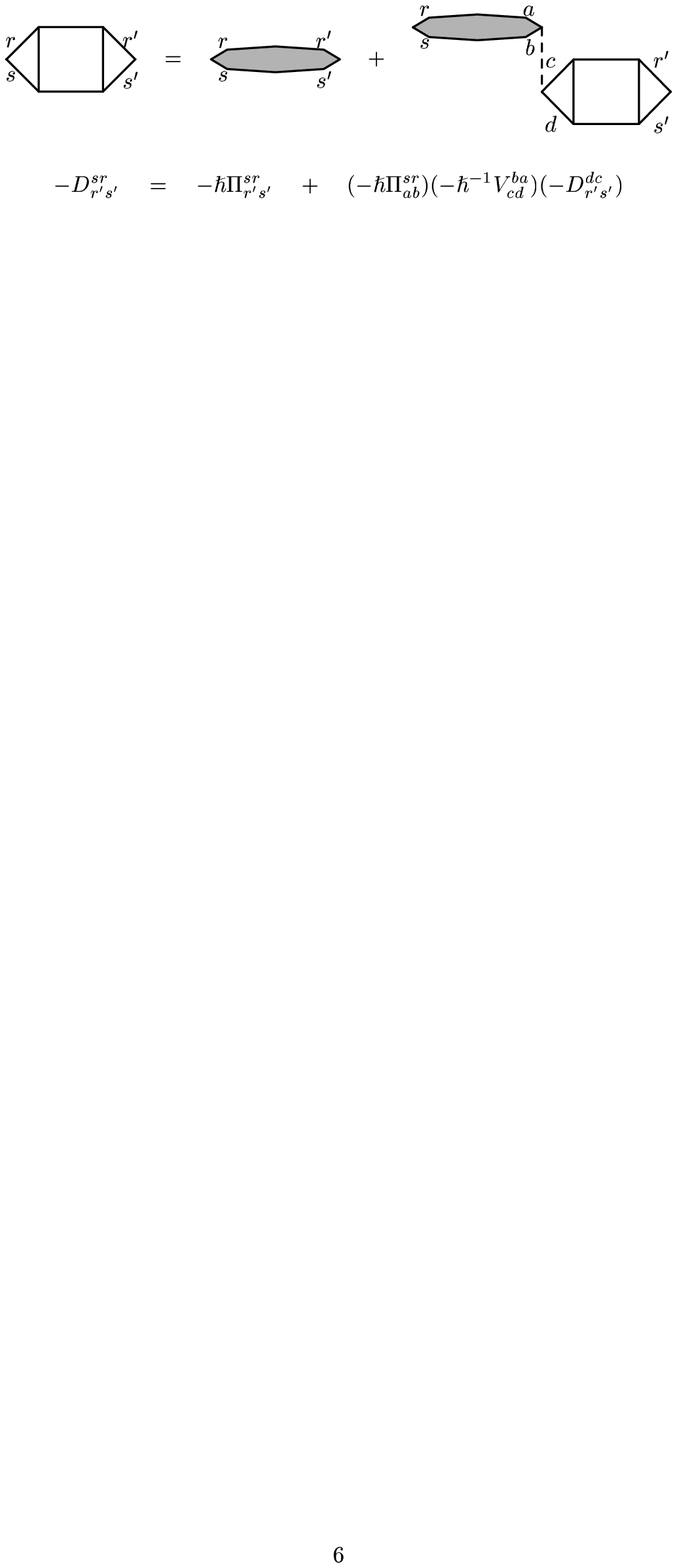}
    \caption{The graphical representation of Eq. \eqref{eq:propeq}. The gray
      polygon represents the proper graphs}
    \label{fig:prop}
  \end{center}
\end{figure}

Similar rearrangements can be carried out for the perturbation series
of the generalized density correlation functions \eqref{eq:d6}.  One
can introduce their proper parts (the polarization parts), the
contribution of those graphs (the polarization graphs) which cannot be
split by two with cutting a single interaction line, that is they are
interaction line irreducible. Then the equations determining these
correlation functions read as:
\begin{equation}
  \label{eq:propeq}
  D^{sr}_{r's'}\komega=\hslash\Pi^{sr}_{r's'}\komega+\Pi^{sr}_{ab}\komega
  V^{ba}_{cd}D^{dc}_{r's'}\komega.
\end{equation}
The Feynman graph corresponding to this equation is shown in Fig.
\ref{fig:prop}, where the generalized density correlation functions
are represented as boxes which can connect to two interaction lines
and their proper parts are represented as grey polygons also able to
connect to two interaction lines. The proper parts satisfy the same
symmetry properties as the generalized correlation functions
(see Eq. \eqref{eq:ds3}).

Above the critical temperature (or for a noncondensed system), where
$\left< a_r (\vec{k}) \right>=0$ for all $\vec{k}$, the anomalous
correlation functions \eqref{eqs:anom} are zero. In this case
the self-energies, which are irreducible by definition are proper as
well and the polarization parts, which are proper by definition are
irreducible as well. In the Bose condensed phase the appearance of the
anomalous averages $\left< a_r (\vec{0}) \right>\neq0$ leads to the
nonzero value of the anomalous correlation functions and to a
situation where the self-energies are no longer proper and the
polarization parts are no longer irreducible. However, they can be
separated such as
\begin{subequations}
  \begin{align}
    \Sigma^{rs}_{\gamma\delta}&=\widetilde{\Sigma}^{rs}_{\gamma\delta}+
    M^{rs}_{\gamma\delta},\label{eq:selfdec}\\
    \Pi^{sr}_{r's'}&=\Pi^{(r)sr}_{r's'}+\Pi^{(s)sr}_{r's'},\label{eq:propdec}
  \end{align}
\end{subequations}
where $\widetilde{\Sigma}$ is the contribution of those self-energy
graphs which are proper, while the graphs contributing to $M$ will be
improper, and similarly $\Pi^{(r)}$ is the contribution of those
polarization graphs which are irreducible as well and $\Pi^{(s)}$ is
the contribution of the reducible polarization graphs. The separations
\eqref{eq:selfdec} and \eqref{eq:propdec} are the starting steps
toward the dielectric formalism. See for the scalar gas Ref.
\cite{SzK,Griffin}.

\begin{figure}[ht]
  \begin{center}
    \includegraphics*[30mm,240mm][190mm,275mm]{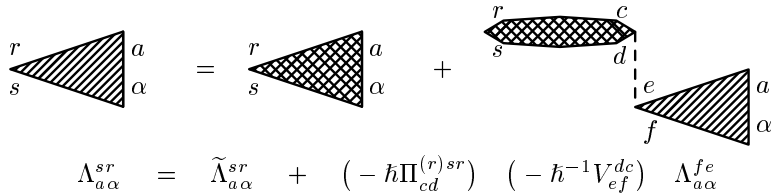}
    \caption{The graphical symbolization of building up of the anomalous
      vertex from proper parts}
    \label{fig:fig_anomvertprop}
  \end{center}
\end{figure}

It can be directly seen from the perturbation series of the anomalous
correlation functions \eqref{eq:anom1} can be decomposed such a way that
\begin{equation}\label{eq:anomdec1}
  A^{sr}_{a\alpha}\komega=\Lambda^{sr}_{c\gamma}\komega
  \Gr^{ca}_{\gamma\alpha}\komega,
\end{equation}
where $\Gr^{ca}_{\gamma\alpha}$ is the one-particle Green's function
and $\Lambda^{sr}_{c\gamma}$ is the anomalous vertex, which is the sum
of the irreducible contributions of those graphs with one incoming
interaction and one incoming particle line. These anomalous vertex
functions can be expressed with the use their proper parts and the
irreducible and proper parts of the density correlation functions as
seen in Fig.  \ref{fig:fig_anomvertprop}.
\begin{equation}\label{eq:anprope}
  \Lambda^{sr}_{a\alpha}\komega=\widetilde{\Lambda}^{sr}_{a\alpha}\komega+
  \Pi^{(r)sr}_{cd}\komega V^{dc}_{ef}\Lambda^{fe}_{a\alpha}\komega.
\end{equation}

\subsection{The determination of the chemical potential and the treatment 
  of the non-interaction self-energy vertex}
\label{sec:chem_pot}

\begin{figure}[ht]
  \begin{center}
    \includegraphics*[30mm,245mm][190mm,275mm]{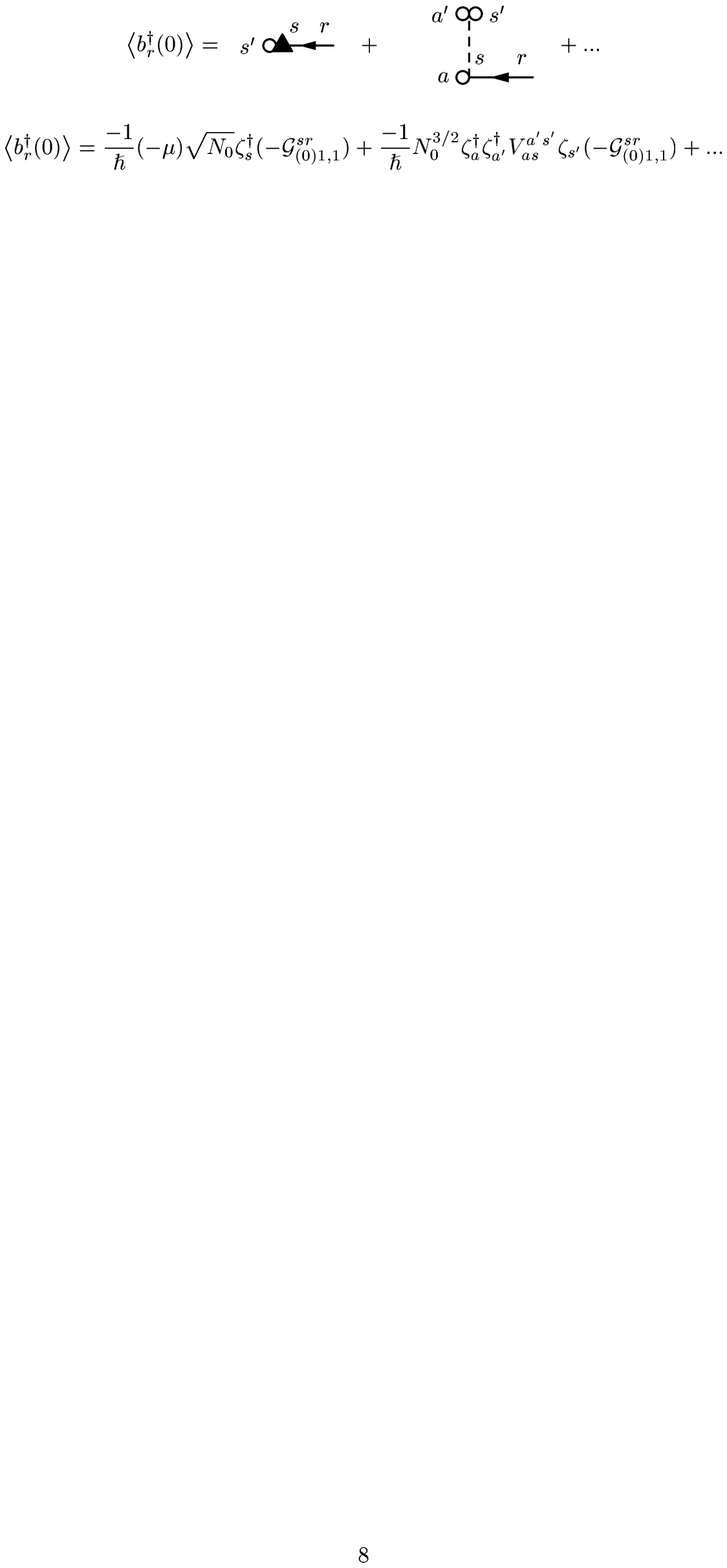}
    \caption{The beginning of the perturbation series of the consistency 
      condition}
    \label{fig:cons_cond}
  \end{center}
\end{figure}

As pointed out earlier, with the introduction of the new set of
creation and destruction operators with Eqs. \eqref{eq:cantr1} and
\eqref{eq:cantr2} one can derive the relation between the condensate
density and the chemical potential from the requirement that
$\left<b_r(\vec{k})\right>=\left<b_r\adj(\vec{k})\right>=0$. Using
perturbation theory for the evaluation of these expressions one can
arrive at equations which can be symbolized as seen in Fig.
\ref{fig:cons_cond}. One can notice that a similar rearrangement is
possible as was made with the Green's functions which leads to:
\begin{equation}
  \label{eq:cons_condref}
    \big<b_r^\gamma(0,0)\big>=\Sigma^s_{0\delta}(0,0)\Gr^{sr}_{\delta\gamma}
    (0,0),
\end{equation}
where $\Sigma^{s}_{0\delta}$ is the sum of those irreducible graphs
which has only one incoming (outgoing) lines. Using the symmetry
properties of the Green's functions one can easily derive that the
consistency condition is equivalent to
\begin{equation}
  \label{eq:cons_condfin}
  \Sigma^{s}_{0\gamma}(0,0)=0.
\end{equation}

\begin{figure}[ht]
  \begin{center}
    \includegraphics*[30mm,245mm][190mm,275mm]{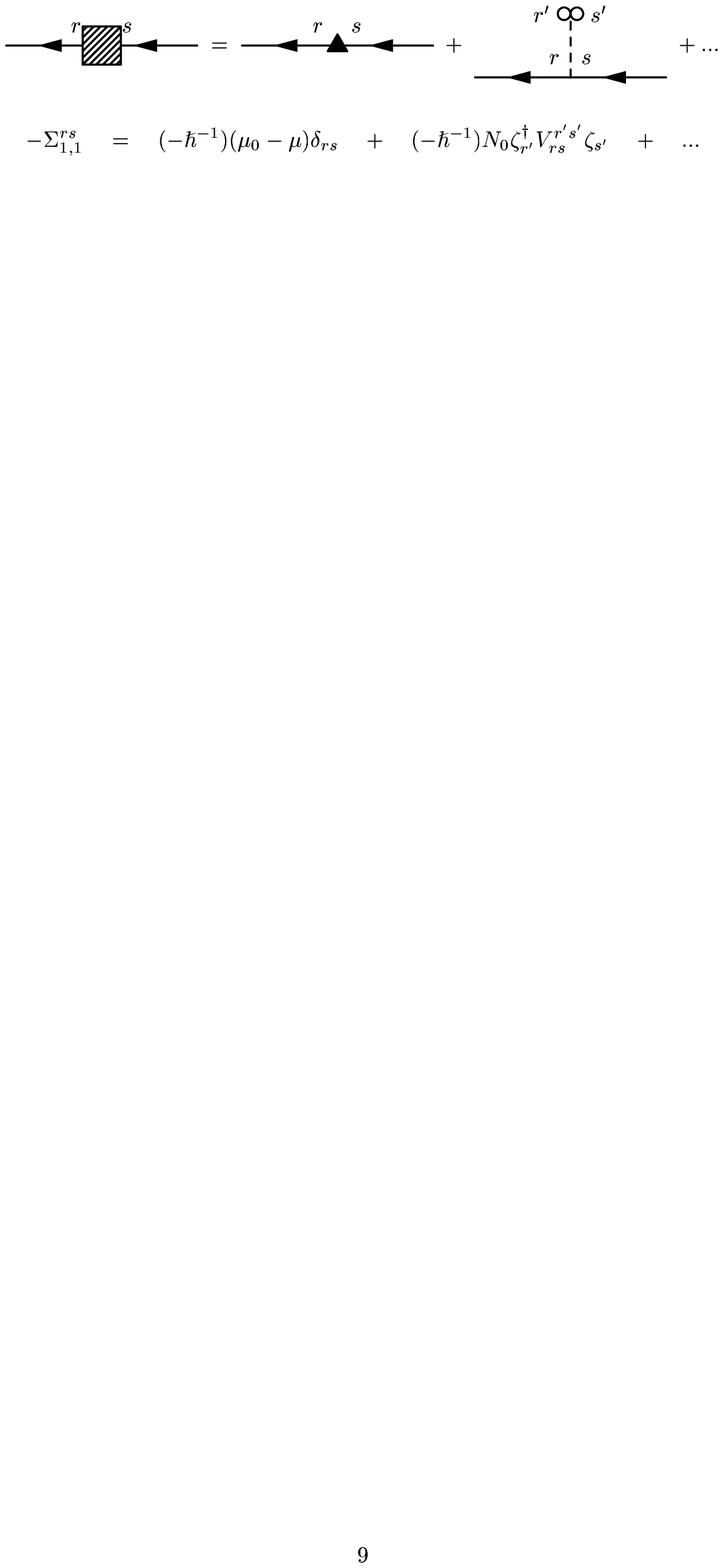}
    \caption{The self-energy with the non interaction vertex}
    \label{fig:fig_twolegged}
  \end{center}
\end{figure}

The non interaction self-energy vertex corresponding to the last term
in the Hamiltonian of Eq. \eqref{eq:ham2} is to be calculated in each
order of perturbation theory. Since this vertex is proper and diagonal
in spin and anomalous indices it is part of the proper self-energy as
illustrated in Fig. \ref{fig:fig_twolegged}.

\subsection{Rotational symmetry and spin transfer decomposition}
\label{sec:rs_std}

Since rotational symmetry along the z axis is not broken the z
component of the total angular momentum is conserved which results for
$\Gr^{rs}_{\gamma\delta}$ and $\Sigma^{rs}_{\gamma\delta}$ in $\gamma
r -\delta s=(\gamma-\delta) \zeta\adj_r(F_z)_{rs}\zeta_s$ or
symbolically:
\begin{figure}[ht!]
  \begin{center}
    \includegraphics*[30mm,235mm][190mm,275mm]{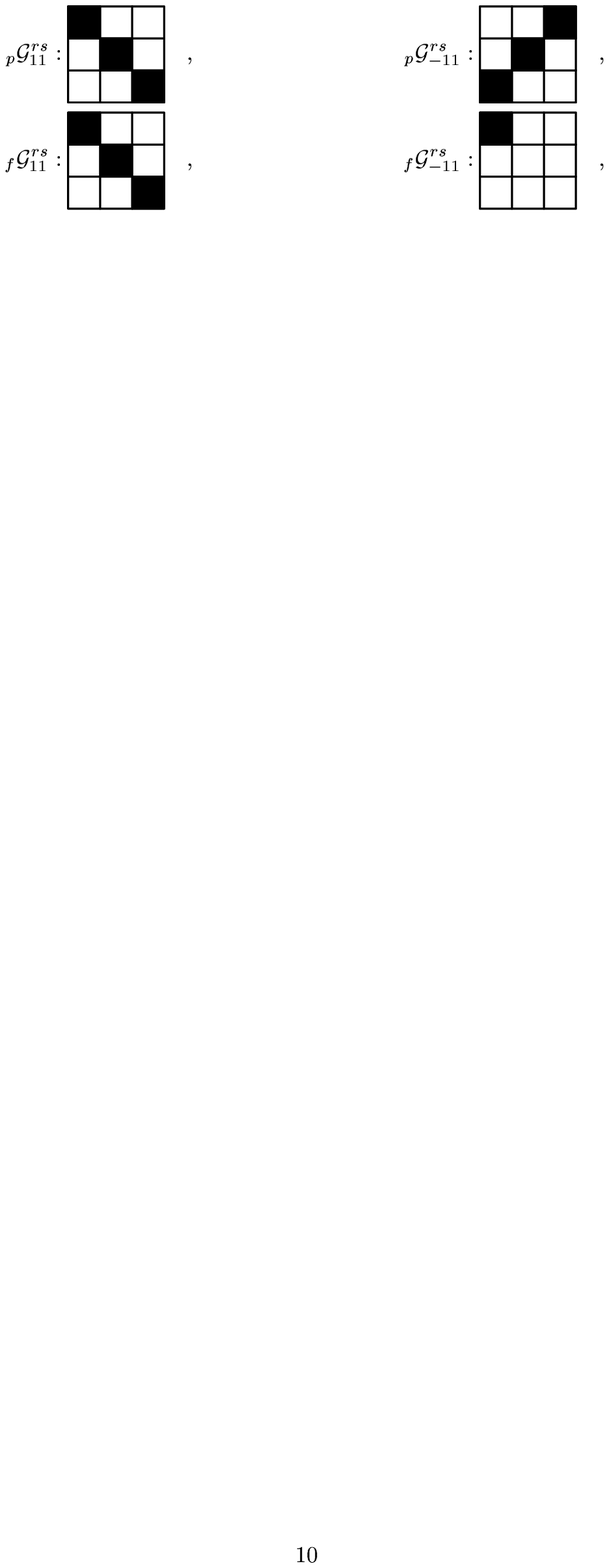}
    \caption{The spin conservation property of the Green's functions.
      Here $_p\Gr$ denotes the Green's functions of the polar case
      while $_f\Gr$ denotes the Green's functions of the ferromagnetic
      case. A black box refers to a non zero element and we used
      $\zeta=(0,1,0)^T$ for the polar case and $\zeta=(1,0,0)^T$ for
      the ferromagnetic case specifying the condensate}
    \label{fig:grsym}
  \end{center}
\end{figure}

As a consequence of these symmetry relations Eq. \eqref{eq:dyson} can
be separated to three equations according to the number $n=[ \gamma
(c-r) + \delta(c-s) ] / 2$, with $c=\zeta\adj_r (F_z)_{rs} \zeta_s$
specifying the spin carried by the propagator relative to the spin of
the atom in the condensate. For this reason let us define the
following matrices for the polar case
\begin{subequations}
  \label{eqs:gmatp}
  \begin{align}
    \gmat[0][\Gr]_{\gamma\delta}&=\left[\begin{array}{c c c}
        \Gr^{00}_{1,1} & \Gr^{00}_{1,-1}\\
        \Gr^{00}_{-1,1} & \Gr^{00}_{-1,-1}
      \end{array}\right]_{\gamma\delta}, &
    \gmat[0][\Sigma]_{\gamma\delta}&=\left[\begin{array}{c c c}
        \Sigma^{00}_{1,1} & \Sigma^{00}_{1,-1}\\
        \Sigma^{00}_{-1,1} & \Sigma^{00}_{-1,-1}
      \end{array}\right]_{\gamma\delta},\label{eq:gmat0p}\\
    \gmat[+][\Gr]_{\gamma\delta}&=\left[\begin{array}{c c c}
        \Gr^{--}_{1,1} & \Gr^{-+}_{1,-1}\\
        \Gr^{+-}_{-1,1} & \Gr^{++}_{-1,-1}
      \end{array}\right]_{\gamma\delta}, &
    \gmat[+][\Sigma]_{\gamma\delta}&=\left[\begin{array}{c c c}
        \Sigma^{--}_{1,1} & \Sigma^{-+}_{1,-1}\\
        \Sigma^{+-}_{-1,1} & \Sigma^{++}_{-1,-1}
      \end{array}\right]_{\gamma\delta},\label{eq:gmat+p}\\
    \gmat[-][\Gr]_{\gamma\delta}&=\left[\begin{array}{c c c}
        \Gr^{++}_{1,1} & \Gr^{+-}_{1,-1}\\
        \Gr^{-+}_{-1,1} & \Gr^{--}_{-1,-1}
      \end{array}\right]_{\gamma\delta}, &
    \gmat[-][\Sigma]_{\gamma\delta}&=\left[\begin{array}{c c c}
        \Sigma^{++}_{1,1} & \Sigma^{+-}_{1,-1}\\
        \Sigma^{-+}_{-1,1} & \Sigma^{--}_{-1,-1}
      \end{array}\right]_{\gamma\delta},\label{eq:gmat-p}
  \end{align}
\end{subequations}
where the indices in the upper left corner correspond to the number $n$.
For the ferromagnetic one let us define the following quantities
\begin{subequations}
  \label{eqs:gmatf}
  \begin{align}
    \gmat[0][\Gr]_{\gamma\delta}&=\left[\begin{array}{c c c}
        \Gr^{++}_{1,1} & \Gr^{++}_{1,-1}\\
        \Gr^{++}_{-1,1} & \Gr^{++}_{-1,-1}
      \end{array}\right]_{\gamma\delta}, &
    \gmat[0][\Sigma]_{\gamma\delta}&=\left[\begin{array}{c c c}
        \Sigma^{++}_{1,1} & \Sigma^{++}_{1,-1}\\
        \Sigma^{++}_{-1,1} & \Sigma^{++}_{-1,-1}
      \end{array}\right]_{\gamma\delta},\label{eq:gmat0f}\\
    \gmat[+][\Gr] & = \Gr^{00}_{1,1}, & \gmat[+][\Sigma]&= \Sigma^{00}_{1,1},
    \label{eq:gmat+f}\\
    \gmat[-][\Gr] & = \Gr^{00}_{-1,-1}, & \gmat[-][\Sigma]&=
    \Sigma^{00}_{-1,-1},\label{eq:gmat-f}\\
    \gmat[Q][\Gr] & = \Gr^{--}_{1,1}, & \gmat[Q][\Sigma]&= \Sigma^{--}_{1,1},
    \label{eq:gmatQ+f}\\
    \gmat[-Q][\Gr] & = \Gr^{--}_{-1,-1}, & \gmat[-Q][\Sigma]&=
    \Sigma^{--}_{-1,-1}.
    \label{eq:gmatQ-f}
  \end{align}
\end{subequations}
The indices in the upper left corner correspond to the number $n$
except for the quadrupolar spin transfer $n=2$ and similarly $-Q$ is
for $n=-2$. (Note, that there is no quadrupolar Green's function for
the polar case because the value of $n$ can not be $\pm2$
here.)  The later modes of the ferromagnetic case ($+,-,Q,-Q$) differ
from all the other modes since for these cases spin conservation
forbids the existence of anomalous Green's functions and self
energies. With these definitions one can explicitly construct the
Green's functions. The propagators corresponding to different spin
transfers will be called different modes in the following. For all
three modes of the polar case ($n=0,+,-$) and for the first mode of
the ferromagnetic case ($n=0$) the Green's functions are
\begin{equation}
  \label{eq:greg}
  \gmat[n][\Gr]_{\alpha\gamma}=\frac{\delta_{\alpha,\gamma}(\alpha i\omega_n+
    \hslash^{-1}\epsilon_{\vec{k}})+\alpha\gamma\Sigma_{-\gamma,-\alpha}}
  {(i\omega_n-\hslash^{-1}\epsilon_{\vec{k}}-\Sigma_{1,1})(i\omega_n+
    \hslash^{-1}\epsilon_{\vec{k}}+\Sigma_{-1,-1})+\Sigma_{-1,1}
    \Sigma_{1,-1}}\equiv\frac{\gmat[n][N]_{\alpha,\gamma}}{\gmat[n][\Delta]}
\end{equation}
where $\epsilon_\vec{k}=\kink-\mu_0$ is introduced. This equation
defines the quantities $\gmat[n][N]_{\alpha\gamma}$ and
$\gmat[n][\Delta]$ for the first four modes. For the last four modes
of the ferromagnetic case one can get
\begin{subequations}
  \label{eqs:gnreg}
  \begin{align}
    \gmat[\pm][\Gr]=\frac{1}{\pm i\omega_n-\hslash^{-1}\epsilon_{\vec{k}}-
      \gmat[\pm][\Sigma]},\label{eq:gnreg+f} \\
    \gmat[\pm Q][\Gr]=\frac{1}{\pm i\omega_n-\hslash^{-1}\epsilon_{\vec{k}}-
      \gmat[\pm Q][\Sigma]}.\label{eq:gnregQf}
  \end{align}
\end{subequations}

Sometimes it will be practical to cast the formulas into a most
concise form. This can be achieved by introducing the following formal
definitions:
\begin{subequations}
  \begin{align}
    \gmat[+][\Gr]_{\gamma\delta}&=\left[\begin{array}{c c c}
        \Gr^{00}_{1,1} & 0\\
        0 & 0
      \end{array}\right]_{\gamma\delta}, &
    \gmat[+][\Sigma]_{\gamma\delta}&=\left[\begin{array}{c c c}
        \Sigma^{00}_{1,1} & 0\\
        0 & 0
      \end{array}\right]_{\gamma\delta},\\
    \gmat[-][\Gr]_{\gamma\delta}&=\left[\begin{array}{c c c}
        0 & 0\\
        0 & \Gr^{00}_{-1,-1}
      \end{array}\right]_{\gamma\delta}, &
    \gmat[-][\Sigma]_{\gamma\delta}&=\left[\begin{array}{c c c}
        0 & 0\\
        0 & \Sigma^{00}_{-1,-1}\\
      \end{array}\right]_{\gamma\delta},
  \end{align}
\end{subequations}
with the help of which equation \eqref{eq:gnreg+f} takes the form of
Eq. \eqref{eq:greg}, i.e., the validity of Eq. \eqref{eq:greg} is
extended for $n=0,+,-$ for the ferromagnetic phase as well. 

The conservation of the z component of the spin for these
$D^{sr}_{r's'}$ correlation functions means that $r-s=r'-s'$ which
holds for the proper parts as well. And as a consequence Eq.
\eqref{eq:propeq} also decouples into three ordinary matrix equations
according to the specific spin transfer.  For this reason it is useful
to define the following matrices:
\begin{itemize}
\item For $0$ spin transfer:
  \begin{subequations}
    \label{eqs:0trans}
    \begin{align}
      &\big(\mat[0][D]\big)_{ab}:=D^{aa}_{bb}=
      \left[\begin{array}{c c c}
          D^{++}_{++} & D^{++}_{00} & D^{++}_{--}\\
          D^{00}_{++} & D^{00}_{00} & D^{00}_{--}\\
          D^{--}_{++} & D^{--}_{00} & D^{--}_{--}
        \end{array}\right]_{ab},\label{0D}\\ 
      &\big(\mat[0][\Pi]\big)_{ab}:=\Pi^{aa}_{bb}=
      \left[\begin{array}{c c c}
          \Pi^{++}_{++} & \Pi^{++}_{00} & \Pi^{++}_{--}\\
          \Pi^{00}_{++} & \Pi^{00}_{00} & \Pi^{00}_{--}\\
          \Pi^{--}_{++} & \Pi^{--}_{00} & \Pi^{--}_{--}
        \end{array}\right]_{ab},\label{0Pi}\\
      &\big(\mat[0][C]\big)_{ab}:=V^{aa}_{bb}=
      \left[\begin{array}{c c c}
          c_n+c_s & c_n & c_n-c_s\\
          c_n & c_n & c_n\\
          c_n-c_s & c_n & c_n+c_s
        \end{array}\right]_{ab},\label{0C}
    \end{align}
  \end{subequations}
  where no automatic summation is understood now. From Eq.
  \eqref{eq:propeq} one obtains
  \begin{equation}
    \label{eq:prop0m}
    \mat[0][D]=\hslash\cn\mat[0][\Pi]+\mat[0][\Pi]\cn\mat[0][C]\cn\mat[0][D].
  \end{equation}

\item For $+1$ spin transfer:
  \begin{subequations}
    \label{eqs:+trans}
    \begin{align}
      &\big(\mat[+][D]\big)_{ab}:=D^{a,a+1}_{b+1,b}=
      \left[\begin{array}{c c}
          D^{0+}_{+0} & D^{0+}_{0-}\\
          D^{-0}_{+0} & D^{-0}_{0-}
        \end{array}\right]_{ab},\label{+D}\\
      &\big(\mat[+][\Pi]\big)_{ab}:=\Pi^{a,a+1}_{b+1,b}=
      \left[\begin{array}{c c}
          \Pi^{0+}_{+0} & \Pi^{0+}_{0-}\\
          \Pi^{-0}_{+0} & \Pi^{-0}_{0-}
        \end{array}\right]_{ab},\label{+Pi}\\ 
      &\big(\mat[+][C]\big)_{ab}:=V^{a,a+1}_{b+1,b}=
      \left[\begin{array}{c c}
          c_s & c_s\\
          c_s & c_s
        \end{array}\right]_{ab},\label{+C}
    \end{align}
  \end{subequations}
  and the resulting matrix equation is:
  \begin{equation}
    \label{eq:prop+m}
    \mat[+][D]=\hslash\cn\mat[+][\Pi]+\mat[+][\Pi]\cn\mat[+][C]\cn\mat[+][D].
  \end{equation}

\item For $-1$ spin transfer:
  \begin{subequations}
    \label{eqs:-trans}
    \begin{align}
      &\big(\mat[-][D]\big)_{ab}:=D^{a+1,a}_{b,b+1}=
      \left[\begin{array}{c c}
          D^{+0}_{0+} & D^{+0}_{-0}\\
          D^{0-}_{0+} & D^{0-}_{-0}
        \end{array}\right]_{ab},\label{-D}\\
      &\big(\mat[-][\Pi]\big)_{ab}:=\Pi^{a+1,a}_{b,b+1}=
      \left[\begin{array}{c c}
          \Pi^{+0}_{0+} & \Pi^{+0}_{-0}\\
          \Pi^{0-}_{0+} & \Pi^{0-}_{-0}
        \end{array}\right]_{ab},\label{-Pi}\\ 
      &\big(\mat[-][C]\big)_{ab}:=V^{a+1,a}_{b,b+1}=
      \left[\begin{array}{c c}
          c_s & c_s\\
          c_s & c_s
        \end{array}\right]_{ab},\label{-C}
    \end{align}
  \end{subequations}
  and similarly 
  \begin{equation}
    \label{eq:prop-m}
    \mat[-][D]=\hslash\cn\mat[-][\Pi]+\mat[-][\Pi]\cn\mat[-][C]\cn\mat[-][D].
  \end{equation}
  
\item For the $\pm2$ spin transfer case from the fact that
  $V^{+-}_{-+}=V^{-+}_{+-}=0$ the following results can be obtained
  \begin{align}
    D^{-+}_{+-} &= \hslash \Pi^{-+}_{+-}\label{eq:propQm}\\
    D^{+-}_{-+} &= \hslash\label{eq:prop-Qm}
  \Pi^{+-}_{-+}.
  \end{align}
\end{itemize}

As a consequence of Eq. \eqref{eq:ds3} $\mat[0][D]$, $\mat[\pm][D]$
(and their proper parts) are symmetric matrices, furthermore
$\mat[0][D](k,i\omega_n)=\mat[0][D](k,-i\omega{_n})$ and
$\mat[+][D](k,i\omega_n)=\mat[-][D](k,-i\omega{_n})$. This means that
the generalized density correlation functions for zero spin transfer
are completely symmetric under time reversal.

Equation \eqref{eq:anprope} also decouples according to the amount of spin
transferred. 
\begin{subequations}
  \label{eqs:anpropde}
  \begin{align}
    \vv[][0][\Lambda]_{\alpha}&=\vv[][0][\widetilde{\Lambda}]_{\alpha}+
    \mat[0][\Pi]^{(r)}\cn\mat[0][C]\cn\vv[][0][\Lambda]_{\alpha},
    \label{eq:anpropde1}\\
    \vv[][\pm][\Lambda]_{\alpha}&=\vv[][\pm][\widetilde{\Lambda}]_{\alpha}+
    \mat[\pm][\Pi]^{(r)}\cn\mat[\pm][C]\cn\vv[][\pm][\Lambda]_{\alpha},
    \label{eq:anpropde2}
  \end{align}
\end{subequations}
where the introduced $\vv[][][\Lambda]_{\alpha}$ vectors are different
for the polar and for the ferromagnetic cases since the allowed spin
projection of the incoming (outgoing) one-particle is determined by
the rule of spin conservation and the spin projection of the
condensate. It results that for a given spin transfer the spin
projection of the incoming (outgoing) particle can take only one value
(others are forbidden by spin conservation). So for the polar case
(where $c=\zeta\adj_r (F_z)_{rs} \zeta_s = 0$) the anomalous vertex
vectors are
\begin{subequations}
  \label{eqs:anomvertvectp}
  \begin{align}
    \vv[p][0][\Lambda]_{\alpha}&=\left(\begin{array}{c}
        \Lambda^{++}_{0\alpha}\\
        \Lambda^{00}_{0\alpha}\\
        \Lambda^{--}_{0\alpha}
      \end{array}\right), &
    \vv[p][0][\Lambda]^{\alpha}&=\left(\begin{array}{c}
        \Lambda^{0\alpha}_{++}\\
        \Lambda^{0\alpha}_{00}\\
        \Lambda^{0\alpha}_{--}
      \end{array}\right),\label{vectvert0pol}\\
    \vv[p][+][\Lambda]_{1}&=\left(\begin{array}{c}
        \Lambda^{0+}_{-,1}\\
        \Lambda^{-0}_{-,1}
      \end{array}\right), &
    \vv[p][+][\Lambda]^{1}&=\left(\begin{array}{c}
        \Lambda^{-,1}_{+0}\\
        \Lambda^{-,1}_{0-}
      \end{array}\right), &
    \vv[p][+][\Lambda]_{-1}&=\left(\begin{array}{c}
        \Lambda^{0+}_{+,-1}\\
        \Lambda^{-0}_{+,-1}
      \end{array}\right), &
    \vv[p][+][\Lambda]^{-1}&=\left(\begin{array}{c}
        \Lambda^{+,-1}_{+0}\\
        \Lambda^{+,-1}_{0-}
      \end{array}\right), \label{vectvert+pol}\\
    \vv[p][-][\Lambda]_{1}&=\left(\begin{array}{c}
        \Lambda^{+0}_{+,1}\\
        \Lambda^{0-}_{+,1}
      \end{array}\right), &
    \vv[p][-][\Lambda]^{1}&=\left(\begin{array}{c}
        \Lambda^{+,1}_{0+}\\
        \Lambda^{+,1}_{-0}
      \end{array}\right), &
    \vv[p][-][\Lambda]_{-1}&=\left(\begin{array}{c}
        \Lambda^{+0}_{-,-1}\\
        \Lambda^{0-}_{-,-1}
      \end{array}\right), &
    \vv[p][-][\Lambda]^{-1}&=\left(\begin{array}{c}
        \Lambda^{-,-1}_{0+}\\
        \Lambda^{-,-1}_{-0}
      \end{array}\right). \label{vectvert-pol}
  \end{align}
\end{subequations}
For the ferromagnetic case (where $c=\zeta\adj_r (F_z)_{rs} \zeta_s =
1$) it is easy to verify that spin conservation forbids any spin index
for $\vv[][+][\Lambda]_{\,-1}$ and for $\vv[][-][\Lambda]_{1}$. These
later vectors can be taken as zero.
\begin{subequations}
  \label{eqs:anomvertvectf}
  \begin{align}
    \vv[f][0][\Lambda]_{\alpha}&=\left(\begin{array}{c}
        \Lambda^{++}_{+\alpha}\\
        \Lambda^{00}_{+\alpha}\\
        \Lambda^{--}_{+\alpha}
      \end{array}\right), &
    \vv[f][0][\Lambda]^{\alpha}&=\left(\begin{array}{c}
        \Lambda^{+\alpha}_{++}\\
        \Lambda^{+\alpha}_{00}\\
        \Lambda^{+\alpha}_{--}
      \end{array}\right),\label{vectvert0fer}\\
    \vv[f][+][\Lambda]_{1}&=\left(\begin{array}{c}
        \Lambda^{0+}_{0,1}\\
        \Lambda^{-0}_{0,1}
      \end{array}\right), &
    \vv[f][+][\Lambda]^{1}&=\left(\begin{array}{c}
        \Lambda^{0,1}_{+0}\\
        \Lambda^{0,1}_{0-}
      \end{array}\right), &
    \vv[f][+][\Lambda]_{-1}&=\vv[][][0], &
    \vv[f][+][\Lambda]^{-1}&=\vv[][][0], \label{vectvert+fer}\\
    \vv[f][-][\Lambda]_{1}&=\vv[][][0], &
    \vv[f][-][\Lambda]^{1}&=\vv[][][0], &
    \vv[f][-][\Lambda]_{-1}&=\left(\begin{array}{c}
        \Lambda^{+0}_{0,-1}\\
        \Lambda^{0-}_{0,-1}
      \end{array}\right), &
    \vv[f][-][\Lambda]^{-1}&=\left(\begin{array}{c}
        \Lambda^{0,-1}_{0+}\\
        \Lambda^{0,-1}_{-0}
      \end{array}\right).\label{vectvert-fer}
  \end{align}
\end{subequations}
for the ferromagnetic case. The symmetry relations of Eq.
\eqref{eq:ansym3} hold for the proper and irreducible parts as well,
which is equivalent to:
\begin{subequations}
  \begin{align}
    \vv[][0][\Lambda]^\alpha\komega=\vv[][0][\Lambda]^\alpha(k,i\omega_n)
    =\vv[][0][\Lambda]_\alpha(k,i\omega_n)&=\vv[][0][\Lambda]_{-\alpha}
    (k,-i\omega_n)=\vv[][0][\Lambda]^{-\alpha}(k,-i\omega_n),
    \label{eq:anomvvsym1}\\
    \vv[][+][\Lambda]^\alpha\komega=\vv[][+][\Lambda]^\alpha(k,i\omega_n)=
    \vv[][+][\Lambda]_\alpha(k,i\omega_n)&=\vv[][-][\Lambda]_{-\alpha}
    (k,-i\omega_n)=\vv[][-][\Lambda]^{-\alpha}(k,-i\omega_n).
    \label{eq:anomvvsym2}
  \end{align}
\end{subequations}

\subsection{Dielectric functions}
\label{sec:dielfn1}

With the definition of the
\begin{subequations}
  \label{eqs:dielfng}
  \begin{align}
    &{\eps}^{sr}_{cd}\komega=\delta^{s}_{d}\delta^{r}_{c}-\Pi^{sr}_{ab}
    \komega V^{ba}_{cd},\label{eq:eps}\\
    &\mat[0][\eps]\komega=\mat[0][1]-\mat[0][\Pi]\komega\c\mat[0][C],
    \label{eq:eps0}\\
    &\mat[\pm][\eps]\komega=\mat[\pm][1]-\mat[\pm][\Pi]\komega\c\mat[\pm][C]
    \label{eq:epspm}
  \end{align}
\end{subequations}
dielectric functions Eqs. \eqref{eq:propeq}, \eqref{eq:prop0m},
\eqref{eq:prop+m} and \eqref{eq:prop-m} can be rewritten as
\begin{subequations}\label{eqs:simpeqs}
  \begin{align}
    &\eps^{sr}_{cd}\komega D^{dc}_{r's'}\komega=\hslash\cn\Pi^{sr}_{r's'}
    \komega,\label{eq:simpeq}\\
    &\mat[0][\eps]\komega\c\mat[0][D]\komega=\hslash\cn\mat[0][\Pi]\komega,
    \label{eq:simpeq0}\\
    &\mat[\pm][\eps]\komega\c\mat[\pm][D]\komega=\hslash\cn\mat[\pm][\Pi]
    \komega.\label{eq:simpeqpm}
  \end{align}
\end{subequations}

\subsection{Interaction propagator}
\label{sec:intprop}

\begin{figure}[ht]
  \begin{center}
    \includegraphics*[30mm,235mm][190mm,275mm]{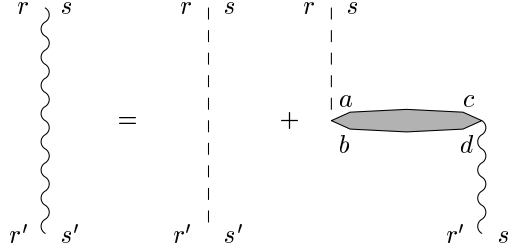}
    \caption{The Feynman graph of the interaction propagator}
    \label{fig:reninter}
  \end{center}
\end{figure}

With the use of the proper parts of the density correlation functions one can
define an interaction propagator:
\begin{equation}
  \label{eq:reninter}
  {\mathcal W}^{rs}_{r's'}=V^{rs}_{r's'}+{\mathcal W}^{rs}_{ab}\Pi^{ba}_{cd}
  V^{dc}_{r's'},
\end{equation}
which can be symbolized as depicted in Fig. \ref{fig:reninter}. Using the
dielectric functions one can get
\begin{subequations}
  \label{eqs:reninterdi}
  \begin{equation}
    {\mathcal W}^{rs}_{ab}\cn\eps^{ba}_{r's'}=V^{rs}_{r's'},
    \label{eq:reninterdi1}    
  \end{equation}
  This equation also splits to parts according to the spin transfer.
  \begin{align}
    \mat[0][{\mathcal W}]\cn\mat[0][\eps]&=\mat[0][C],\label{eq:reninterdi2}\\
    \mat[\pm][{\mathcal W}]\cn\mat[\pm][\eps]&=\mat[\pm][C],
    \label{eq:reninterdi3}\\
    {\mathcal W}^{+-}_{-+}&={\mathcal W}^{-+}_{+-}=0.\label{eq:reninterdi4}
  \end{align}
\end{subequations}

\subsection{Proper Green's function and irreducible polarization part}
\label{sec:progreirpol}

With a simple substitution one can verify that both Eq.
\eqref{eq:dyson} and Eq. \eqref{eq:reninterdi1} have the following
partition property.  If one decomposes the self-energy and the proper
part of \eqref{eq:d6} into two
\begin{align}
  \Sigma^{rs}_{\gamma\delta}=\Sigma^{(1)rs}_{\gamma\delta}+\Sigma^{(2)rs}_{
    \gamma\delta},\label{sfelb}\\
  \Pi^{sr}_{r's'}=\Pi^{(1)sr}_{r's'}+\Pi^{(2)sr}_{r's'},\label{pfelb}
\end{align}
then one can define propagators corresponding e.g. to the first part
of these quantities
\begin{align}
  &\Gr^{(1)rs}_{\gamma\delta}=\Gr^{rs}_{(0)\gamma\delta}+\Gr^{rr'}_{(0)\gamma
    \sigma}\Sigma^{(1)r's'}_{\sigma\rho}\Gr^{(1)s's}_{\rho\delta},
  \label{gfel1}\\
  &{\mathcal W}^{(1)rs}_{r's'}=V^{rs}_{r's'}+{\mathcal W}^{(1)rs}_{ab}
  \Pi^{(1)ba}_{cd}V^{dc}_{r's'}
  \label{vfel1}
\end{align}
in such a way that
\begin{align}
  &\Gr^{rs}_{\gamma\delta}=\Gr^{(1)rs}_{\gamma\delta}+\Gr^{(1)rr'}_{\gamma
    \sigma}\Sigma^{(2)r's'}_{\sigma\rho}\Gr^{s's}_{\rho\delta},
  \label{gfel2}\\
  &{\mathcal W}^{rs}_{r's'}={\mathcal W}^{(1)rs}_{r's'}+{\mathcal W}^{rs}_{ab}
  \Pi^{(2)ba}_{cd}{\mathcal W}^{(1)dc}_{r's'}\label{vfel2}
\end{align}
is fulfilled at the same time. Specially if one decomposes the
self-energies by \eqref{eq:selfdec} and the proper graphs by
\eqref{eq:propdec} it defines the proper Green's functions
\begin{equation}
  \label{eq:propgr}
  \widetilde{\Gr}^{rs}_{\gamma\delta}=\Gr^{rs}_{(0)\gamma\delta}+
  \Gr^{rr'}_{(0)\gamma\sigma}\widetilde{\Sigma}^{r's'}_{\sigma\rho}
  \widetilde{\Gr}^{s's}_{\rho\delta}
\end{equation}
and an effective potential
\begin{equation}
  \label{eq:shpot}
  W^{rs}_{r's'}=V^{rs}_{r's'}+W^{rs}_{ab}\Pi^{(r)ba}_{cd}V^{dc}_{r's'}
\end{equation}
such a way that
\begin{subequations}
  \begin{align}
    \label{eq:propgrtogr}
    \Gr^{rs}_{\gamma\delta}&=\widetilde{\Gr}^{rs}_{\gamma\delta}+
    \widetilde{\Gr}^{rr'}_{\gamma\sigma}M^{r's'}_{\sigma\rho}
    \Gr^{s's}_{\rho\delta},\\
    \label{eq:shpottointerprop}
    {\mathcal W}^{rs}_{r's'}&=W^{rs}_{r's'}+{\mathcal W}^{rs}_{ab}
    \Pi^{(s)ba}_{cd}W^{dc}_{r's'}
  \end{align}
\end{subequations}
is fulfilled. The earlier discussed symmetry properties hold for these
decomposed parts as well (since these are defined by a class of
graphs) and as consequence, Eqs. \eqref{eq:propgr},
\eqref{eq:propgrtogr} and Eqs.  \eqref{eq:shpot},
\eqref{eq:shpottointerprop} split to matrix equations in the same way
as Eq. \eqref{eq:dyson} and Eq.  \eqref{eq:propeq} did. It is
convenient to define the regular part of the dielectric functions with
the
\begin{subequations}
  \label{eqs:regdiel}
  \begin{align}
    \eps^{(r)sr}_{cd}&=\delta^s_d\delta^r_c-\Pi^{(r)sr}_{ab}V^{ba}_{cd}
    \label{eq:regdiel}\\
    \mat[0][\eps]^{(r)}&=\mat[0][1]-\mat[0][\Pi]^{(r)}\c\mat[0][C],
    \label{eq:regeps0}\\
    \mat[\pm][\eps]^{(r)}&=\mat[\pm][1]-\mat[\pm][\Pi]^{(r)}\c\mat[\pm][C]
    \label{eq:regepspm}
  \end{align}
\end{subequations}
equations which can be used to express the effective potential as:
\begin{subequations}
  \label{eqs:shpotwdiel}
  \begin{align}
    \mat[0][W]\cn\mat[0][\eps]^{(r)}&=\mat[0][C],\label{eq:shpotwdiel1}\\
    \mat[\pm][W]\cn\mat[\pm][\eps]^{(r)}&=\mat[\pm][C],\label{eq:shpotwdiel2}\\
    W^{+-}_{-+}&=W^{-+}_{+-}=0.
  \end{align}
\end{subequations}

\subsection{Improper self-energy and singular polarization}
\label{sec:impselsipol}

\begin{figure}[ht]
  \begin{center}
    \includegraphics*[30mm,255mm][190mm,275mm]{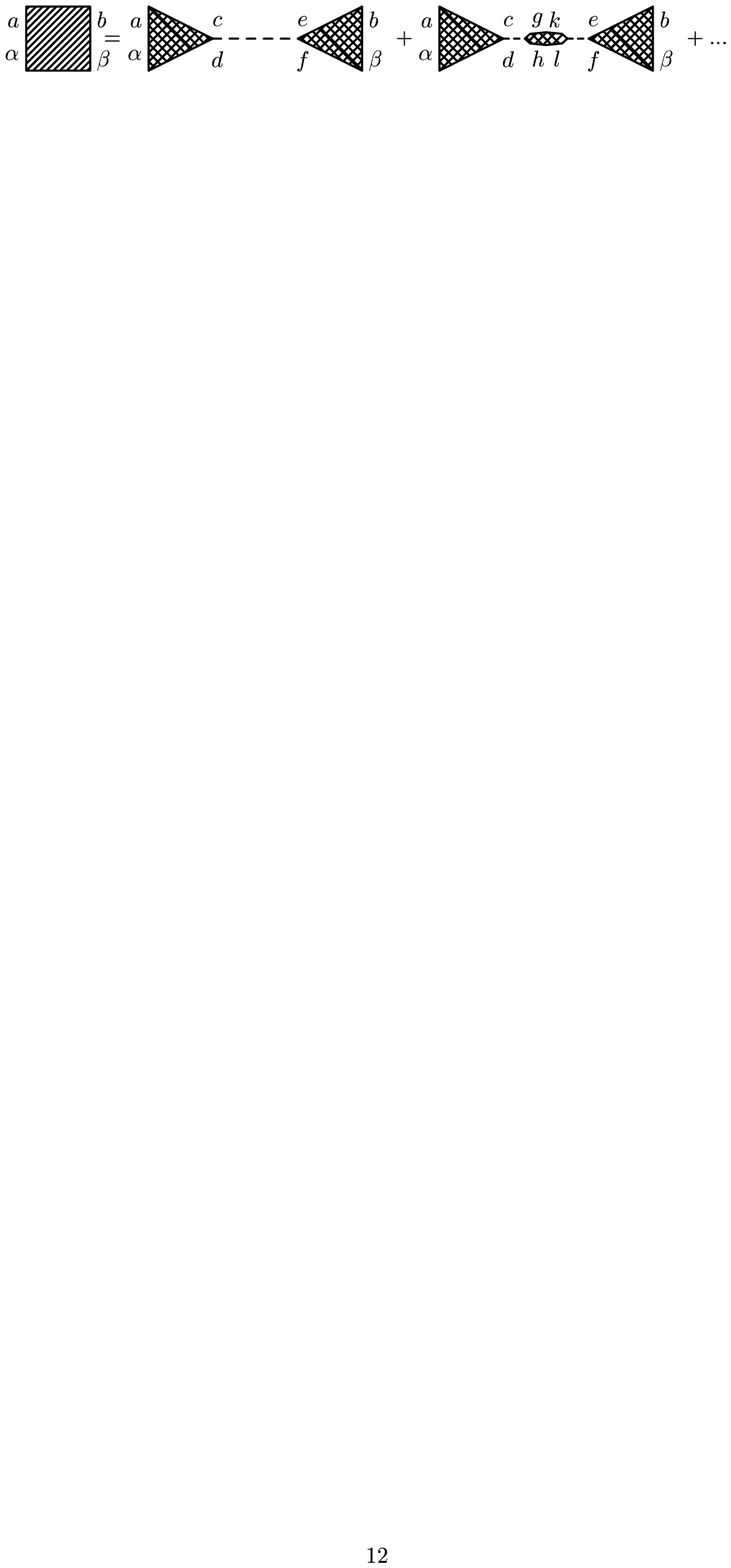}
    \caption{The structure of the perturbation series of the improper self-energy}
    \label{fig:impself}
  \end{center}
\end{figure}

With the use of the irreducible and proper anomalous vertex functions
and the effective potential one can construct the improper self
energies (as seen in Fig. \ref{fig:impself}.):
\begin{subequations}
  \label{eqs:anomself}
  \begin{align}
    \hslash\cn M^{ab}_{\alpha\beta}\komega&=
    \widetilde{\Lambda}^{a\alpha}_{cd}\komega
    W^{dc}_{ef}\komega\widetilde{\Lambda}^{fe}_{b\beta}\komega,
    \label{eq:impself}\\
    \hslash\cn\gmat[0][M]_{\alpha\beta}&=\vv[][0][\widetilde{\Lambda}]^{
      \alpha}\cn\mat[0][W]\cn\vv[][0][\widetilde{\Lambda}]_{\beta}=
    \vv[][0][\widetilde{\Lambda}]^{\alpha}\cn\mat[0][C]\cn\mat[0][\eps]^{
      (r)^{-1}}\cn\vv[][0][\widetilde{\Lambda}]_{\beta},
    \label{eq:anomselmat0}\\
    \hslash\cn\gmat[\pm][M]_{\alpha\beta}&=\vv[][\pm][\widetilde{\Lambda}]^{
      \alpha}\cn\mat[\pm][W]\cn\vv[][\pm][\widetilde{\Lambda}]_{\beta}=\vv[][
    \pm][\widetilde{\Lambda}]^{\alpha}\cn\mat[\pm][C]\cn\mat[\pm][\eps]^{
      (r)^{-1}}\cn\vv[][\pm][\widetilde{\Lambda}]_{\beta}.
    \label{eq:anomselmatpm}
  \end{align}
  It is easy to verify that $W^{-+}_{+-}=0$ which means that for the
  ferromagnetic case the improper self-energy in the $\pm2$
  (quadrupolar) spin transfer mode
  \begin{equation}
    \gmat[\pm Q][M]=0.\label{eq:anomselmatpmQ}
  \end{equation}
\end{subequations}

\begin{figure}[ht]
  \begin{center}
    \includegraphics*[30mm,255mm][190mm,275mm]{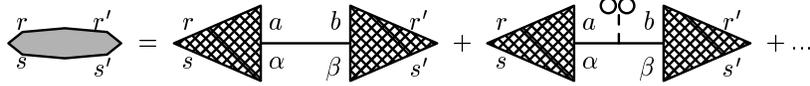}
    \caption{The diagrammatic structure of the singular polarization}
    \label{fig:singpol}
  \end{center}
\end{figure}

The singular polarization can be expressed as well with the use of the
proper Green's functions and the proper anomalous vertices as seen in
Fig. \ref{fig:singpol}. and is of the form:
\begin{subequations}
  \label{eqs:singpol}
  \begin{align}
    \hslash\cn\Pi^{(s)sr}_{r's'}\komega&=
    \widetilde{\Lambda}^{sr}_{a\alpha}
    \komega\widetilde{\Gr}^{ab}_{\alpha\beta}\komega\widetilde{\Lambda}
    ^{b\beta}_{r's'}\komega,\label{eq:singpol}\\
    \hslash\cn\mat[0][\Pi]^{(s)}&=\gmat[0][\widetilde{\Gr}]_{\alpha\beta}
    \cn\vv[][0][\widetilde{\Lambda}]_{\alpha}\circ\vv[][0][\widetilde{
      \Lambda}]^{\beta},\label{eq:singpolmat0}\\
    \hslash\cn\mat[\pm][\Pi]^{(s)}&=\gmat[\pm][\widetilde{\Gr}]_{
      \alpha\beta}\cn\vv[][\pm][\widetilde{\Lambda}]_{\alpha}\circ
    \vv[][\pm][\widetilde{\Lambda}]^{\beta},\label{eq:singpolmat-}
  \end{align}
\end{subequations}
where the circle denotes the diadic product operation. For the polar
case the singular polarization in the $\pm2$ spin transfer mode is
equal to zero since there is no corresponding proper Green's function.
For the ferromagnetic case e.g. for the $+2$ spin transfer mode the
singular polarization is:
\begin{equation}
  \label{eq:singpolferQ}
  \hslash\cn{_f\Pi^{(s)-+}_{+-}}={_f\widetilde{\Lambda}^{-+}_{-,1}}\cn
  {_f\widetilde{\Gr}^{--}_{11}}\cn{_f\widetilde{\Lambda}^{-,1}_{+-}}.
\end{equation}

As for the scalar case \cite{SzK,Griffin} one can see intermediate
states corresponding to collective excitations in the perturbation
series of the one-particle propagators and at the same time one can
identify one-particle intermediate states in the perturbation series
of the correlation functions of the collective modes. However there is
a great difference between the scalar and the spinor models. The
rotational invariance of the interaction potential results in that
$V^{-+}_{+-}=V^{+-}_{-+}=0$ which leads to the proper nature of the
quadrupolar spin density correlation functions. This combined with the
fact that in the polar case there is no anomalous vertex with $\pm2$
spin transfer results in the prohibition of the coupling to any of the
one-particle correlation functions in this phase.

\subsection{Couplings amongst the correlation functions}
\label{sec:gcons}

With the help of the quantities discussed in the previous subsection
the properties of the density correlation functions can be further
investigated. For this reason let us introduce the following vectors:
$\vvec{\xi}_1 = (1,1,1)\adj / \sqrt{3}$, $\vvec{\xi}_2 = (1,0,-1)\adj
/ \sqrt{2}$, $\vvec{\xi}_3 = (1,-2,1)\adj / \sqrt{6}$ for the zero
spin transfer mode and $\vvec{\chi}_1 = (1,1)\adj/ \sqrt{2}$ and
$\vvec{\chi}_2 = (1,-1)\adj/\sqrt{2}$ for the $\pm1$ spin transfer
mode. It is easy to verify that $\{\vvec{\xi}_1, \vvec{\xi}_2,
\vvec{\xi}_3\}$ and $\{\vvec{\chi}_1, \vvec{\chi}_2\}$ are orthogonal
and normalized basis sets in the linear vector spaces with dimension
three and two, respectively. The density correlation functions
\eqref{eq:d1},\eqref{eq:d2},\eqref{eq:d3} and \eqref{eq:d4} can be
expressed as follows:
\begin{subequations}
  \label{eq:dwdm}
  \begin{gather}
    D_{nn}=3\left(\vvec{\xi}_1\adj\cn\mat[0][D]\cn\vvec{\xi}_1\right), \quad 
    D_{zz}=2\left(\vvec{\xi}_2\adj\cn\mat[0][D]\cn\vvec{\xi}_2\right), \quad
    D_{nz}=\sqrt{6}\left(\vvec{\xi}_1\adj\cn\mat[0][D]\cn\vvec{\xi}_2\right),\\
    D_{++}=4\left(\vvec{\chi}_1\adj\cn\mat[+][D]\cn\vvec{\chi}_1\right), \quad
    D_{--}=4\left(\vvec{\chi}_1\adj\cn\mat[-][D]\cn\vvec{\chi}_1\right).
  \end{gather}
\end{subequations}
Since $\mat[0][D]$ is a symmetric matrix $D_{nz}=D_{zn}$ holds.

Multiplying Eq. \eqref{eq:prop+m} with $2\vvec{\chi}_1$ from both
sides and using that $\mat[+][C] = 2 c_s \vvec{\chi}_1 \circ
\vvec{\chi}_1$ one obtains
\begin{equation}
  \label{eq:D++eq1}
  D_{++}=\hslash\Pi_{++}+\frac{c_s}{2}\Pi_{++}D_{++},
\end{equation}
where $\Pi_{++}=4\vvec{\chi}_1\adj\mat[+][\Pi]\vvec{\chi}_1$. A
similar equation can be derived to $D_{--}$. There are both regular
and singular contributions to these propagators.

One can get a closed system of equations for the zero spin transfer
correlation functions of Eq. \eqref{eq:d1}, Eq. \eqref{eq:d2} and Eq.
\eqref{eq:d3}. To this end first we multiply Eq. \eqref{eq:prop0m} with
$\sqrt{3}\vvec{\xi}_1$ from both sides. We further use that
$\mat[0][C] = 3c_n \vvec{\xi}_1 \circ \vvec{\xi}_1+2c_s \vvec{\xi}_2
\circ \vvec{\xi}_2$. The resulting equations sound as
\begin{subequations}
  \begin{equation}
    \label{eq:Dnneq1}
    D_{nn}=\hslash\Pi_{nn}+c_n\Pi_{nn}D_{nn}+c_s\Pi_{nz}D_{nz},
  \end{equation}
  where $\Pi_{nn}=3\vvec{\xi}_1\mat[0][\Pi]\vvec{\xi}_1$ and $\Pi_{nz}
  = \sqrt{6} \vvec{\xi}_1 \mat[0][\Pi] \vvec{\xi}_2$. Multiplying
  \eqref{eq:prop0m} with $\sqrt{2}\vvec{\xi}_2$ from both sides one
  gets
  \begin{equation}
    \label{eq:Dzzeq1}
    D_{zz}=\hslash\Pi_{zz}+c_n\Pi_{nz}D_{nz}+c_s\Pi_{zz}D_{zz},
  \end{equation}
  with $\Pi_{zz}=2\vvec{\xi}_2\mat[0][\Pi]\vvec{\xi}_2$.  The third
  equation can be obtained by multiplying Eq.  \eqref{eq:prop0m} with
  $\sqrt{3}\vvec{\xi}_1$ from the left hand side and with
  $\sqrt{2}\vvec{\xi}_2$ from the right hand side which leads to
  \begin{equation}
    \label{eq:Dnzeq1}
    D_{nz}=\hslash\Pi_{nz}+c_n\Pi_{nn}D_{nz}+c_s\Pi_{nz}D_{zz}.
  \end{equation}
\end{subequations}
These coupled equations can be solved for the correlation functions.
The polarization part $\Pi_{nz}$ written out in detail reads as
\begin{equation}
  \label{eq:Pnz1}
  \Pi_{nz}=\Pi^{++}_{++}-\Pi^{--}_{--}+\Pi^{00}_{++}-\Pi^{00}_{--}.
\end{equation}
It is obviously zero if the system is invariant under spin reflection,
i.e., when those matrix elements coincide which can be get from each
other by reverting a $+$ to a $-$ and vice versa. This condition is
fulfilled for both cases in the symmetric phase, moreover for the
polar case it even holds throughout the condensed phase. With
$\Pi_{nz}=0$ Eqs.  \eqref{eq:Dnneq1} and \eqref{eq:Dzzeq1} are
independent, giving the solutions
\begin{subequations}
  \label{eqs:sepmodes}
  \begin{align}
    D_{nn}&=\frac{\hslash\Pi_{nn}}{1-c_n\Pi_{nn}},\\
    D_{nz}&=0,\\
    D_{zz}&=\frac{\hslash\Pi_{zz}}{1-c_s\Pi_{zz}}\label{eq:Dzzsr}
  \end{align}
\end{subequations}
and leading to two separate excitation spectra. The singular part of
$\Pi_{zz}$ can be cast with Eq.  \eqref{eq:singpolmat0} to the form
\begin{equation}
  \label{eq:singpolzzpol}
  \hslash\Pi^{(s)}_{zz}=2\left(\vvec{\xi}_2
  \vv[][0][\widetilde{\Lambda}]_\alpha\right)\left(
  \vv[][0][\widetilde{\Lambda}]^\beta\vvec{\xi}_2
  \right)\gmat[0][\widetilde{\Gr}]_{\alpha\beta},
\end{equation}
which is zero if spin reflection symmetry is present, since the
anomalous vertex vector $\vv[p][0][\widetilde{\Lambda}]_\alpha$ is
orthogonal to $\vvec{\xi}_2$ in this case.

In the ferromagnetic phase the spectra of collective excitations
corresponding to spin density waves and density waves with $n=0$ are
coupled. Above the critical temperature this coupling vanishes and the
two modes will be independent. In the polar phase spin density
fluctuations will always be independent of the particle density
fluctuations, furthermore $\Pi_{zz}^{(s)} = 0$ which means that the
spin density correlation function \eqref{eq:Dzzsr} has only regular
contribution.

In the polar phase further analysis leads to results important for the
following. First note that, when the system exhibits the spin
reflection symmetry, $\mat[0][D]$ and $\mat[0][\Pi]$ have only 4
independent elements (instead of 6 characteristic to a symmetric
matrix). These independent elements are $\Pi^{++}_{++} =
\Pi^{--}_{--}$, $\Pi^{--}_{++} = \Pi^{++}_{--}$, $\Pi^{00}_{++} =
\Pi^{++}_{00} = \Pi^{00}_{--} = \Pi^{--}_{00}$ and $\Pi^{00}_{00}$.
Second, these can be cast into a more practical form with the help of
the orthogonal matrix
\begin{equation}
  \label{eq:orthtr}
  \mat[0][{\mathcal O}]=\left[\begin{array}{c c c}
      \frac{\sqrt{3}}{3} & \frac{\sqrt{3}}{3} & \frac{\sqrt{3}}{3}\\
      \frac{\sqrt{2}}{2} & 0 & -\frac{\sqrt{2}}{2}\\
      \frac{\sqrt{6}}{6} & -\frac{\sqrt{6}}{3} & \frac{\sqrt{6}}{6}
    \end{array}\right].
\end{equation}
Its rows are built from the vectors $\{\vvec{\xi}_1, \vvec{\xi}_2,
\vvec{\xi}_3\}$. Performing the transformation leads to
\begin{subequations}
  \begin{align}
    \mat[0][C]'&:=\mat[0][{\mathcal O}]\c\mat[0][C]\c\mat[0][{\mathcal O}]^T=
    \left[\begin{array}{c c c}
        3c_n & 0 & 0\\
        0 & 2c_s & 0\\
        0 & 0 & 0
      \end{array}\right],\label{eq:intmatp}\\
    \mat[0][\Pi]^{(x)'}&:=\mat[0][{\mathcal O}]\c\mat[0][\Pi]^{(x)}\c
    \mat[0][{\mathcal O}]^T=
    \left[\begin{array}{c c c}
        \frac{\Pi_{nn}^{(x)}}{3} & 0 & \frac{\Pi_{13}^{(x)}}{3\sqrt{2}}\\
        0 & \frac{\Pi_{zz}^{(x)}}{2} & 0\\
        \frac{\Pi_{13}^{(x)}}{3\sqrt{2}} & 0 & \frac{\Pi_{33}^{(x)}}{6}
      \end{array}\right],\label{eq:propmat0p}
  \end{align}
\end{subequations}
with $\Pi_{13}^{(x)}=3 \sqrt{2} \vvec{\xi}_1 \mat[0][\Pi]^{(x)}
\vvec{\xi}_3$ and $\Pi_{33}^{(x)}=6 \vvec{\xi}_3 \mat[0][\Pi]^{(x)}
\vvec{\xi}_3$, where $x$ can be $r$ or $s$ corresponding to the
regular or the singular parts, respectively.  Furthermore, according
to \eqref{eq:singpolzzpol} $\Pi_{zz}^{(s)}=0$.

The total
polarization matrix therefore becomes
\begin{equation}
  \label{eq:fullpolmat}
    \mat[0][\Pi]'=\left[\begin{array}{c c c}
        \frac{\Pi_{nn}^{(r)}+\Pi_{nn}^{(s)}}{3} & 0 & 
        \frac{\Pi_{13}^{(r)}+\Pi_{13}^{(s)}}{3\sqrt{2}}\\
        0 & \frac{\Pi_{zz}^{(r)}}{2} & 0\\
        \frac{\Pi_{13}^{(r)}+\Pi_{13}^{(s)}}{3\sqrt{2}} & 0 & 
        \frac{\Pi_{33}^{(r)}+\Pi_{33}^{(s)}}{6}
      \end{array}\right].
\end{equation}
Using Eqs. \eqref{eq:intmatp} and \eqref{eq:fullpolmat} the dielectric
function \eqref{eq:eps0}, after the transformation, can be cast into
the form
\begin{equation}
  \label{eq:dielmatp}
  \mat[0][\eps]'=\left[\begin{array}{c c c}
      1-c_n\left(\Pi_{nn}^{(r)}+\Pi_{nn}^{(s)}\right) & 0 & 0\\
      0 & 1-c_s\Pi_{zz}^{(r)} & 0\\
      -\frac{c_n}{\sqrt{2}}\left(\Pi_{13}^{(r)}+\Pi_{13}^{(s)}\right) & 0 & 1
      \end{array}\right].
\end{equation}
We arrive at the important result, that the determinant of the
dielectric function, which is invariant under such transformations,
reads as
\begin{subequations}
  \begin{equation}
    \label{eq:detepsp}
    \det\mat[0][\eps]=\det\mat[0][\eps]'=
    \left[1-c_n\left(\Pi_{nn}^{(r)}+\Pi_{nn}^{(s)}\right)
    \right]\left(1-c_s\Pi_{zz}^{(r)}\right).
  \end{equation}
  This determinant factorize into two in agreement with the separation
  of the density and spin density fluctuations. Furthermore it shares
  one factor with its regular counterpart since
  \begin{equation}
    \label{eq:detepsrp}
    \det\mat[0][\eps]^{(r)}=\left(1-c_n\Pi_{nn}^{(r)}\right)\left(1-c_s
      \Pi_{zz}^{(r)}\right),  
  \end{equation}
\end{subequations}
which can be obtained by taking the determinant of Eq.
\eqref{eq:dielmatp} after setting all $\Pi^{(s)}=0$.

Another important consequence of the spin reflection symmetry is that
the interaction propagator \eqref{eq:reninterdi2} and its regular
part \eqref{eq:shpotwdiel1} has the same structure as the interaction
matrix \eqref{eq:intmatp}:
\begin{subequations}
  \label{eqs:renint}
  \begin{equation}
    \label{eq:reninterp}
    \mat[0][{\mathcal W}]'=
    \left[\begin{array}{c c c}
        3{\mathcal C}_n & 0 & 0\\
        0 & 2{\mathcal C}_s & 0\\
        0 & 0 & 0
      \end{array}\right],\qquad
    \mat[0][W]'=
    \left[\begin{array}{c c c}
        3{\mathcal C}_n^{(r)} & 0 & 0\\
        0 & 2{\mathcal C}_s^{(r)} & 0\\
        0 & 0 & 0
      \end{array}\right],
  \end{equation}
  with
  \begin{align}
    {\mathcal C}_n &= \frac{c_n}{1-c_n\Pi_{nn}},&
    {\mathcal C}_n^{(r)} &= \frac{c_n}{1-c_n\Pi_{nn}^{(r)}}\label{eq:renCn}\\
    {\mathcal C}_s &= \frac{c_s}{1-c_s\Pi_{zz}},&
    {\mathcal C}_s^{(r)} &= \frac{c_s}{1-c_s\Pi_{zz}^{(r)}}.\label{eq:renCs}
  \end{align}
\end{subequations}

Those correlation functions which have singular polarization parts
correspond to fluctuations belonging both to the condensate and to the
noncondensate, while those which have no singular proper part only
belong to the noncondensate (they can not couple to any of the
one-particle modes). So in the \textit{polar phase} the spin density
fluctuations described by $D_{zz}$ (which are independent from the
particle density fluctuations described by $D_{nn}$) belong only to
the noncondensate. The singular polarization parts of the quadrupolar
modes are also zero (since there is no corresponding Green's function
and the anomalous vertices of the $\pm2$ spin transfer are also zero)
meaning that these type of fluctuations also belong only to the
noncondensate.  The proper parts $\Pi_{nn}$, $\Pi_{++}$ and $\Pi_{--}$
have singular contributions too with intermediate one-particle states
$\gmat[0][\widetilde{ \Gr}]_{ \alpha \beta}$, $\gmat[+][\widetilde{
  \Gr}]_{ \alpha \beta}$ and $\gmat[-][\widetilde{ \Gr}]_{ \alpha
  \beta}$, respectively. These modes are coupled to the condensate.
In the \textit{ferromagnetic} phase $D_{zz}$ and $D_{nn}$ are coupled
(their cross correlation $D_{nz}$ do not vanish either), their
denominators are common. All of their proper parts have singular
contributions resulting in their coupling to the one-particle
excitations with $\gmat[0][\widetilde{\Gr}]$.  These fluctuations thus
belong both to the condensate and noncondensate. The correlation
functions of the spin waves (both the $\pm1$ and $\pm2$ spin transfer
modes) $D_{++}$, $D_{--}$, $D^Q_{++}$ and $D^Q_{--}$ have singular
proper parts connecting them with the propagators $\gmat[+][
\widetilde{\Gr} ]$, $\gmat[-][ \widetilde{\Gr} ]$, $\gmat[Q][
\widetilde{\Gr} ]$ and $\gmat[-Q][ \widetilde{\Gr} ]$, respectively.
Note, that there are no improper self-energies for the quadrupolar
spin transfer modes, consequently their proper Green's functions are
the same as the full propagators of these two modes.

\subsection{Coupling between the Green's functions and the correlation
  functions and the spectra of excitations}
\label{sec:det_Green}

A) Let us first consider the modes with $n=0,+,-$. With the
decomposition \eqref{eq:selfdec} of the self-energies and with a
straightforward calculation starting from Eq. \eqref{eq:greg} one
arrives at the expression
\begin{subequations}
  \label{eqs:grwprop}
  \begin{equation}
    \label{eq:grwprop}
    \gmat[n][\Gr]_{\alpha\gamma}\equiv\frac{\gmat[n][N]_{\alpha\gamma}}
      {\gmat[n][\Delta]}=\frac{\gmat[n][\widetilde{N}]_{\alpha\gamma}
      +\alpha\gamma\gmat[n][M]_{-\gamma,-\alpha}}{
      \gmat[n][\widetilde{\Delta}]-
      \gmat[n][\widetilde{N}]_{\sigma\tau}\gmat[n][M]_{\tau\sigma}-\det
      \gmat[n][M]},
  \end{equation}
  both for the polar and ferromagnetic phases. Here we introduced the
  \begin{align}
    \gmat[n][\widetilde{N}]_{\alpha\gamma}&=\delta_{\alpha\gamma} (\alpha
    i\omega_n+\hslash^{-1}\epsilon_{\vec{k}})+\alpha\gamma\gmat[n][
    \widetilde{\Sigma}]_{-\gamma,-\alpha},\\
    \gmat[n][\widetilde{\Delta}]&=(i\omega_n-\hslash^{-1}\epsilon_{\vec{k}}-
    \gmat[n][\widetilde{\Sigma}]_{11})(i\omega_n+\hslash^{-1}
    \epsilon_{\vec{k}}+\gmat[n][\widetilde{\Sigma}]_{-1,-1})+
    \gmat[n][\widetilde{\Sigma}]_{-1,1}\gmat[n][\widetilde{\Sigma}]_{1,-1}
  \end{align}
\end{subequations}
quantities, with $\gmat[n][\widetilde{\Gr}]_{\alpha\gamma} =
\gmat[n][\widetilde{N}]_{\alpha\gamma} / \gmat[n][\widetilde{\Delta}]$
and $\det \gmat[n][M]=\gmat[n][M]_{11} \gmat[n][M]_{-1,-1} -
\gmat[n][M]_{1,-1} \gmat[n][M]_{-1,1}$. With the help of Eqs.
\eqref{eq:simpeq0} and \eqref{eq:simpeqpm} the density correlation
functions with $n=0,+,-$ can be brought to the form
\begin{subequations}
  \label{eqs:dcf}
  \begin{equation}
    \label{eq:dcf}
    \mat[n][D] = \hslash \frac{\mat[n][E] \cn \mat[n][\Pi]}{\det
      \mat[n][\eps]},
  \end{equation}
  by writing
  \begin{equation}
    \label{eq:ninveps}
    \mat[n][E]=\det\left(\mat[n][\eps]\right) \cn \mat[n][\eps]^{-1}.
  \end{equation}
\end{subequations}

An important relationship can be shown in the following way.  First
applying the decomposition $\mat[n][\eps] = \mat[n][\eps]^{(r)} +
\mat[n][\eps]^{(s)}$, with $\mat[n][\eps]^{(s)} = -\mat[n][\Pi]^{(s)}
\cn \mat[n][C]$, to $\det \mat[n][\eps]$ leads to
\begin{equation}
  \label{eq:ced1}
  \det \mat[n][\eps] = \det\left(
  \mat[n][\eps]^{(r)}\right) \det\left(1-\mat[n][\Pi]^{(s)}\cn\mat[n][W]
\right).
\end{equation}
With the help of Eqs. \eqref{eq:singpolmat0} and
\eqref{eq:singpolmat-} and a straightforward but rather lengthy
algebraic manipulation one can find that
\begin{multline}
  \label{eq:detunf}
  \det\left(1-\mat[n][\Pi]^{(s)}\cn\mat[n][W]\right)=\det\left[\mat[][1]-
    \hslash^{-1}\gmat[n][\widetilde{\Gr}]_{\alpha\beta}
    \vv[][n][\widetilde{\Lambda}]_\alpha\circ
    \vv[][n][\widetilde{\Lambda}]^\beta\cn\mat[n][W]\right]=
  1-\hslash^{-1}\gmat[n][\widetilde{\Gr}]_{\alpha\beta}
  \vv[][n][\widetilde{\Lambda}]_\alpha\cn
  \mat[n][W]\cn\vv[][n][\widetilde{\Lambda}]^\beta\\
  +\hslash^{-1}\det\big(\gmat[n][\widetilde{\Gr}]\big)
  \det\big(\vv[][n][\widetilde{\Lambda}]_\alpha\cn\mat[n][W]\cn
  \vv[][n][\widetilde{\Lambda}]^\beta\big).
\end{multline}
Inserting it to Eq. \eqref{eq:ced1} and using Eqs.
\eqref{eq:anomselmat0} and \eqref{eq:anomselmatpm} and that $\det
\gmat[n][\widetilde \Gr]=-\gmat[n][\widetilde \Delta]^{-1}$ and
comparing the result with the denominator of the Green's functions
\eqref{eq:grwprop} one can arrive at the basic connection
\begin{equation}
  \label{eq:match}
  \gmat[n][\widetilde \Delta]\det\mat[n][\eps]
  = \gmat[n][\Delta] \det \mat[n][\eps]^{(r)}.
\end{equation}

Changing to retarded correlation functions, which is done in the usual
way (by analytically continuing in frequency) \cite{FW}, the
elementary excitations of the system are determined by the poles of the
corresponding correlation functions. The spectra of the one-particle
excitations (quasiparticles) are given by the poles of the
Green's functions, or equivalently by the equation
\begin{equation}
  \label{eq:elexc2}
  \gmat[n][\Delta] =0.
\end{equation}
The spectra of collective excitations are determined by the poles of
the density correlation functions, or equivalently by the equation
\begin{equation}
  \label{eq:collexc2}
  \det \mat[n][\eps] = 0.
\end{equation}

Equation \eqref{eq:match} shows that the Green's functions with spin
transfer $n=0,+,-$ can be arranged to have the same denominator as the
density correlation functions with the same spin transfer. This means
that if $\det \mat[n][\eps]$ has a zero (the corresponding density
correlation function, $\mat[n][D]$, has a pole) than
$\gmat[n][\Delta]$ must have a zero ($\gmat[n][\Gr]$ must have a pole)
there as well in general. The zero spin transfer mode in the polar
case is an exception, where both the dielectric function and its
regular part factorize and share some of their zeroes (see Eqs.
\eqref{eq:detepsp} and \eqref{eq:detepsrp}). In this case the zero of
the left hand side coming from $\det \mat[0][\eps]$ satisfies the
equation with the zero coming from $\det \mat[0][\eps]^{(r)}$ of the
right hand side, instead of $\gmat[0][\Delta]$. Hence, this zero do
not appear amongst the poles of $\gmat[0][\Gr]$. This further means
that density correlation functions with $0$ and $\pm1$ spin transfer
has common denominators with the Green's functions corresponding to
the same spin transfer for both the polar and the ferromagnetic cases,
except the $D_{zz}$ spin density correlation function in the polar
case, which do not couple to any of the Green's functions.

B) Concerning the quadrupolar spin waves ($n=\pm2$ modes) let us
consider first the ferromagnetic phase. Both the full Green's function
and the correlation function are proper (see Eq.
\eqref{eq:anomselmatpmQ}, \eqref{eq:propQm} and \eqref{eq:prop-Qm})
for such spin transfers. Consequently one can subtitute in Eq.
\eqref{eq:singpolferQ} the full Green's function leading to
\begin{equation}
  \label{eq:qdcf}
  D^{Q}_{\pm\pm}=4\left(\Pi^{(r)\mp\pm}_{\pm\mp}+\hslash^{-1}
    {_f\widetilde{\Lambda}^{\mp\pm}_{-,\pm1}}\cn
    {_f\Gr^{--}_{\pm1,\pm1}}\cn{_f\widetilde{\Lambda}^{-,\pm1}_{\pm\mp}}
  \right).
\end{equation}
The one-particle excitations then appear also as collective ones.

In the polar case there is no Green's function with $n=\pm2$, see Sec.
\ref{sec:rs_std} therefore these type of collective excitations, do
not show up as one-particle elementary excitations.

Thus in the polar case the collective and one-particle excitation
spectra do not match fully. There are excitation modes in this phase,
which do not belong to the condensate. In the ferromagnetic case all
of the collective excitation modes are connected to corresponding
Green's functions, meaning that all types of density fluctuations are
governed by condensate dynamics.

\section{Bogoliubov approximation}
\label{sec:bogoapp}

In this section we discuss the simplest approximation in the framework
of the dielectric formalism, namely the Bogoliubov approximation. This
approximation is a very low temperature calculation, it neglects all
terms coming from the non-condensate density. This approximation was
studied by other authors as well \cite{Ho2,OM} with other techniques.
With the Green's function method one can give the form of the
correlation functions as well not just the frequency of the modes.

\begin{figure}[ht]
  \begin{center}
    \includegraphics*[30mm,235mm][190mm,275mm]{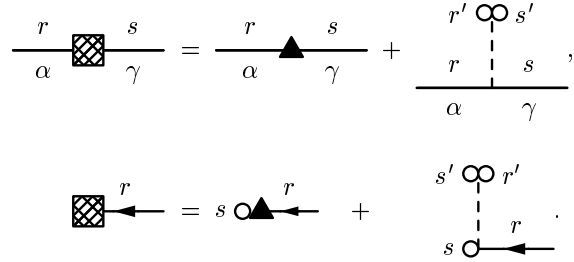}
    \caption{The Feynman graphs of the proper self-energies and the tadpole diagrams in the Bogoliubov approximation}
    \label{fig:psb}
  \end{center}
\end{figure}

The proper self-energies in the Bogoliubov approximation are
\begin{subequations}
  \label{eqs:psb}
  \begin{align}
    \widetilde{\Sigma}^{rs}_{\alpha\gamma}&=\hslash^{-1}\left[ (\mu_0-\mu)
      \delta_{rs}\delta_{\alpha\gamma}+N_0\zeta_{r'}\adj V^{r's'}_{rs}
      \zeta_{s'}\delta_{\alpha\gamma}\right],\label{eq:psb1}\\
    \widetilde{\Sigma}^r_{01}&=\hslash^{-1}\sqrt{N_0}\left[
      (-\mu)\zeta_r+N_0\zeta_{s'}\adj\zeta_s\adj V^{s'r'}_{sr}\zeta_{r'}
    \right].\label{eq:psb2}
  \end{align}
\end{subequations}
The corresponding Feynman-graphs can be seen in Fig. \ref{fig:psb}.

\begin{figure}[ht]
  \begin{center}
    \includegraphics*[30mm,255mm][190mm,275mm]{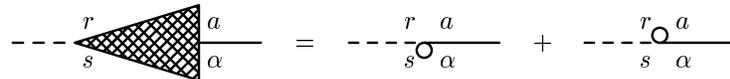}
    \caption{The proper anomalous vertex of the Bogoliubov approximation}
    \label{fig:pavb}
  \end{center}
\end{figure}

The regular polarization is zero in the Bogoliubov approximation and
the proper anomalous vertex reads as
\begin{equation}
  \label{eq:pavb}
  \widetilde{\Lambda}^{sr}_{a\alpha}=\sqrt{N_0}\left[\delta_{ra}
    \delta_{\alpha,-1}\zeta_s+\delta_{sa}\delta_{\alpha,1}\zeta_r\adj\right],
\end{equation}
which can be graphically represented as seen in Fig. \ref{fig:pavb}.

After calculating the proper Green's functions the improper
self-energy and the singular polarization can be constructed from Eqs.
\eqref{eq:impself} and \eqref{eq:singpol}. With their explicit form
known the correlation functions can be expressed. These calculations
have to be made separately for the polar and for the ferromagnetic
cases.

\subsection{Polar case}
\label{sec:pcb}

In the polar case $c_s >0$; the condensate spinor in this phase can be
chosen to $\zeta_r=\delta_{r,0}$. The chemical potential can be
calculated using Eq. \eqref{eq:cons_condfin} which results in
$\mu=N_0c_n$. Since the proper self-energy is zero (except the
diagonal $\mu_0$ terms) the proper Green's functions will be:
\begin{equation}
  \label{eq:pgbp}
  \widetilde{\Gr}^{rs}_{\alpha\gamma}\komega=\frac{\delta_{rs}
    \delta_{\alpha\gamma}}{\alpha i\omega_n-\hslash^{-1}\kink}.
\end{equation}
This is similar to the free propagator of Eq. \eqref{eq:freeprop} but
the $\mu_0$ parameter has been canceled out. The anomalous vertex
vectors are:
\begin{equation}
  \label{eq:avvbp}
    \vv[p][0][\widetilde{\Lambda}]_\alpha=\left(\begin{array}{c}
        0\\
        \sqrt{N_0}\\
        0
      \end{array}\right),\quad
    \vv[p][+][\widetilde{\Lambda}]_1=\left(\begin{array}{c}
        0\\
        \sqrt{N_0}
      \end{array}\right),\quad
    \vv[p][+][\widetilde{\Lambda}]_{-1}=\left(\begin{array}{c}
        \sqrt{N_0}\\
        0
      \end{array}\right).
\end{equation}
With the use of them the singular polarization matrices and the
improper self-energies can be expressed using Eqs. \eqref{eqs:singpol}
and \eqref{eqs:anomself}. The singular polarization matrices are
\begin{equation}
  \label{eq:spbp}
  \mat[0][\Pi]^{(s)}=\left[\begin{array}{c c c}
      0 & 0 & 0\\
      0 & \Pi_S & 0\\
      0 & 0 & 0
      \end{array}\right],\qquad
  \mat[\pm][\Pi]^{(s)}=\left[\begin{array}{c c}
      \Pi_\mp & 0 \\
      0 & \Pi_\pm
      \end{array}\right],
\end{equation}
where $\Pi_\pm=N_0/(\pm i\hslash\omega_n-\kink)$ and
$\Pi_S=\Pi_++\Pi_-$ were introduced. The improper self-energy matrices are:
\begin{equation}
  \label{eq:isebp}
  \mat[0][M]=\hslash^{-1}\left[\begin{array}{c c}
      N_0c_n & N_0c_n\\
      N_0c_n & N_0c_n
    \end{array}\right],\qquad
  \mat[\pm][M]=\hslash^{-1}\left[\begin{array}{c c}
      N_0c_s & N_0c_s\\
      N_0c_s & N_0c_s
    \end{array}\right].  
\end{equation}
Putting all together to Eq. \eqref{eq:grwprop} the Green's functions
will be
\begin{subequations}
  \label{eqs:gfbp}
\begin{align}
  \gmat[0][\Gr]_{\alpha\gamma}&=\frac{\delta_{\alpha\gamma}(\alpha i\omega_n
    +\hslash^{-1}\kink)+\alpha\gamma \hslash^{-1}N_0c_n}{(i\omega_n)^2-
    \hslash^{-2}\kink(\kink+2N_0c_n)},\label{eq:gfbp1}\\
  \gmat[\pm][\Gr]_{\alpha\gamma}&=\frac{\delta_{\alpha\gamma}(\alpha i\omega_n
    +\hslash^{-1}\kink)+\alpha\gamma \hslash^{-1}N_0c_s}{(i\omega_n)^2-
    \hslash^{-2}\kink(\kink+2N_0c_s)}.\label{eq:gfbp2}
\end{align}
\end{subequations}
By the use of Eqs. \eqref{eqs:dcf} the correlation functions
\eqref{0D}, \eqref{+D} and \eqref{-D} can be cast to a form, after
multiplying both the numerator and the denominator with
$\gmat[n][{\widetilde \Delta}]=(i \omega_n - \hslash^{-1} \kink) (i
\omega_n + \hslash^{-1} \kink)$, where the denominators are common
with the corresponding Green's functions in agreement with Eq.
\eqref{eq:match})
\begin{subequations}
  \label{eqs:dcfbp}
  \begin{align}
    \mat[0][D]&=\frac{\hslash^{-1}2N_0\kink}{(i\omega_n)^2-\hslash^{-2}\kink
      (\kink+2N_0c_n)}\left[\begin{array}{c c c}
        0 & 0 & 0\\
        0 & 1 & 0\\
        0& 0 & 0
      \end{array}\right],\label{eq:dc0fbp}\\
    \mat[\pm][D]&=\frac{1}{(i\omega_n)^2-\hslash^{-2}\kink(\kink+2N_0c_s)}
    \left[\begin{array}{c c}
        \mp N_0[i\omega_n\mp\hslash^{-1}(\kink+N_0c_s)]&-c_s\hslash^{-1}N_0^2\\
        -c_s\hslash^{-1}N_0^2&\pm N_0[i\omega_n\pm\hslash^{-1}(\kink+N_0c_s)]
      \end{array}\right].\label{eq:dcpmfbp}
  \end{align}
\end{subequations}
With the use of Eqs. \eqref{eq:dwdm} the particle number and spin
density correlation functions can also be calculated. The result is
that $D_{nn}=2\hslash^{-1}N_0\kink/[(i\omega_n)^2 - \hslash^{-2} \kink
(\kink+ 2 N_0 c_n)]$ while $D_{zz}=D_{nz}=0$. For spin waves
$D_{++}=D_{--}=4\hslash^{-1}N_0\kink/[(i\omega_n)^2 - \hslash^{-2}
\kink (\kink+ 2 N_0 c_s)]$. From Eq. \eqref{eq:qdcf} and Eq.
\eqref{eq:singpol} $D^{Q}_{\pm\pm}=0$ as mentioned before.

\subsection{Ferromagnetic case}
\label{sec:fcb}

In the ferromagnetic case $c_s<0$. One can choose the condensate
spinor as $\zeta_r=\delta_{r,+}$. The calculation is analogous to that
made in the previous subsection. The chemical potential (from Eq.
\eqref{eq:cons_condfin}) is $\mu=N_0(c_n+c_s)\equiv N_0 g$.  The
$\widetilde{\Sigma}^{rs}_{11}$ proper self-energy has 3 different
components in its diagonal (other self-energy components are zero)
which results in the following proper Green's functions:
\begin{subequations}
  \label{eqs:pgbf}
  \begin{align}
    \widetilde\Gr^{++}_{\alpha\gamma}\komega&=\frac{\delta_{\alpha\gamma}}
    {\alpha i\omega_n-\hslash^{-1}\kink},\label{eq:pbgf+}\\
    \widetilde\Gr^{00}_{\alpha\gamma}\komega&=\frac{\delta_{\alpha\gamma}}
    {\alpha i\omega_n-\hslash^{-1}(\kink-N_0c_s)},\label{eq:pbgf0}\\
    \widetilde\Gr^{--}_{\alpha\gamma}\komega&=\frac{\delta_{\alpha\gamma}}
    {\alpha i\omega_n-\hslash^{-1}(\kink-2N_0c_s)}.\label{eq:pbgf-}
  \end{align}
\end{subequations}
There are two things to note. The first one is that the $\mu_0$
parameter has cancelled out here as well, the second one is that the
remaining self-energies (in the last two proper Green's functions) are
negative since $c_s<0$ in the ferromagnetic case. The anomalous vertex
vectors are:
\begin{equation}
  \label{eq:avvbf}
  \vv[f][0][\widetilde{\Lambda}]_\alpha=\left(\begin{array}{c}
        \sqrt{N_0}\\
        0\\
        0
      \end{array}\right),\quad
    \vv[f][+][\widetilde{\Lambda}]_1=\left(\begin{array}{c}
        \sqrt{N_0}\\
        0
      \end{array}\right),\quad
    \vv[f][+][\widetilde{\Lambda}]_{-1}=\vv[][][0].
\end{equation}
The singular polarization matrices are:
\begin{subequations}
  \label{eqs:spbf}
  \begin{equation}
    \label{eq:spbf1}
    \mat[0][\Pi]^{(s)}=\left[\begin{array}{c c c}
        \Pi^{(s)++}_{++} & 0 & 0\\
        0 & 0 & 0\\
        0 & 0 & 0
      \end{array}\right],\quad
    \mat[+][\Pi]^{(s)}=\left[\begin{array}{c c}
        \Pi^{(s)0+}_{+0} & 0 \\
        0 & 0
      \end{array}\right],\quad
    \mat[-][\Pi]^{(s)}=\left[\begin{array}{c c}
        \Pi^{(s)+0}_{0+} & 0 \\
        0 & 0
      \end{array}\right],
  \end{equation}
  with
  \begin{align}
    \Pi^{(s)++}_{++}&=\hslash^{-1}N_0(\widetilde{\Gr}^{++}_{11}+
    \widetilde{\Gr}^{++}_{-1,-1}),\label{eq:spbf2}\\
    \Pi^{(s)0+}_{+0}&=\hslash^{-1}N_0\widetilde{\Gr}^{00}_{1,1},
    \label{eq:spbf3}\\
    \Pi^{(s)+0}_{0+}&=\hslash^{-1}N_0\widetilde{\Gr}^{00}_{-1,-1}.
    \label{eq:spbf4}
  \end{align}
  The singular polarization for the $\pm2$ spin transfer reads as:
  \begin{equation}
    \label{eq:spqbf}
    \Pi^{(s)\mp\pm}_{\pm\mp}=\hslash^{-1}N_0\widetilde{\Gr}^{--}_{\pm1,\pm1}.
  \end{equation}
\end{subequations}

The improper self-energies are:
\begin{equation}
  \label{eq:isebf}
  \mat[0][M]=\hslash^{-1}\left[\begin{array}{c c}
      N_0g & N_0g\\
      N_0g & N_0g
    \end{array}\right],\quad
  \gmat[+][M]=M^{00}_{11}=\hslash^{-1}N_0c_s,\quad
  \gmat[Q][M]=M^{--}_{11}=0,
\end{equation}
where $g=c_n+c_s$. For the ferromagnetic case the resulting Green's
functions are (for the later ones see Eqs. \eqref{eqs:gnreg}):
\begin{subequations}
  \label{eqs:gfbf}
  \begin{align}
    \gmat[0][\Gr]_{\alpha\gamma}&=\Gr^{++}_{\alpha\gamma}=\frac{
      \delta_{\alpha\gamma}(\alpha i\omega_n+\hslash^{-1}\kink)+\alpha\gamma
      \hslash^{-1}N_0g}{(i\omega_n)^2-\hslash^{-2}\kink(\kink+2N_0g)},
    \label{eq:gf0bf}\\
    \gmat[+][\Gr]_{11}&=\Gr^{00}_{11}=\frac{1}{i\omega_n-\hslash^{-1}\kink},
    \label{eq:gfpbf}\\
    \gmat[Q][\Gr]_{11}&=\Gr^{--}_{11}=\frac{1}{i\omega_n-\hslash^{-1}
      (\kink-2N_0c_s)}.\label{eq:gfqbf}
  \end{align}
\end{subequations}
For spin transfers $n=0,+,-$ the correlation functions \eqref{0D},
\eqref{+D} and \eqref{-D} can be evaluated from Eqs. \eqref{eqs:dcf}
and with the use of the polarization matrices \eqref{eqs:spbf}. Both
the numerators and denominators are multiplied with the corresponding
$\gmat[n][{\widetilde \Delta}]$, leading to
\begin{subequations}
  \label{eqs:dcfbf}
  \begin{align}
    \mat[0][D]&=\frac{2\hslash^{-1}N_0\kink}{(i\omega_n)^2-\hslash^{-2}\kink(
      \kink+2N_0g)}\left[\begin{array}{c c c}
        1 & 0 & 0\\
        0 & 0 & 0\\
        0 & 0 & 0
      \end{array}\right],\label{eq:dc0fbf}\\
    \mat[+][D]&=\frac{N_0}{i\omega_n-\hslash^{-1}\kink}\left[\begin{array}{c c}
        1 & 0\\
        0 & 0
        \end{array}\right],\label{eq:dcpfbf}\\
    \mat[-][D]&=\frac{N_0}{-i\omega_n-\hslash^{-1}\kink}
    \left[\begin{array}{c c}
        1 & 0\\
        0 & 0
      \end{array}\right].\label{eq:dcmfbf}\\
  \end{align}
  The correlation functions with $n=\pm2$ obtained directly from Eqs.
  \eqref{eq:propQm} and \eqref{eq:prop-Qm} with polarization functions
  \eqref{eq:spqbf} read as
  \begin{equation}
    D^{\mp\pm}_{\pm\mp}=\frac{N_0}{\pm i\omega_n-\hslash^{-1}(\kink-2N_0c_s)}
      .\label{eq:dcqpmfbf}
    \end{equation}
\end{subequations}
The particle number and spin density correlation functions from Eqs.
\eqref{eq:dwdm} are:
\begin{subequations}
  \label{eqs:dcnbf}
  \begin{align}
    D_{nn}&=D_{zz}=D_{nz}=\frac{2\hslash^{-1}N_0\kink}{(i\omega_n)^2-
      \hslash^{-2}\kink(\kink+2N_0g)},\label{eq:dcnbf1}\\
    D_{++}&=\frac{2N_0}{i\omega_n-\hslash^{-1}\kink},\label{eq:dcnbf2}\\
    D_{--}&=\frac{2N_0}{-i\omega_n-\hslash^{-1}\kink},\label{eq:dcnbf3}\\
    D^Q_{\pm\pm}&=\frac{4N_0}{\pm i\omega_n-\hslash^{-1}(\kink-2N_0c_s)}.
    \label{eq:dcnbf4}
  \end{align}
\end{subequations}

\subsection{Collective excitations in the Bogoliubov approximation}
\label{sec:CBA}

The spectra of collective excitations can be expressed using Eq.
\eqref{eq:collexc2} (for the $0$ and $\pm1$ spin transfer modes) or
equivalently by the zeroes of the denominators of the appropriate
retarded correlation functions. For the polar case this results in
\begin{subequations}
  \label{eqs:spbp}
  \begin{align}
    {^0\omega}&=\hslash^{-1}\sqrt{\kink(\kink+2N_0c_n)}
    \mathop{\rightarrow}_{k\rightarrow0}\sqrt{\frac{N_0c_n}{M}}k,
    \label{eq:spb0p}\\
    {^\pm\omega}&=\hslash^{-1}\sqrt{\kink(\kink+2N_0c_s)}
    \mathop{\rightarrow}_{k\rightarrow0}\sqrt{\frac{N_0c_s}{M}}k.
    \label{eq:spbpmp}
  \end{align}
\end{subequations}
There are no other excitation modes for the polar case in this
approximation. These modes have linear spectrum and are Goldstone
modes. The ${^0\omega}$ mode belongs to the ${^0\Gr_{\alpha\gamma}}$
Green's functions and the $D_{nn}$ ($\mat[0][D]$) correlation
functions, while the ${^\pm\omega}$ mode belongs to the
${^\pm\Gr_{\alpha\gamma}}$ Green's functions and the $D_{\pm\pm}$
($\mat[\pm][D]$) correlation functions.

For the ferromagnetic case the excitational energies are at:
\begin{subequations}
  \label{eqs:spbcf}
  \begin{align}
    {^0\omega}&=\hslash^{-1}\sqrt{\kink(\kink+2N_0g)}
    \mathop{\rightarrow}_{k\rightarrow0}\sqrt{\frac{N_0g}{M}}k,
    \label{eq:spbc0f}\\
    {^\pm\omega}&=\pm\hslash^{-1}\kink,\label{eq:spbcpmf}\\
    {^\pm_Q\omega}&=\pm\hslash^{-1}(\kink-2N_0c_s).\label{eq:spbcqf}
  \end{align}
\end{subequations}
The first (${^0\omega}$) mode, which is responsible to density (and
spin density) fluctuations, belongs to the ${^0\Gr_{\alpha\gamma}}$
Green's functions and the connected $D_{nn}$, $D_{zz}$ and $D_{nz}$
($\mat[0][D]$) correlation functions. This is a linear Goldstone mode.
The next (${^\pm\omega}$) modes are responsible for ordinary spin
waves and belong to the ${^\pm\Gr_{11}}=\Gr^{00}_{\pm\pm}$ Green's
functions and the $D_{\pm\pm}$ ($\mat[\pm][D]$) correlation functions.
These are also Goldstone modes but with a quadratic dispersion. The
last (${^\pm_Q\omega}$) modes describe quadrupolar spin waves. They
belong to the $^\pm_Q\Gr_{11}=\Gr^{--}_{\pm\pm}$ Green's functions and
the $D^Q_{\pm\pm}$ ($D^{\mp\pm}_{\pm\mp}$) correlation functions.
These are non Goldstone modes and they start with a gap. The
frequencies found in the Bogoliubov approximation agree with those of
Refs. \cite{Ho2,OM} obtained in a different manner.

\section{Random Phase Approximation}
\label{sec:RPA}

\begin{figure}[ht]
  \begin{center}
    \includegraphics*[30mm,240mm][190mm,275mm]{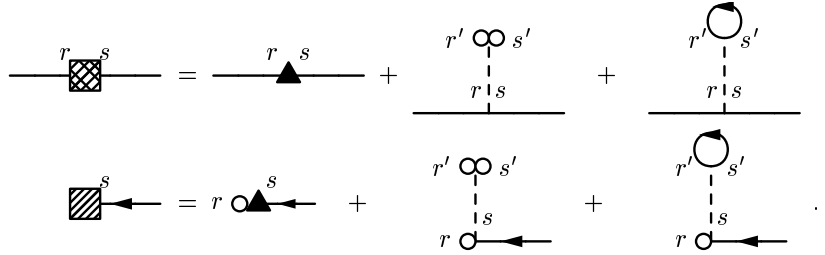}
    \caption{The graphical representation of the proper self-energies and the tadpole graphs in the RPA}
    \label{fig:psr}
  \end{center}
\end{figure}

The simplest way to take into account the appearance of the
noncondensed atoms, leading to a damping mechanism, is to add a
Hartree term to the Bogoliubov proper self-energies (compare Fig.
\ref{fig:psb} with Fig. \ref{fig:psr}) and to choose the regular
polarization graphs as bubbles. Some results of this approximation
were published earlier in \cite{SzSz1}.

Therefore the proper self-energies of the random phase approximation
can be graphically represented as seen in Fig. \ref{fig:psr} and their
contributions are
\begin{subequations}
  \label{eqs:psr}
  \begin{align}
    \widetilde{\Sigma}^{rs}_{\alpha\gamma}&=\hslash^{-1}\left[(\mu_0-\mu)
      \delta_{rs}\delta_{\alpha\gamma}+N_0\zeta_{r'}\adj V^{r's'}_{rs}
      \zeta_{s'}+H^{s'r'}V^{r's'}_{rs}\right],\label{eq:psr1}\\
    \widetilde{\Sigma}^s_{01}&=\hslash^{-1}\sqrt{N_0}\left[(-\mu)\zeta_r\adj
      \delta_{rs}+N_0\zeta_{r'}\zeta_r\adj V^{r's'}_{rs}+H^{s'r'}V^{r's'}_{rs}
    \right].\label{eq:psr2}
  \end{align}
\end{subequations}
Here the $H^{sr}$ notation is introduced for the contribution of the
Hartree term:
\begin{equation}
  \label{eq:hterm}
  H^{rs}=\lim_{\eta=0}\int\frac{d^3q}{(2\pi)^3}\frac{1}{\beta\hslash}
  \sum_{i\nu_n}\left(-\widetilde{\Gr}^{rs}_{11}(\vec{q},i\nu_n)\right)
  e^{i\nu_n\eta}.
\end{equation}

The proper self-energies are independent of the wave-numbers and
frequencies and diagonal both in the Roman and Greek indices. The
proper Green's functions are diagonal in the Roman and Greek indices
as well. The approximation is done in a self-consistent way, namely
the internal propagators used are proper Green's functions in the
Hartree approximation.

\begin{figure}[ht]
  \begin{center}
    \includegraphics*[30mm,255mm][190mm,275mm]{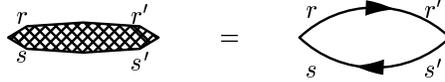}
    \caption{The Feynamn graph of the regular polarization function in the
      random phase (Hartree) approximation}
    \label{fig:regpolrpa}
  \end{center}
\end{figure}

The regular polarization will be a bubble (with proper Green's
functions) as seen in Fig. \ref{fig:regpolrpa}. Its contribution is:
\begin{equation}
  \label{eq:bubbleg}
  -\hslash\Pi^{(r)sr}_{r's'}\komega=\int\frac{d^3q}{(2\pi)^3}
  \frac{1}{\beta\hslash}\sum_{i\nu_n}\widetilde{\Gr}^{r,r'}_{11}(\vec{q},
  i\nu_n)\widetilde{\Gr}^{s's}_{11}(\vec{k}+\vec{q},i\omega_n+i\nu_n).
\end{equation}
Since our approximate proper Green's functions are diagonal in their
spin indices ($r=r'$ and $s=s'$) the regular polarization matrices
will be diagonal as well, with
\begin{equation}
  \label{eq:regpolrpa}
  \mat[0][\Pi]^{(r)}=\left[\begin{array}{c c c}
      \Pi^{(r)++}_{++} & 0 & 0\\
      0 & \Pi^{(r)00}_{00} & 0 \\
      0 & 0 & \Pi^{(r)--}_{--}
    \end{array}\right],\quad
  \mat[+][\Pi]^{(r)}=\left[\begin{array}{c c}
      \Pi^{(r)0+}_{+0} & 0\\
      0 & \Pi^{(r)-0}_{0-}
    \end{array}\right].
\end{equation}

The anomalous vertex remains the same as it was in the Bogoliubov
approximation, see Fig \ref{fig:pavb} and Eq.  \eqref{eq:pavb}. With
the building blocks determined we can now turn our attention to the
construction of the correlation functions.

\subsection{Polar case}
\label{sec:pcr}

Using the interaction \eqref{eq:pot3} with $c_s > 0$ and $\zeta_r =
\delta_{r,0}$ the self-energies \eqref{eqs:psr} in the polar case
\begin{subequations}
  \label{eqs:psrp}
  \begin{align}
    \widetilde{\Sigma}^{++}_{11}&=\hslash^{-1}\left[\mu_0+\left(H^{++}
      -H^{--}\right)c_s\right],\label{eq:psrp1}\\
    \widetilde{\Sigma}^{00}_{11}&=\hslash^{-1}\mu_0,\label{eq:psrp2}\\
    \widetilde{\Sigma}^{--}_{11}&=\hslash^{-1}\left[\mu_0-\left(H^{++}
      -H^{--}\right)c_s\right].\label{eq:psrp3}
  \end{align}
\end{subequations}

Since in the polar case the system has zero magnetisation the Hartree
terms should satisfy $H^{++} = H^{--}$ and as a consequence the proper
Green's functions are the same as they were in the Bogoliubov
approximation Eq. \eqref{eq:pgbp}. The situation is similar to that in
the scalar case, the chemical potential cancels the Hartree terms
\cite{SzK}.

The anomalous vertex vectors are given by Eq. \eqref{eq:avvbp}, and so
the singular polarization matrices will also be the same as for the
Bogoliubov approximation Eq. \eqref{eq:spbp}. The regular polarization
matrices are as follows:
\begin{equation}
  \label{eq:regpolrpap}
  \mat[0][\Pi]^{(r)}=\left[\begin{array}{c c c}
      \Pi_0 & 0 & 0\\
      0 & \Pi_0 & 0 \\
      0 & 0 & \Pi_0
    \end{array}\right],\quad
  \mat[+][\Pi]^{(r)}=\left[\begin{array}{c c}
      \Pi_0 & 0\\
      0 & \Pi_0
    \end{array}\right]
\end{equation}
in accordance with the spin reflection symmetry discussed in Sec.
\ref{sec:gcons}. Here the notation $\Pi_0 \equiv \Pi^{(r)++}_{++} =
\Pi^{(r)00}_{00} = \Pi^{(r)--}_{--}$ is introduced. The interaction
propagators are then obtained from Eqs. \eqref{eq:shpotwdiel1} and
\eqref{eq:shpotwdiel2},
\begin{subequations}
  \label{eqs:renintrp}
  \begin{equation}
    \label{eq:reninterprp}
    \mat[0][W]=
    \left[\begin{array}{c c c}
        {\mathcal C}^{(r)}_n+{\mathcal C}^{(r)}_s & {\mathcal C}^{(r)}_n & 
        {\mathcal C}^{(r)}_n-{\mathcal C}^{(r)}_s\\
        {\mathcal C}^{(r)}_n & {\mathcal C}^{(r)}_n & {\mathcal C}^{(r)}_n\\
        {\mathcal C}^{(r)}_n-{\mathcal C}^{(r)}_s & {\mathcal C}^{(r)}_n &
        {\mathcal C}^{(r)}_n-{\mathcal C}^{(r)}_s
      \end{array}\right],\qquad
    \mat[+][W]=
    \left[\begin{array}{c c}
        {\mathcal C}^{(r)}_s & {\mathcal C}^{(r)}_s \\
        {\mathcal C}^{(r)}_s & {\mathcal C}^{(r)}_s
      \end{array}\right],
  \end{equation}
  with
  \begin{equation}
    \label{eq:renCsrp}
    {\mathcal C}^{(r)}_n\komega = \frac{c_n}{1-3c_n\Pi_0\komega},\qquad
    {\mathcal C}^{(r)}_s\komega = \frac{c_s}{1-2c_s\Pi_0\komega}.
  \end{equation}
\end{subequations}
With Eq. \eqref{eqs:anomself} the improper self-energies become
\begin{equation}
  \label{eq:iserp}
  \mat[0][M]=\hslash^{-1}\left[\begin{array}{c c}
      N_0{\mathcal C}^{(r)}_n & N_0{\mathcal C}^{(r)}_n\\
      N_0{\mathcal C}^{(r)}_n & N_0{\mathcal C}^{(r)}_n
    \end{array}\right],\qquad
  \mat[\pm][M]=\hslash^{-1}\left[\begin{array}{c c}
      N_0{\mathcal C}^{(r)}_s & N_0{\mathcal C}^{(r)}_s\\
      N_0{\mathcal C}^{(r)}_s & N_0{\mathcal C}^{(r)}_s
    \end{array}\right].  
\end{equation}
These can be used in Eqs. \eqref{eqs:grwprop} to arrive at the Green's
functions in RPA. After multiplying both the numerator and denominator
with $\det\mat[n][\eps]^{(r)}$ we obtain
\begin{subequations}
  \label{eqs:gfrp}
\begin{align}
  \gmat[0][\Gr]_{\alpha\gamma}&=\frac{\delta_{\alpha\gamma}(\alpha i\omega_n
    +\hslash^{-1}\kink)(1-3c_n\Pi_0)+\alpha\gamma \hslash^{-1}N_0c_n}
  {[(i\omega_n)^2-\hslash^{-2}\kink^2](1-3c_n\Pi_0)-2\hslash^{-2}\kink N_0c_n}
  ,\label{eq:gfrp1}\\
  \gmat[\pm][\Gr]_{\alpha\gamma}&=\frac{\delta_{\alpha\gamma}(\alpha i\omega_n
    +\hslash^{-1}\kink)(1-2c_s\Pi_0)+\alpha\gamma \hslash^{-1}N_0 c_s}
  {[(i\omega_n)^2-\hslash^{-2}\kink^2](1-2c_s\Pi_0)-2\hslash^{-2}\kink N_0c_s}.
  \label{eq:gfrp2}
\end{align}  
\end{subequations}

Calculating the regular dielectric functions from Eqs.
\eqref{eqs:regdiel} and using Eqs. \eqref{eqs:dcf} the correlation
function matrices could be explicitly given. But in the polar case one
can use Eqs. \eqref{eqs:sepmodes} to calculate directly the relevant
correlation functions in the zero spin transfer, with $\Pi_{nn} = 3
c_n \Pi_0 +c_n \Pi_S$ and $\Pi_{zz} = 2 c_s \Pi_0$ leading to
\begin{subequations}
  \label{eqs:rdcrp}
  \begin{align}
    D_{nn} & =\hslash\frac{3\Pi_0[(i\omega_n)^2-\hslash^{-2}\kink^2]+
      2\hslash^{-2}N_0\kink}{[(i\omega_n)^2-\hslash^{-2}\kink^2](1-3c_n\Pi_0)
      -2\hslash^{-2}\kink N_0c_n},\label{eq:rdcrp1}\\
    D_{zz} & =\hslash\frac{2\Pi_0}{1-2c_s\Pi_0},\label{eq:rdcrp2}\\
    D_{nz} & = 0.
  \end{align}
  With Eq. \eqref{eq:D++eq1} and with $\Pi_{\pm\pm}=4\Pi_0+2\Pi_S$ the
  correlation function of the spin waves will become
  \begin{equation}
    \label{eq:rdcrp3}
    D_{\pm\pm}=\hslash\frac{4\Pi_0[(i\omega_n)^2-\hslash^{-2}\kink^2]+
      2\hslash^{-2}N_0 \kink}{[(i\omega_n)^2-\hslash^{-2}\kink^2]
      (1-2c_s\Pi_0)-2\hslash^{-2}\kink N_0c_s}.
  \end{equation}
  The correlation functions of the quadrupolar spin waves come from
  Eq. \eqref{eq:qdcf} with a result of
  \begin{equation}
    \label{eq:rdcrp4}
    D^{Q}_{\pm\pm}=\hslash4\Pi_0.
  \end{equation}
\end{subequations}
One can easily verify that the denominators of the appropriate
correlation functions match as was shown at the general formalism.

\subsection{Ferromagnetic case}
\label{sec:fcr}

In this phase $c_s<0$ and $\zeta_r=\delta_{r,+}$ and with the
interaction \eqref{eq:pot3} the self-energies \eqref{eqs:psr} become
\begin{subequations}
  \label{eqs:psrf}
  \begin{align}
    \widetilde{\Sigma}^{++}_{11}&=\hslash^{-1}\mu_0,\label{eq:psrf1}\\
    \widetilde{\Sigma}^{00}_{11}&=\hslash^{-1}\left(\mu_0-{\mathcal M}
      c_s\right),\label{eq:psrf2}\\
    \widetilde{\Sigma}^{--}_{11}&=\hslash^{-1}\left(\mu_0-2{\mathcal M}
      c_s\right),\label{eq:psrf3}
  \end{align}
  where the newly introduced quantity
  \begin{equation}
    \label{eq:psrf4}
    {\mathcal M}=N_0+H^{++}-H^{--}
  \end{equation}
\end{subequations}
is the total magnetisation of the system. It describes the number of
particles responsible for the magnetic mean-field. Note that since the
Hartree terms are momentum independent the proper self-energies will
also be.  Therefore the proper Green's functions for the ferromagnetic
case read as
\begin{subequations}
  \label{eqs:pgrf}
    \begin{align}
    \widetilde\Gr^{++}_{\alpha\gamma}\komega&=\frac{\delta_{\alpha\gamma}}
    {\alpha i\omega_n-\hslash^{-1}\kink},\label{eq:pgrf1}\\
    \widetilde\Gr^{00}_{\alpha\gamma}\komega&=\frac{\delta_{\alpha\gamma}}
    {\alpha i\omega_n-\hslash^{-1}(\kink-{\mathcal M}c_s)},\label{eq:pgrf2}\\
    \widetilde\Gr^{--}_{\alpha\gamma}\komega&=\frac{\delta_{\alpha\gamma}}
    {\alpha i\omega_n-\hslash^{-1}(\kink-2{\mathcal M}c_s)}.\label{eq:pgrf3}
  \end{align}
\end{subequations}
These proper Green's functions are similar to their Bogoliubov
counterparts only the momentum independent mean field has changed with
the appearance of the noncondensed particles with spin projection $+$
and $-$.  The chemical potential here do not fully cancels the proper
self-energies leaving behind the energy shift due to the magnetic mean
field.

The anomalous vertex vectors are given by Eqs. \eqref{eq:avvbf}. One
can express the singular polarization matrices from Eqs.
\eqref{eqs:singpol} for $n=0,+,-$, and they are of the same form as
their Bogoliubov counterparts \eqref{eq:spbf1} with elements
\eqref{eq:spbf2}, \eqref{eq:spbf3} and \eqref{eq:spbf4} where the
proper Green's functions \eqref{eqs:pgrf} are to be used. The
singular polarization for the $\pm2$ spin transfer can also be brought
to the form \eqref{eq:spqbf} with \eqref{eq:pgrf3} as the proper
Green's function,
\begin{equation}
  \label{eq:spqrf}
  \Pi^{(s)\mp\pm}_{\pm\mp}=\hslash^{-1}N_0
  \widetilde{\Gr}^{--}_{\pm1,\pm1}.
\end{equation}

Using Eqs. \eqref{eqs:regdiel} and Eq. \eqref{eq:regpolrpa} the
effective potential $W$ can be constructed. Inserting it to Eqs.
\eqref{eqs:anomself} the improper self-energies will read as
\begin{subequations}
  \label{eqs:iserf}
  \begin{equation}
    \label{eq:iserf1}
    \mat[0][M]=\frac{\hslash^{-1}N_0\rho}{\det\mat[0][\eps]^{(r)}}\left[
      \begin{array}{c c}
        1 & 1\\
        1 & 1
      \end{array}\right],\quad
    \mat[+][M]=\frac{\hslash^{-1}N_0c_s}{\det\mat[+][\eps]^{(r)}}\left[
      \begin{array}{c c}
        1 & 0\\
        0 & 0
      \end{array}\right],\quad
    \gmat[Q][M]=M^{--}_{11}=0,
  \end{equation}
  where the following notations are introduced to simplify the
  equations:
  \begin{align}
    \rho&=c_n+c_s-c_nc_s(4\Pi^{(r)--}_{--}+\Pi^{(r)00}_{00}),\\
    \det\mat[0][\eps]^{(r)}&=\left(1-c_n\Pi^{(r)}_{nn}\right)\left(1-c_s
      \Pi^{(r)}_{zz}\right)-c_nc_s{\Pi^{(r)}_{nz}}^2,\\
    \det\mat[+][\eps]^{(r)}&=1-\frac{c_s}{2}\Pi^{(r)}_{++},
  \end{align}
  with
  \begin{align}
    \Pi^{(r)}_{nn} & = \Pi^{(r)++}_{++}+\Pi^{(r)00}_{00}+\Pi^{(r)--}_{--},\\
    \Pi^{(r)}_{zz} & = \Pi^{(r)++}_{++}+\Pi^{(r)--}_{--},\\
    \Pi^{(r)}_{nz} & = \Pi^{(r)++}_{++}-\Pi^{(r)--}_{--},\\
    \Pi^{(r)}_{++} & = 2\left(\Pi^{(r)0+}_{+0}+\Pi^{(r)+0}_{0+}\right).
  \end{align}
\end{subequations}
From Eqs. \eqref{eqs:grwprop} and from Eq. \eqref{eq:gnregQf} the
Green's functions are
\begin{subequations}
  \label{eqs:gfrf}
  \begin{align}
    \gmat[0][\Gr]_{\alpha\gamma}&=\Gr^{++}_{\alpha\gamma}=\frac{
      \delta_{\alpha\gamma}(\alpha i\omega_n+\hslash^{-1}\kink)
      \det\mat[0][\eps]^{(r)}+\alpha\gamma\hslash^{-1}N_0\rho}{\left[
        (i\omega_n)^2-\hslash^{-2}\kink^2\right]\det\mat[0][\eps]^{(r)}-
      2\hslash^{-2}\kink N_0\rho},\label{eq:gf0rf}\\
    \gmat[+][\Gr]_{11}&=\Gr^{00}_{11}=\frac{\det\mat[+][\eps]^{(r)}}{\left[i
        \omega_n-\hslash^{-1}\left(\kink-{\mathcal M}c_s\right)\right]
      \det\mat[+][\eps]^{(r)}-\hslash^{-1}N_0c_s},
    \label{eq:gfprf}\\
    \gmat[Q][\Gr]_{11}&=\Gr^{--}_{11}=\frac{1}{i\omega_n-\hslash^{-1}
      (\kink-2{\mathcal M}c_s)}.\label{eq:gfqrf}
  \end{align}
\end{subequations}

Equations \eqref{eq:dcf} and \eqref{eq:qdcf} give the correlation
function matrices which combined with Eqs.  \eqref{eq:dwdm} result in
\begin{subequations}
  \label{eqs:dcnrf}
  \begin{align}
    D_{nn}&=\hslash\frac{\left(\Pi^{(r)++}_{++}+\Pi^{(s)++}_{++}\right)\left[
        1-c_s\left(\Pi^{(r)00}_{00}+4\Pi^{(r)--}_{--}\right)\right]+
      \Pi^{(r)00}_{00}+\Pi^{(r)--}_{--}\left(1-c_s\Pi^{(r)00}_{00}\right)}{
      \det\mat[0][\eps]},\label{eq:dcnrf1}\\
    D_{zz}&=\hslash\frac{\left(\Pi^{(r)++}_{++}+\Pi^{(s)++}_{++}\right)\left[
        1-c_n\left(\Pi^{(r)00}_{00}+4\Pi^{(r)--}_{--}\right)\right]
      +\Pi^{(r)--}_{--}\left(1-c_n\Pi^{(r)00}_{00}\right)}{
      \det\mat[0][\eps]},\label{eq:dcnrf2}\\
    D_{nz}&=\hslash\frac{\Pi^{(r)++}_{++}+\Pi^{(s)++}_{++}-\Pi^{(r)--}_{--}}{
      \det\mat[0][\eps]},\label{eq:dcnrf3}\\
    D_{++}&=2\hslash\frac{\Pi^{(r)0+}_{+0}+\Pi^{(s)0+}_{+0}+\Pi^{(r)+0}_{0+}}{
      \det\mat[+][\eps]},
    \label{eq:dcnrf4}\\
    D^Q_{++}&=4\hslash\left(\Pi^{(r)-+}_{+-}+\Pi^{(s)-+}_{+-}\right).
    \label{eq:dcnbf5}
  \end{align}
  Here
  \begin{align}
    \det\mat[0][\eps]&=\left(1-c_n\Pi_{nn}\right)
    \left(1-c_s\Pi_{zz}\right)-c_nc_s\Pi_{nz}^2,\\
    \det\mat[+][\eps]&=1-\frac{c_s}{2}\Pi_{++},
  \end{align}
\end{subequations}
where $\Pi_{nn} = \Pi^{(r)}_{nn} + \Pi^{(s)++}_{++}$, $\Pi_{zz} =
\Pi^{(r)}_{zz} + \Pi^{(s)++}_{++}$, $\Pi_{nz} = \Pi^{(r)}_{nz} +
\Pi^{(s)++}_{++}$ and $\Pi_{++} = \Pi^{(r)}_{++} + \Pi^{(s)0+}_{+0}$.
Equations \eqref{eq:dcnrf1}, \eqref{eq:dcnrf2}, \eqref{eq:dcnrf3} and
\eqref{eq:dcnrf4} can be cast to a form having common denominators
with the corresponding Green's functions by multiplying both their
numerators and denominators with the apropriate $\gmat[n][{\widetilde
  \Delta}]$. Only writig down the resulting denominators:
\begin{subequations}
  \label{eqs:denomirf}
  \begin{align}
    \gmat[0][{\widetilde \Delta}] \det\mat[0][\eps] & = [(i\omega_n)^2 -
    \hslash^{-2}\kink^2]\det\mat[0][\eps]^{(r)}-2\hslash^{-1}\kink N_0 \rho\\
    \gmat[+][{\widetilde \Delta}] \det\mat[+][\eps] & = [(i\omega_n) -
    \hslash^{-1}(\kink-{\mathcal M}c_s)]\det\mat[+][\eps]^{(r)}-
    \hslash^{-1}N_0 c_s
  \end{align}
\end{subequations}

For the rest of the paper we will deal with the retarded correlation
functions which can be obtained in the usual way by analytically
continuing in frequencies (see e.g. Ref.\cite{FW}). First however we
will discuss shortly the static properties.

\subsection{Static properties of the spin-1 Bose gas in the random phase approximation}
\label{sec:sptsBg1}

We will study the equation of state and connect it to the density
autocorrelation function through the compressibility sum rule. The
equation of state is investigated by choosing $\mu,N$ as conjugated
variables. The number of particles $N$ is given by
\begin{equation}
  \label{eq:numpart}
  N=N_0+H^{ss},
\end{equation}
where $H^{sr}$ is defined by Eq. \eqref{eq:hterm}. In the ordered
phase Eqs. \eqref{eq:cons_condfin}, \eqref{eq:psrf4} and
\eqref{eq:numpart} provide a second relationship
\begin{equation}
  \label{eq:chpot1}
  \mu=c_n N + c_s {\mathcal M}.
\end{equation}
It is convenient to introduce the following quantities: the thermal
wavelength $\lambda=\hslash(\beta/2M)^{1/2}$, the critical particle
density of the ideal gas $N_c = 3\Gamma\left({\scriptstyle
    \frac{3}{2}} \right)\zeta({\scriptstyle \frac{3}{2}})/[(2\pi)^2
\lambda^3]$, dimensionless interaction strengths $\epsilon_{n,s} = N_c
\beta |c_{n,s}|$, relative total magnetisation $m = {\mathcal M}/N_c$,
relative particle number $x = N/N_c$ and dimensionless chemical
potential $u = \beta\mu$. Here $\Gamma(s)$ is the Gamma function and
$\zeta(s)$ is the Riemann-zeta function.

Both for the polar and ferromagnetic cases in the symmetric phase
(where no condensate is present and the total magnetisation is zero)
the equation of state reads as
\begin{equation}
  \label{eq:statesym}
  x=\frac{1}{\zeta({\scriptstyle \frac{3}{2}})}F\left({\scriptstyle
      \frac{3}{2}},x\epsilon_n-u\right),
\end{equation}
where
\begin{equation}
  \label{eq:Bei}
  F(s,\gamma)=\frac{1}{\Gamma(s)}\int_0^\infty\frac{t^{s-1}}{e^{t+\gamma}-1}dt
\end{equation}
is the Bose-Einstein integral.

In the condensed phase for the polar case the chemical potential is
determined by Eq. \eqref{eq:cons_condfin} which leads to
\begin{equation}
  \label{eq:stateconpol}
  u=\epsilon_n x.
\end{equation}
Equations \eqref{eq:statesym} and \eqref{eq:stateconpol} can be used
to give the isotherms $u(x)$ of the polar case. Equation
\eqref{eq:statesym} is valid when $x<1$ and Eq. \eqref{eq:stateconpol}
is to be used if $x>1$. These isotherms show the character of a
continuous phase transition.

In the condensed phase for the ferromagnetic case the chemical
potential can be calculated also from Eq. \eqref{eq:cons_condfin}
\begin{subequations}
  \label{eqs:stateconfer}
  \begin{equation}
    \label{eq:stateconfer}
    u=\epsilon_n x-\epsilon_s m(x)=\epsilon_s[\kappa x - m(x)],
  \end{equation}
  where $\kappa=\epsilon_n/\epsilon_s$ and $m(x)$ is the relative
  total magnetisation as a function of the relative total density
  given by the
  \begin{equation}
    \label{eq:magnet}
    m = x - \frac{1}{3 \zeta\left({\scriptstyle \frac{3}{2}}\right)}\left[
      F\left({\scriptstyle \frac{3}{2}}, m\epsilon_s\right)+2
      F\left({\scriptstyle \frac{3}{2}}, 2m\epsilon_s\right)\right]
  \end{equation}
\end{subequations}
equation. The isotherms are given by Eqs. \eqref{eq:statesym},
\eqref{eq:stateconfer} and \eqref{eq:magnet}. The solution of Eq.
\eqref{eq:magnet} is unique if $x>1$ but at $x<1$ a second solution
emerges also with nonzero magnetization and both solutions can be
continued down to $x_v<1$, where they coincide and vanish. This means
that the chemical potential has three solutions between $x=x_v$ and
$x=1$. (One of them with zero magnetization at a given density.) In
the applications $\epsilon_s \ll 1$, so the Bose-Einstein integral can
be well approximated by $F({\scriptstyle \frac{3}{2}}, \alpha) \approx
\zeta({\scriptstyle \frac{3}{2}})-2\sqrt{\pi\alpha}$ leading to
\begin{equation}
  \label{eq:spp1}
  x_v = 1-\frac{(1+2\sqrt{2})^2\pi\epsilon_s}{9\zeta({\scriptstyle
      \frac{3}{2}})^2}.
\end{equation}
At $x=1$ the two solutions are at
\begin{equation}
  \label{eq:twosols}
  m_{1,2}(x=1) = \left\{\begin{array}{c}
        \frac{4(1+2\sqrt{2})^2\pi\epsilon_s}{9\zeta({\scriptstyle
      \frac{3}{2}})^2},\\
  0.
  \end{array}\right.
\end{equation}
These isotherms show the character of a very weak first order phase
transition. A similar equation of state with a first order transition
was derived in \cite{FRSzG} for scalar condensates but in a different
model approximation. In that work the first order character of the
phase transition was caused by the exchange interaction (Fock term)
which is known to lead to such behaviour \cite{SG}. The situation is
interesting, since for scalar particles the RPA (Hartree
approximation) gave a continuous phase transition as here for the
polar case \cite{SzK,Griffin}.

Since the above behaviour of the equation of state is not common a
further consistency check is required.  The compressibility sum rule
is a candidate, since it relates the static correlation functions to
the derivatives of the equation of state \cite{Nozieres1,FRSzG},
namely
\begin{equation}
  \label{eq:compsumrule}
  \left(\frac{\partial N}{\partial \mu}\right)_T = \hslash^{-1} D_{nn}
  (k\rightarrow0,\omega=0).
\end{equation}
The inverse of the left hand side of Eq. \eqref{eq:compsumrule} can be
easily calculated, since
\begin{equation}
  \label{eq:lhscsr}
  \left(\frac{\partial \mu}{\partial N}\right)_T = \frac{1}{\beta N_c}
  u'(x)=\frac{1}{\beta N_c}\left[\kappa\epsilon_s-m'(x)\epsilon_s\right].
\end{equation}
The derivative $m'(x)$ can be expressed from Eq. \eqref{eq:magnet}
with the identity $\partial_\alpha F(s,\alpha)=-F(s-1,\alpha)$,
leading to
\begin{equation}
  \label{eq:csrf}
  \left(\frac{\partial \mu}{\partial N}\right)_T = \frac{c_n + c_s -c_n c_s
    \left[\Pi^{(r)00}_{00}(0,0)+4\Pi^{(r)--}_{--}(0,0)\right]}{1-c_s \left[
      \Pi^{(r)00}_{00}(0,0)+4\Pi^{(r)--}_{--}(0,0)\right]},
\end{equation}
where we used that $\Pi^{(r)00}_{00}(0,0)=-\beta N_c F({\scriptstyle
  \frac{1}{2}}, m\epsilon_s) /(3\zeta({\scriptstyle \frac{3}{2}}))$
and $\Pi^{(r)--}_{--}(0,0)=-\beta N_c F({\scriptstyle \frac{1}{2}},
2m\epsilon_s) /(3\zeta({\scriptstyle \frac{3}{2}}))$.  It is
straightforward to verify that the static limit of $\hslash/D_{nn}$ is
just the right hand side of Eq. \eqref{eq:csrf} in agreement with Eq.
\eqref{eq:compsumrule}. The point $x_v$ where the two solutions $m_1$
and $m_2$ coincide is the point where the graph of $u(x)$ has a
vertical tangent. The point where the graph of $u(x)$ has a horizontal
tangent is at where the density autocorrelation function has its pole
at $q=0$ and $\omega=0$. This happens at
\begin{equation}
  \label{eq:spp2}
  x_h=1-\frac{\kappa^2-2\kappa}{(\kappa-1)^2}\frac{(1+2\sqrt{2})^2\pi
    \epsilon_s}{9\zeta({\scriptstyle \frac{3}{2}})^2}.
\end{equation}
For $x>x_h$ (in the condensed phase) the static correlation functions
have no real positive pole in the $q$ variable.

\subsection{Collective excitations in the random phase approximation}
\label{sec:cerpa}

The excitation spectra can be obtained in the usual way as the poles
of the retarded Green's functions and correlation functions. The real
parts of the poles correspond to the energies while the imaginary
parts correspond to the damping (the inverse of their lifetimes) of
the excitations.

Our further calculations can be done more appropriately in a
dimensionless form. For this reason let us further introduce the
following characteristic lengths: $\xi^B_{n,s} = \hslash(4 M
N_0|c_{n,s}|)^{-1/2}$ the mean field correlation lengths associated
with the interaction strength $c_n$ and $c_s$, and $\xi' =
M/(4\pi\hslash^2N_0\beta)$ the length scale of the critical
fluctuations. The calculation will be limited to the intermediate
temperature region. This region is defined by the $\xi^B_{n,s} \gg
\xi', \lambda$ conditions. In this temperature range the $k \lambda \ll
1$ condition can be fulfilled for the physically interesting wave
numbers and the contribution of the bubble diagrams can be taken as
perturbation for the frequencies considered.

A dimensionless frequency can be introduced with
\begin{equation}
  \label{eq:dfr}
  \Omega=\frac{\hslash\omega}{\kink}.
\end{equation}

The singular polarization functions in the polar case take the
following form:
\begin{subequations}
  \label{eqs:spdl}
  \begin{equation}
    c_{n,s}\Pi_{S}\kaomega=\left(\frac{\lambda}{\xi^B_{n,s}}\right)^2
    \frac{1}{(k\lambda)^2}\frac{1}{\Omega^2-1}.\label{eq:spdl1}
  \end{equation}
  While in the ferromagnetic case they read as
  \begin{align}
    |c_{n,s}|\Pi^{(s)++}_{++}\kaomega&=\left(\frac{\lambda}{\xi^B_{n,s}}
    \right)^2\frac{1}{(k\lambda)^2}\frac{1}{\Omega^2-1}\label{eq:spdl2},\\
    |c_s|\Pi^{(s)0+}_{+0}\kaomega&=\frac{\gamma_0}{(k\lambda)^2(\Omega-1)
      -{\widetilde \gamma}}\label{eq:spdl3},\\
    |c_s|\Pi^{(s)-+}_{+-}\kaomega&=\frac{\gamma_0}{(k\lambda)^2(\Omega-1)
      -2{\widetilde \gamma}},\label{eq:spdl4}
  \end{align}
with
\begin{align}
  \gamma_0 & = \beta N_0 |c_s|, \label{eq:spdl5}\\
  {\widetilde \gamma}&=\beta{\mathcal M}|c_s|. \label{eq:spdl6}
\end{align}
\end{subequations}

For $k \lambda \ll 1$ and $|\Omega k \lambda|\ll 1$ the regular
polarization function (the bubble) in the polar case can be
approximated as
\begin{subequations}
  \label{eqs:rpdl}
  \begin{equation}
    \label{eq:rpdl1}
    c_{n,s}\Pi_0\kaomega \approx \frac{(\lambda\xi')}{(\xi^B_{n,s})^2}
    \frac{1}{k\lambda}\frac{i}{2} \ln\frac{\Omega-1}{\Omega+1},
  \end{equation}
  while in the ferromagnetic case they read as
  \begin{align}
    c_{n,s}\Pi^{(r)++}_{++}\kaomega &\approx \frac{(\lambda\xi')}
    {(\xi^B_{n,s})^2}\frac{1}{k\lambda}\frac{i}{2}
    \ln\frac{\Omega-1}{\Omega+1},\label{eq:rpdl2}\\
    c_{n,s}\Pi^{(r)00}_{00}\kaomega &\approx \frac{(\lambda\xi')}
    {(\xi^B_{n,s})^2}\frac{1}{k\lambda}\frac{i}{2}
    \ln\frac{\Omega-1+2i\frac{\sqrt{\widetilde \gamma}}{k\lambda}}
    {\Omega+1+2i\frac{\sqrt{\widetilde \gamma}}{k\lambda}},\label{eq:rpdl3}\\
    c_{n,s}\Pi^{(r)--}_{--}\kaomega &\approx \frac{(\lambda\xi')}
    {(\xi^B_{n,s})^2}\frac{1}{k\lambda}\frac{i}{2}
    \ln\frac{\Omega-1+2i\frac{\sqrt{2\widetilde \gamma}}{k\lambda}}
    {\Omega+1+2i\frac{\sqrt{2\widetilde \gamma}}{k\lambda}}.\label{eq:rpdl4}
  \end{align}
\end{subequations}
These results can be obtained with the Mittag-Leffler expansion of
their spectral functions similarly as for the scalar case \cite{SzK}.
The approximation of the spin wave polarization functions can be
obtained in a similar way, but due to their different symmetry
properties compared with density polarization functions another limit
is to be taken in the spectral function. The calculation is outlined
in the appendix.  The results are
\begin{subequations}
  \label{eqs:rpdlsw}
  \begin{equation}
    \label{eq:rpdlsw1}
    |c_s|\left(\Pi^{(r)0+}_{+0}+\Pi^{(r)-0}_{0-}\right) = \frac{C_- - C_+}
    {\widetilde \gamma}+ \left[\frac{2C_0 + C_+ + C_-}{{\widetilde \gamma}^2}
      + \frac{\Omega}{{\widetilde \gamma}^2}(C_- - C_+)+\frac{D_- - D_+}{
        {\widetilde \gamma}^3}\right](k\lambda)^2 + {\mathcal O}\left(
      (k\lambda)^4 \right),
  \end{equation}
with
  \begin{align}
    C_0&=\frac{\lambda\xi'}{(\xi^B_s)^2}\frac{2\Gamma\left(\frac{3}{2}\right)
      F\left(\frac{3}{2}\big|\widetilde\gamma\right)}{\pi},\label{C0def}\\
    C_+&=\frac{\lambda\xi'}{(\xi^B_s)^2}\frac{\Gamma\left(\frac{3}{2}\right)
      F\left(\frac{3}{2}\big|0\right)}{\pi},\label{C+def}\\
    C_-&=\frac{\lambda\xi'}{(\xi^B_s)^2}\frac{\Gamma\left(\frac{3}{2}\right)
      F\left(\frac{3}{2}\big|2\widetilde\gamma\right)}{\pi},\label{C-def}\\
    D_+&=\frac{\lambda\xi'}{(\xi^B_s)^2}\frac{4\Gamma\left(\frac{5}{2}\right)
      F\left(\frac{5}{2}\big|0\right)}{3\pi},\\
    D_-&=\frac{\lambda\xi'}{(\xi^B_s)^2}\frac{4\Gamma\left(\frac{5}{2}\right)
      F\left(\frac{5}{2}\big|2\widetilde\gamma\right)}{3\pi}
  \end{align}
\end{subequations}
and ${\widetilde \gamma}$ is given by \eqref{eq:spdl6}.

In the case of linear dispersion $\Omega \gg 1$ can be assumed (while
still satisfying the $|\Omega k \lambda| \ll 1$ condition). For the
ferromagnetic state we further restrict ourselves to $\sqrt{\widetilde
  \gamma} \ll |\Omega k\lambda|$. We will see that this later
condition is equivalent with $c_n/|c_s|\gg1$, which is fulfilled in
the applications. In this case the regular polarization functions with
zero spin transfer \eqref{eqs:rpdl} can be further approximated by
\begin{equation}
  \label{eq:rpdla}
    |c_{n,s}|\Pi^{(r)}\kaomega \approx -\frac{(\lambda\xi')}{(\xi^B_{n,s})^2}
    \frac{1}{k\lambda}\frac{i}{\Omega}
\end{equation}
both for the polar and ferromagnetic states. Since we are interested
only in the low momentum behaviour of the dispersion curve $\Omega$
will be searched as a power series of the $(k\lambda)$ variable.

{\it Polar case, density mode ($n=0$)}: The spectrum of these types of
excitations can be determined by the poles of \eqref{eq:rdcrp1}. The
corresponding equation is
\begin{equation}
  \label{eq:deqp1}
  1-3c_n \Pi_0-c_n\Pi_S=0.
\end{equation}
Using Eqs. \eqref{eq:spdl1} and \eqref{eq:rpdla} and substituting
$\Omega=a_{-1} (k\lambda)^{-1} + {\mathcal O}((k\lambda)^0)$, the
resulting equation for the $a_{-1}$ coefficient reads as:
\begin{equation}
  \label{eq:deqp2}
  a_{-1}^2+3 i \frac{\lambda\xi'}{(\xi^B_n)^2}a_{-1}-\left(\frac{\lambda}
    {\xi^B_n}\right)^2=0.
\end{equation}
Solving this quadratic equation (and assuming, that the imaginary part
is small) one arrives at
\begin{equation}
  \label{eq:drd1}
  \Omega=\pm\frac{1}{k\xi^B_n}-i\frac{3}{2}\frac{\xi'}{k(\xi^B_n)^2}.
\end{equation}
Returning to the more familiar variables one obtains
\begin{equation}
  \label{eq:drd2}
  \omega = \pm\sqrt{\frac{N_0c_n}{M}} k -i\frac{3c_n M}{4\pi\hslash^3\beta} k
\end{equation}
for the beginning of the dispersion curve.

{\it Polar case, spin density mode ($n=0$)}: The excitation spectrum is
obtained as the poles of \eqref{eq:rdcrp2}. The corresponding equation
is
\begin{equation}
  \label{eq:sdeqp1}
  1-2c_s \Pi_0=0.
\end{equation}
With the use of Eq. \eqref{eq:rpdla} the resulting equation for
$\Omega$ reads as:
\begin{equation}
  \label{eq:sdeqp2}
  1+2 i \frac{\xi'}{(\xi^B_s)^2}\frac{1}{\Omega k}=0.
\end{equation}
The solution is at
\begin{equation}
  \label{eq:sdrd1}
  \Omega=-i\frac{2\xi'}{k(\xi^B_s)^2},
\end{equation}
or equivalently at
\begin{equation}
  \label{eq:sdrd2}
  \omega = -i\frac{c_s M}{\pi\hslash^3\beta} k
\end{equation}
for the beginning of the dispersion curve. Note that such a mode did
not appear in the Bogoliubov approximation, because there $D_{zz}$ is
identically zero.

{\it Polar case, spin wave mode ($n=\pm1$)}: The poles of the
autocorrelation function \eqref{eq:rdcrp3} determine the spectrum of
these types of excitations. The corresponding equation is
\begin{equation}
  \label{eq:swqp1}
  1-2c_s \Pi_0-c_s\Pi_S=0.
\end{equation}
Using Eqs. \eqref{eq:spdl1} and \eqref{eq:rpdla} and substituting
$\Omega=a_{-1} (k\lambda)^{-1} + {\mathcal O}((k\lambda)^0)$, the
resulting equation for the $a_{-1}$ coefficient in this case reads as
\begin{equation}
  \label{eq:swqp2}
  a_{-1}^2+2 i \frac{\lambda\xi'}{(\xi^B_s)^2}a_{-1}-\left(\frac{\lambda}
    {\xi^B_s}\right)^2=0.
\end{equation}
Whence
\begin{equation}
  \label{eq:swrd1}
  \Omega=\pm\frac{1}{k\xi^B_s}-i\frac{\xi'}{k(\xi^B_s)^2},
\end{equation}
or equivalently
\begin{equation}
  \label{eq:swrd2}
  \omega = \pm\sqrt{\frac{N_0c_s}{M}} k -i\frac{c_s M}{2\pi\hslash^3\beta} k.
\end{equation}

{\it Ferromagnetic case, density, spin density mode ($n=0$)}: The
poles of the autocorrelation functions \eqref{eq:dcnrf1} or
\eqref{eq:dcnrf2} or \eqref{eq:dcnrf3} determine the spectrum of these
excitations, which is given by the equation
\begin{equation}
  \label{eq:deqf1}
  \left(1-c_n\Pi_{nn}\right)\left(1-c_s\Pi_{zz}\right)-
  c_nc_s\Pi_{nz}^2=0.
\end{equation}
Substituting the approximations \eqref{eq:spdl2} and \eqref{eq:rpdla}
and $\Omega = a_{-1} (k\lambda)^{-1}+{\mathcal O}( (k\lambda)^0)$, the
resulting equation for the $a_{-1}$ coefficient reads as
\begin{equation}
  \label{eq:deqf2}
  \alpha_n \alpha_s a_{-1}^3+ i \alpha(3 \alpha_s-2 \alpha_n)a_{-1}^2-
  (\alpha_s - \alpha_n - 6 \alpha^2)a_{-1}+i5\alpha=0,
\end{equation}
with $\alpha=\xi'/\lambda$, $\alpha_{n,s}=(\xi^B_{n,s}/\lambda)^2$.
In the discussed temperature region $\alpha_{s} \gg \alpha_{n} \gg
\alpha^2$. The solution of the equation can be calculated
perturbatively, with results:
\begin{equation}
  \label{eq:drdf1}
  \Omega=\pm\sqrt{\frac{\alpha_s-\alpha_n}{\alpha_n \alpha_s}}\frac{1}
  {k\lambda}-i\frac{\alpha}{2}\frac{3 \alpha_s^2 + 2 \alpha_n^2}
  {\alpha_n \alpha_s(\alpha_s-\alpha_n)}\frac{1}{k\lambda}
\end{equation}
from which the frequency
\begin{equation}
  \label{eq:drdf2}
  \omega=\pm\sqrt{\frac{N_0(c_n+c_s)}{M}}k-i\frac{M}{4\pi\hslash^3\beta}
  \frac{3c_n^2+2c_s^2}{c_n+c_s}k.
\end{equation}

We can see now, that the $\sqrt{\widetilde \gamma}\ll |\Omega k
\lambda|$ condition is equivalent with the $1 \ll 2 c_n/|c_s|$
condition, which is well satisfied.

{\it Ferromagnetic case, spin wave mode with $n=\pm1$}: The excitation
spectrum of the ferromagnetic spin waves is obtained from the poles of
the autocorrelation function \eqref{eq:dcnrf4}. The equation to be
solved is
\begin{equation}
  \label{eq:sweqf1}
  1-\frac{c_s}{2}\left[\Pi^{(r)0+}_{+0}+\Pi^{(r)-0}_{0-}+\Pi^{(s)0+}_{+0}
  \right]=0.
\end{equation}
Substituting the approximations Eq. \eqref{eq:spdl3} and Eq.
\eqref{eq:rpdlsw1} and that $\Omega = a_0 + {\mathcal O}( (k\lambda)^2
)$ leads to
\begin{equation}
  \label{eq:sweqf2}
  1-\frac{\gamma_0}{\widetilde \gamma}+\frac{C_- - C_+}{\widetilde \gamma}+
  \frac{(k\lambda)^2}{{\widetilde \gamma}^2}\left[(1-a_0)\gamma_0 + 2C_0 + 
    C_+ + C_- + \frac{a_0}{\widetilde \gamma}(C_- - C_+) + \frac{D_- - D_+}
    {\widetilde \gamma}\right] + {\mathcal O}\left( (k\lambda)^4\right) = 0
\end{equation}

The equation for ${\mathcal O}( (k\lambda)^0 )$ is an identity, since
${\widetilde \gamma} = \gamma_0 + \beta |c_s|(H^{++}-H^{--})$. The
$a_0$ constant can be determined from ${\mathcal O}( (k\lambda)^2 )$
order terms
\begin{equation}
  \label{eq:swsf1}
  a_0 = \frac{ {\widetilde \gamma}(2C_0 + C_+ + C_- + \gamma_0)-D_+ + D_-}
  {{\widetilde \gamma}^2}.
\end{equation}
The frequency of the spin wave excitation then is a quadratic one
\begin{subequations}
  \label{eqs:swef}
  \begin{equation}
    \label{eq:swef1}
    \omega=\frac{\hslash k^2}{2M^*},
  \end{equation}
  with the effective mass
  \begin{equation}
    \label{eq:swef2}
    M^* = \frac{M}{a_0}.
  \end{equation}
\end{subequations}
The Bogoliubov solution \eqref{eq:spbcpmf} can be regained in the
$T\rightarrow0$ limit, since $C_0,C_+,C_-,D_+,D_-\propto T$,
${\widetilde \gamma}$ and $\gamma_0$ tend to $N_0(T=0)|c_s|/k_B T$.

{\it Ferromagnetic case, quadrupolar spin wave mode}: Since the
autocorrelation function \eqref{eq:dcnbf5} is proper, so the spectrum
of the quadrupolar spin waves is determined by the poles of the
polarization function $\Pi^{-+}_{+-}$. The singular polarization
\eqref{eq:spqrf} has its pole at
\begin{equation}
  \label{eq:sqwf1}
  \Omega = 1+\frac{2{\widetilde \gamma}}{(k\lambda)^2}.
\end{equation}
The regular polarization function $\Pi^{(r)-+}_{+-}$ has a logarithmic
singularity in the same place. The corresponding excitational frequency is
\begin{equation}
  \label{eq:sqwf2}
  \omega=\frac{2{\mathcal M}|c_s|}{\hslash} + \frac{\hslash k^2}{2M}.
\end{equation}

Some final remark is appropriate about the validity of the calculation
of these last two modes. The approximation \eqref{eq:rpdlsw1} is valid
if $(k\lambda)^2/{\widetilde \gamma}\ll 1$. This is equivalent with
the $k\ll\sqrt{4M{\mathcal M}|c_s|}/\hslash$ condition, meaning that
the approximation for the bubble breaks down near the transition
temperature, where the magnetisation vanishes.

\section{Conclusions}
\label{sec:conc}

In the high temperature phase density and spin fluctuations do not
couple to each other and to the Green's functions. The latter appear
only in the intermediate states in the perturbational expansion for
the correlation functions describing density and spin fluctuations.
Correspondingly one can experience independent excitation branches of
one-particle and collective type. As discussed in the present paper in
detail collective and one-particle excitations can hybridize in the
Bose-Einstein condensed phase due to the symmetry breaking, which is
different in the polar and in the ferromagnetic phase resulting in
different couplings. In the polar phase, however such hybridization
does not occur for spin modes characterized by spin transfers zero and
$\pm2$. The general results have been demonstrated in the RPA scheme.
By treating the regular polarization contributions as perturbations,
we obtained damping for a number of modes determined first in the
Bogoliubov approximation \cite{Ho2,OM}. In the ferromagnetic phase for
the transverse spin mode, whose energy has been shown to agree with
the free particle kinetic energy in the Bogoliubov approximation, it
is found that the eigenfrequency remains proportional to $k^2$ in RPA,
but with an effective mass, which approaches the mass of the atoms in
the zero temperature limit. Moreover the gap in the quadrupolar spin
mode gets a temperature dependent correction, namely it is
proportional to the magnetization which tends to the condensate
density in the zero temperature limit and then the gap coincides with
that of the Bogoliubov approximation. These transverse and quadrupolar
spin modes are found free from damping.

There are thermal excitations for which the bubble graphs can not be
treated as perturbations as shown earlier in the scalar case
\cite{SzK,FRSzG}. In the case of a spinor condensate one expects even
a multitude of such excitations, already in RPA. Their investigation
will be the subject of a forthcoming paper.

The calculations in this paper has been made for a homogeneous system.
In experiments with alkali atoms the gas sample is confined in an
optical trap which can be modelled by a harmonic potential. The
inhomogeneous nature of the trapped system results in that the
algebraic equations presented here are to be changed to coupled
integral equations but their main structure remains the same. We note
also that experiments can be designed where the local speed of sound
can be measured directly making the results obtained for a homogeneous
system also experimentally relevant.

\section{Acknowledgements}

The present work has been partially supported by the Hungarian
Research National Foundation under grant No. OTKA T029552.

\appendix

\section{The approximations of the contributions of the bubble graphs}

In this appendix we briefly outline the analytical properties of the
contribution of the bubble graphs encountered and outline the
approximation \eqref{eq:rpdlsw1}. Some of these results are known from
earlier works \cite{SzK}.

In the random phase approximation the self-energies of the internal
lines are wavenumber and frequency independent, therefore the
contribution of the bubble graph \eqref{eq:bubbleg}, depicted in Fig.
\ref{fig:regpolrpa} can be cast to the form
\begin{equation}
  \label{eq:bubcont1}
  \Pi^{(r)sr}_{r's'}\komega=-\frac{1}{\hslash}\int\frac{d^3 q}{(2\pi)^3}
  \frac{n^0(\epsilon_{\vec{k}+\vec{q}}+\hslash{\widetilde \Sigma}^{s's})
    -n^0(\epsilon_{\vec{q}}+\hslash{\widetilde \Sigma}^{rr'})}{i\omega_n-
    \hslash^{-1}(\epsilon_{\vec{k}+\vec{q}}-\epsilon_{\vec{q}}) - \Delta
    {\widetilde \Sigma}},
\end{equation}
with
\begin{equation}
  \label{eq:delsi}
  \Delta{\widetilde \Sigma} = {\widetilde \Sigma}^{s's} - 
  {\widetilde \Sigma}^{rr'}.
\end{equation}
The nondiagonal elements of $\Pi^{(r)sr}_{r's'}$ are zero, as stated
below Eq. \eqref{eq:bubbleg}, the same is true for the proper
self-energies. In the appendix the automatic summation over repeated
indices do not apply.

Changing to retarded polarization functions with analytically
continuing in frequency, the imaginary part plays the role of the
spectral function:
\begin{equation}
  \label{eq:spfn1}
  \Pi^{(r)sr}_{rs}\kaomega=-\frac{1}{\pi}\int\frac{\Im
    \Pi^{(r)sr}_{rs}(\vec{k},\omega')}{\omega-\omega'}d\omega'.
\end{equation}
The imaginary part can be brought to the form
\begin{equation}
  \label{eq:spfn2}
  \Im\Pi^{(r)sr}_{rs}(\vec{k},\omega) = \frac{m^2}{4\pi\hslash^4 k \beta}
    \left[\int_\infty^{y_0} dz \frac{2 z}{e^{z^2+\gamma_b}-1}-
    \int_\infty^{y_1} dz \frac{2 z}{e^{z^2+\gamma_f}-1}\right],
\end{equation}
with
\begin{align}
  y_0 & = \frac{k\lambda}{2}\left(\Omega-1-\frac{\Delta\gamma}{(k\lambda)^2}
    \right),\label{eq:y0}\\
  y_1 & = \frac{k\lambda}{2}\left(\Omega+1-\frac{\Delta\gamma}{(k\lambda)^2}
  \right),\label{eq:y1}\\
  \gamma_f & = \hslash \beta {\widetilde \Sigma}^{ss},\\
  \gamma_b & = \hslash \beta {\widetilde \Sigma}^{rr},\\
  \Delta \gamma & = \gamma_f - \gamma_b = \hslash \beta \Delta {\widetilde
    \Sigma}
\end{align}
and $\Omega$ defined in Eq. \eqref{eq:dfr}. Introducing the function
\begin{equation}
  \label{eq:R}
  R_\gamma(z) = \int_{-\infty}^{\infty} \frac{2x}{e^{x^2+\gamma}-1}
  \frac{1}{x-z} dx, \quad \text{for }\Im z>0,
\end{equation}
which is to be continued to the whole complex plane, the function
$\Pi^{(r)sr}_{rs}(\vec{k},\omega)$ can be represented as
\begin{equation}
  \label{eq:spfn3}
  \Pi^{(r)sr}_{rs}(\vec{k},\omega) = \frac{m^2}{(4\pi)^2\hslash^4k\beta}
  \left[\int_\infty^{y_0} R_{\gamma_b}(z)dz - \int_\infty^{y_1}
    R_{\gamma_f}(z)dz\right], \quad \text{for }\Im \omega>0.
\end{equation}

The analytic structure of $R_\gamma(z)$ is important for the
approximations, it can be found in Ref. \cite{SzK}. If the
self-energies of the forward and backward propagating one-particle
lines are equal, i.e. in the polar case and for the $n=0$ mode in the
ferromagnetic case, $\Delta\gamma=0$ and the expression
\eqref{eq:spfn3} reduces that of Ref. \cite{SzK}. If the conditions
above Eq. \eqref{eqs:rpdl} are fulfilled, the leading contribution of
the Mittag-Leffler representation of $R_\gamma(z)$, reading as
\begin{equation}
  \label{eq:MLR}
  R_\gamma(z)=2\sqrt{\pi}\zeta({\scriptstyle \frac{1}{2}})-i\pi z-
  \frac{2\pi}{\sqrt{\gamma}}\frac{z}{z+i\sqrt{\gamma}}+4\pi i\sum_{n=1}^\infty
  \frac{1}{a_n^2-b_n^2}\left[\frac{a_n^2+\gamma}{z+a_n}-\frac{b_n^2+\gamma}{z+b_n}+(a_n-b_n)\left(
      \frac{\gamma}{a_nb_n}-1\right)\right],
\end{equation}
with
\begin{align}
a_n&=i r_n e^{i\frac{\varphi_n}{2}},& b_n&=i r_n e^{-i\frac{\varphi_n}{2}},\\  
r_n&=\sqrt[4]{\gamma^2+4n^2\pi^2}, & \varphi_n&=\arctan\frac{2n\pi}{\gamma},
\end{align}
can then be used to arrive at the approximating formulas Eqs.
\eqref{eqs:rpdl}, see Ref. \cite{SzK}.

If the self-energies of the forward and backward propagating
one-particle lines of the bubble graph are different, the upper limits
of the integral \eqref{eq:spfn3} will diverge when $k$ goes to zero,
see Eqs.  \eqref{eq:y0} and \eqref{eq:y1}. If one is interested in the
long wavelength dynamics the asymptotic series of $R_\gamma(z)$ can
then be used, namely
\begin{equation}
  \label{eq:Ras}
  R_\gamma(z) \sim -\frac{2}{z^2}\sum_n\frac{\Gamma\left(n+\frac{3}{2}\right)
    F\left(n+\frac{3}{2}|\gamma\right)}{z^{2n}},
\end{equation}
where $\Gamma(s)$ is the Gamma function and $F(s|\gamma)$ is the
Bose-Einstein integral \eqref{eq:Bei} \cite{SzK}. With the help of Eq.
\eqref{eq:spfn3} and Eq. \eqref{eq:Ras}, the approximation
\eqref{eq:rpdlsw1} can be obtained.

\end{document}